\documentclass[preprint]{aastex}

\shorttitle{Empirical optical k-Corrections for $z \le 0.7$}
\shortauthors{Westra et al.}

\providecommand{\includeIDLfigP}[2][\columnwidth]{\includegraphics[angle=90, width=#1, trim=6pt 16pt 18pt 15pt, clip]{#2}}
\providecommand{\includeIDLfigPcustom}[6][\columnwidth]{\includegraphics[angle=90, width=#1, trim=#2 #3 #4 #5, clip]{#6}}
\newcommand{\dn}{\ifmmode{D_{n}4000}\else{$D_{n}4000$}\fi}

\newcommand{\ha}{\ifmmode{\mathrm{H\alpha}}\else{H$\alpha$}\fi}
\newcommand{\hb}{\ifmmode{\mathrm{H\beta}}\else{H$\beta$}\fi}
\newcommand{\hc}{\ifmmode{\mathrm{H\gamma}}\else{H$\gamma$}\fi}
\newcommand{\hd}{\ifmmode{\mathrm{H\delta}}\else{H$\delta$}\fi}

\newcommand{\oii}{[{\sc Oii}]}
\newcommand{\oiii}{\ifmmode{\mbox{[{\sc O\,iii}]}}\else{[{\sc O\,iii}]}\fi}

\newcommand{\Oii}{\oii~$\lambda\lambda$3726,3728}
\newcommand{\Oiii}{\oiii~$\lambda\lambda$4959,5007}
\newcommand{\Oiiia}{\oiii~$\lambda$4959}
\newcommand{\Oiiib}{\oiii~$\lambda$5007}
\newcommand{\nii}{\ifmmode{\mbox{[{\sc N\,ii}]}}\else{[{\sc N\,ii}]}\fi}
\newcommand{\Nii}{\nii~$\lambda\lambda$6550,6585}

\newcommand{\sii}{[{\sc S\,ii}]}
\newcommand{\Sii}{\sii~$\lambda\lambda$6733,6718}

\newcommand{\kms}{\ifmmode{\mathrm{km\,s^{-1}}}\else{km\,s$^{-1}$}\fi}
\newcommand{\perMpcSq}{\ifmmode{\mathrm{Mpc^{-2}}}\else{Mpc$^{-2}$}\fi}
\newcommand{\perMpcQube}{\ifmmode{\mathrm{Mpc^{-3}}}\else{Mpc$^{-3}$}\fi}

\newcommand{\MpcQ}{\ifmmode{\mathrm{Mpc^{3}}}\else{Mpc$^{3}$}\fi}
\newcommand{\pc}{\ifmmode{\mathrm{pc}}\else{pc}\fi}
\newcommand{\uJy}{\ifmmode{\mathrm{\mu Jy}}\else{$\mu$Jy}\fi}
\newcommand{\Jy}{\ifmmode{\mathrm{Jy}}\else{Jy}\fi}
\newcommand{\Myr}{\ifmmode{\mathrm{Myr}}\else{Myr}\fi}
\newcommand{\Gyr}{\ifmmode{\mathrm{Gyr}}\else{Gyr}\fi}
\newcommand{\kmsMpc}{\ifmmode{\mathrm{km\,s^{-1}\,Mpc^{-1}}}\else{km\,s$^{-1}$\,Mpc$^{-1}$}\fi}
\newcommand{\ergs}{\ifmmode{\mathrm{erg\,s^{-1}}}\else{erg\,s$^{-1}$}\fi}

\newcommand{\fluxunits}{\ergs\,\ifmmode{\mathrm{cm^{-2}}}\else{cm$^{-2}$}\fi}
\newcommand{\ergsPerAng}{\fluxunits\,\ifmmode{\mathrm{\AA^{-1}}}\else{\AA$^{-1}$}\fi}
\newcommand{\ergsPerHz}{\fluxunits\,\ifmmode{\mathrm{Hz^{-1}}}\else{Hz$^{-1}$}\fi}
\newcommand{\Msunyr}{\ifmmode{\mathrm{M_\odot\,yr^{-1}}}\else{M$_\odot$\,yr$^{-1}$}\fi}
\newcommand{\lumDens}{\ergs\,\ifmmode{\mathrm{Mpc^{-3}}}\else{Mpc$^{-3}$}\fi}
\newcommand{\sfDens}{\Msunyr\,\ifmmode{\mathrm{Mpc^{-3}}}\else{Mpc$^{-3}$}\fi}

\providecommand{\pow}[2][10]{#1^{#2}}
\providecommand{\eqref}[1]{[\ref{#1}]}

\providecommand{\giCol}{\ifmmode(g-i)\else$(g-i)$\fi}
\providecommand{\grCol}{\ifmmode(g-r)\else$(g-r)$\fi}
\providecommand{\cb}{CB07}

\newenvironment{kxtable}[2]{\begin{deluxetable}{lcccc}\tablewidth{0pt}\tablecaption{#1}\tablehead{& $#2^0$ & $#2^1$ & $#2^2$ & $#2^3$}\startdata}{\enddata\end{deluxetable}}{}

\begin{document}

\defcitealias{Chilingarian10}{CMZ10}

\title{Empirical optical k-corrections for redshifts $\le$ 0.7}

\author{Eduard Westra\altaffilmark{1},
  Margaret~J.~Geller\altaffilmark{1},
  Michael~J.~Kurtz\altaffilmark{1},
  Daniel~G.~Fabricant\altaffilmark{1},
  Ian~Dell'Antonio\altaffilmark{2}}

\altaffiltext{1}{Smithsonian Astrophysical Observatory, 60 Garden
Street, Cambridge, MA 02138, USA}

\altaffiltext{2}{Brown University, Department of Physics, Box 1843,
Providence, RI 02912, USA}

\email{ewestra@cfa.harvard.edu}

\begin{abstract}
The Smithsonian Hectospec Lensing Survey (SHELS) is a
magnitude-limited spectroscopically complete survey for $R \le 21.0$
covering 4\,\sq{}\degr{}. SHELS provides a large sample (15,513) of
flux-calibrated spectra. The wavelength range covered by the spectra
allows empirical determination of k-corrections for the $g$ and $r$
bands from $z = 0$ to $\sim 0.68$ and 0.33, respectively, based on
large samples of spectra. We approximate the k-corrections using only
two parameters in a standard way: \dn{} and redshift, $z$. We use
\dn{} rather than the standard observed galaxy color because \dn{} is
a redshift-independent tracer of the stellar population of the
galaxy. Our approximations for the k-corrections using \dn{} are as
good as those based on observed galaxy color \grCol{} ($\sigma$ of the
scatter is $\sim0.08$\,mag). The approximations for the k-corrections
are available in an online calculator. Our results agree with
previously determined analytical approximations from single stellar
population (SSP) models fitted to multiband optical and near-infrared
photometry for galaxies with a known redshift. Galaxies with the
smallest \dn{}---the galaxies with the youngest stellar
populations---are always attenuated and/or contain contributions from
older stellar populations. We use simple single SSP fits to the SHELS
spectra to study the influence of emission lines on the
k-correction. The effects of emission lines can be ignored for
rest-frame equivalent widths (REWs) $\lesssim 100$\,\AA{} depending on
required photometric accuracy. We also provide analytic approximations
to the k-corrections determined from our model fits for $z \le 0.7$ as
a function of redshift and \dn{} for $ugriz$ and $UBVRI$ ($\sigma$ of
the scatter is typically $\sim0.10$\,mag, the root-mean-square
typically $\sim0.15$\,mag). Again, the approximations using \dn{} are
as good those based on a suitably chosen observed galaxy color. We
provide all analytical approximations in an online calculator.
\end{abstract}

\keywords{Galaxies}

\section{Introduction}
\label{sec:intro}

The expansion of the Universe shifts the spectrum of galaxies redward
with respect to the observer. Observing galaxies at different
cosmological distances in a fixed bandpass thus leads to a different
absolute magnitude in the bandpass after correcting for luminosity
distance. To compare the photometric properties of galaxies at
different redshifts, one needs to bring the photometry of these
galaxies onto a common system. The correction that places the absolute
magnitude on a common system is the k-correction
\citep[after][]{Hubble36}. The k-correction is usually defined with
respect to photometric observations of a galaxy in its rest frame,
i.e., $z=0$.\footnote{To remove large systematic uncertainties from the
k-correction, sometimes the choice for a non-rest frame is made
\citep[e.g.,][]{Blanton07}.} Accurate k-corrections have increasing
importance as cosmological observations become more precise. Here, we
provide k-corrections based on \dn{}---a spectroscopic indicator that
is a measurement of the strength of the 4000\,\AA{}, or
Balmer break---and redshift accurate to 0.1\,mag with negligible
systematic errors.

The k-correction $K_{QR}$ is the correction in magnitudes between the
observed magnitude $m_R$ in the filter $R$ and true rest-frame absolute
magnitude $M_Q$ in the filter $Q$ at distance
modulus $\mathrm{DM}(z)$ \citep[e.g.,][]{Hogg02,Blanton07}:
\begin{equation}
  m_R = M_Q + \mathrm{DM}(z) + K_{QR}(z),
  \label{eq:kcor}
\end{equation}
where $\mathrm{DM}(z) = 5 \log (d_L/10\,\pc) - 5$ and $d_L$ is the
luminosity distance. We consider the case where filter $Q$ is equal to
filter $R$.

To determine the k-correction of a galaxy, one approach is to model
the spectral energy distribution (SED) of the galaxy over the
wavelengths necessary to determine the k-correction. The quality of
the k-correction is then limited by the adequacy of the models; when
using photometric SEDs rather than spectra, the quality is limited by
the number (and width) of the photometric bands used. A second
approach is direct determination of the k-correction from the observed
spectrum of the galaxy. Using observed spectra for calculating the
k-corrections removes the need for model assumptions and empirically
addresses all of the physics required to determine the SED of a
galaxy.

Recently, \citet[][hereafter
\citetalias{Chilingarian10}]{Chilingarian10} approximated the
k-corrections for nine optical and near-infrared filters ($ugrizYJHK$)
with analytical functions of only two parameters: redshift and
observed galaxy color. They used the measured redshift to fit a set of
SSP models based on {\sc pegase.2}
\citep{Fioc97} to the photometry in the nine bands. From the fitted
models, they derived k-corrections for the filter set.

Here, we discuss k-corrections determined directly from spectra from
the Smithsonian Hectospec Lensing Survey. SHELS is a
spectroscopic survey covering 4\,\sq{}\degr{} on the sky to a limiting
magnitude $R = 21.0$ \citep[][Kurtz et al. 2011, in
preparation]{Geller05}. Instead of observed galaxy color, we use the
\dn{} of the galaxy together with redshift to parametrize the
k-correction.

\dn{} is the ratio of average flux red- and blueward of the so-called
4000\,\AA{} break, or Balmer break
\begin{equation}
  \dn = \frac{
    \int _{4000} ^{4100} f_\lambda d\lambda}{
    \int _{3850} ^{3950} f_\lambda d\lambda},
  \label{eq:d4000}
\end{equation}
where $f_\lambda$ is the rest-frame flux density of the galaxy and
each wavelength is the rest-frame wavelength. We indicate the regions
where we measure \dn{} (Fig.~\ref{fig:cb07models}) with respect to the
Charlot \& Bruzual (private communication 2010) (hereafter \cb{})
models at three different redshifts. We also show the $g$ and $r$ band
throughput curves from the SDSS Web site\footnote{See
\url{http://www.sdss.org/dr7/instruments/imager/index.html}.}.

The wavelength range covered by the Hectospec spectra determines the
choice of filters for which we can determine the k-corrections. We can
determine k-corrections only for bandpasses that satisfy the
``Goldilocks principle'': if the bandpass is too blue, we can
determine the rest-frame magnitude, not the observed-frame magnitude;
if the bandpass is too red, we can determine the observed-frame
magnitude, not the rest-frame magnitude. We need both observed- and
rest-frame magnitudes to determine the k-correction. Two filters that
satisfy this principle are the SDSS $g$ and $r$ bands. We derive
these k-corrections (and analytic approximations) in
Section~\ref{sec:fitting}.

We discuss the SHELS spectroscopic data in
Section~\ref{sec:thedata}. In Section~\ref{sec:method}, we describe
the methods used to calculate the k-corrections, the sample selection,
and the use of \dn{} as an indicator for galaxy type. We present the
k-corrections in Section~\ref{sec:fitting}. We compare our results
with our SSP fits based on \cb{} to the previous work of
\citetalias{Chilingarian10} in Section~\ref{sec:chilcomp}. We discuss
the accuracy of our \dn{} in Section~\ref{sec:unc}. In
Section~\ref{sec:el} we discuss the impact of emission lines on the
k-correction. We provide the k-corrections for SDSS bandpasses $ugriz$
and the Johnson-Cousins $UBVRI$ derived from our model fits to the
spectra in Section~\ref{sec:modkcor}. We summarize our results in
Section~\ref{sec:summary}.

Throughout this article we assume a flat universe with $H_0 =
71$\,\kmsMpc{}, $\Omega_{\rm m} = 0.27$, and $\Omega_{\Lambda} =
0.73$. All quoted magnitudes are on the AB-system \citep{Oke83}.

\section{The data}
\label{sec:thedata}
Here, we describe the observations of SHELS. We first describe the
imaging and spectroscopic data. We then address the effects of fixed
solid-angle aperture sizes on our measurements. We conclude with the
selection criteria of our sample of spectra from which we determine
our k-corrections.

\subsection{The observations}
\label{sec:fieldselection}

We draw our data from an expanded SHELS catalog \citep[cf.][who use a
previous version of the
catalog]{Fabricant05,Geller05,Kurtz07,Geller10,Westra10,Woods10}. The
main differences with the prior catalog are: (1) we select fainter
targets, because the survey now reaches $R = 21.0$, instead of $R = 20.3$;
(2) we have more redshifts available (15,513 unique redshifts compared
with 11,447 previously); and (3) we have updated the photometry. We
next summarize the galaxy catalog construction. We discuss the details
in a forthcoming article (Kurtz et al. 2010, in preparation).

We constructed the SHELS galaxy catalog from the $R$-band source list
for the F2 field of the Deep Lens Survey \citep{Wittman02,Wittman06}.
The imaging of F2 is extremely deep; the imaging has a 1\,$\sigma$
surface brightness limit of 28.7\,mag/$\square\arcsec$ in $R$. The
effective exposure time with the MOSAIC I imager \citep{Muller98} on
the Kitt Peak National Observatory Mayall 4\,m telescope is about
14,500\,s. All exposures are observed in less than $0\farcs9$ seeing between
1999 November and 2004 November. \citet{Wittman06} describe the
reduction pipeline.  The 4.2\,\sq{}\degr{} F2 field is centered at
$\alpha = 09^h19^m32\fs4$ and $\delta =
+30\degr{}00\arcmin{}00\arcsec{}$. We exclude regions around bright
stars ($\sim$\,5\,\% of the total survey), resulting in an effective
area of 4.0\,\sq{}\degr{}. We use surface brightness and magnitude in
combination with information from the latest SDSS data release
\citep[DR7;][]{SDSS7} to separate stars from galaxies. This selection
removes some active galactic nuclei (AGNs). The final catalog contains
20,116 galaxies with a total $R$-band magnitude of $\leq$ 21.0.

We acquired spectra for the galaxies with the Hectospec fiber-fed
spectrograph \citep{Fabricant98,Fabricant05} on the MMT on various
nights during the period from 2004 April 13 through 2009 December
15. The Hectospec observation planning software \citep{Roll98} enables
efficient acquisition of a magnitude-limited sample.

The SHELS spectra cover the wavelength range of $\lambda = 3700 -
9150$\,\AA{} with a resolution of $\sim$6\,\AA{}. Exposure times
ranged from 0.75\,hr to 2\,hr for the lowest surface brightness
objects in the survey. We reduced the data with the standard Hectospec
pipeline \citep{Mink07} and derived redshifts with RVSAO
\citep{Kurtz98} with templates constructed for this purpose
\citep{Fabricant05}. The 1468 unique pairs of repeat observations
imply a mean internal error of 56\,\kms{} for absorption-line objects
and 21\,\kms{} for emission-line objects \citep[see
also][]{Fabricant05}.

We photometrically calibrate the spectra by scaling the flux density
obtained by the 1\farcs5 diameter fiber to the total $R$
band. \citet{Fabricant08} describe the technique used for the
Hectospec spectra in detail. We can photometrically calibrate the data
from this fiber-fed spectrograph because it has a particularly stable
instrument response.  This method assumes that the inner 1\farcs5 of a
galaxy is representative for the entire galaxy. See
Section~\ref{sec:apeffect} for a more detailed discussion of aperture
effects.

SHELS includes 20,116 galaxies to the limiting apparent magnitude. The
integral completeness of the redshift survey is 96.5\,\%, 93.8\,\%,
and 77.3\,\% to $R = 20.3$, 20.6, and 21.0,
respectively. Figure~\ref{fig:completeness} shows the integral and
differential completeness of SHELS as a function of total magnitude.

\subsection{Aperture effects}
\label{sec:apeffect}

The Hectospec fibers subtend 1\farcs5 on the sky, sampling only the
inner region of a galaxy. For a given galaxy, the fraction of light
contained in a fixed angular aperture increases with increasing
redshift. \citet{Kewley05} study the influence of fixed aperture size
on the determination of star formation rate, metallicity, and
attenuation in a galaxy. Their conclusion is that when the aperture
contains more than 20\,\% of the light, the central region is a good
representation of the entire galaxy. Like \citet{Woods10}, we use the
$R$-band data to compute the fraction of galaxies where at least
20\,\% of the light is contained in the Hectospec fiber aperture. The
majority of our galaxies fulfill this requirement
(Fig.~\ref{fig:lightfrac}).

\citet{Fabricant05} find no difference between the \dn{} determined
from the SDSS spectra and the \dn{} from the Hectospec spectra (their
Fig.~12). The median redshift of the SDSS is $z\sim0.1$: at this
redshift, at least 20\,\% of the light of a galaxy is captured in an
SDSS fiber (Fig.~\ref{fig:lightfrac}).

Galaxies with less than 20\,\% of their light captured by the
Hectospec fibers are predominantly at low redshifts. A small
difference in color between the nuclear region and the entire galaxy
only leads to a small uncertainty in the k-correction, due to the low
redshift; at low redshifts, the 4000\,\AA{}-break has not yet moved out
of the filter (Fig.~\ref{fig:cb07models}), resulting only in small
k-corrections.

We also assume that the galaxy color measured in a fixed aperture on
the sky is a good representation of the galaxy. The color obviously
varies with aperture in a galaxy with a bulge and an extended disk. To
investigate the effect of our simplifying assumption, we compare
\grCol{} for three different apertures. Figure~\ref{fig:colcomp} shows
the comparison between \grCol{} derived from the SDSS fiber spectra
covering the inner 3\arcsec{} of a galaxy and from the Hectospec fiber
spectra subtending the inner 1\farcs5 of a galaxy. The figure also
shows a comparison between the \grCol{} from the SDSS Petrosian
magnitudes, which include nearly all the galaxy light
\citep{Blanton01,Yasuda01}, and \grCol{} determined from the Hectospec
fiber spectra. The larger scatter in the lower diagram reflects the
large uncertainties in the Petrosian magnitudes.

We find no significant trend for \grCol{} within different apertures
as a function of redshift. There is a slight offset (median of
$\grCol_\mathrm{fiber,SDSS} - \grCol_\mathrm{Hecto} = -0.02$ and
$\grCol_\mathrm{petro,SDSS} - \grCol_\mathrm{Hecto} = -0.05$), smaller
than the median uncertainty for each of the individual SDSS $g$ and
$r$ magnitudes. This offset occurs because the outer parts of a galaxy
is bluer than the bulge \citep[see also][]{Kauffmann03}. The effect on
the determination of the k-corrections is negligible.

We conclude that \dn{} and \grCol{} determined from a Hectospec
spectrum are reasonable representations of the \dn{} and \grCol{}
obtained from a spectrum that covers the entire galaxy.

\subsection{Sample selection}
\label{sec:selection}

To derive the best k-corrections, we select a subsample of spectra
from the sample of 15,513 galaxies with unique redshifts. We retain
the spectra of galaxies satisfying all of the following criteria:
\begin{itemize}
\item The spectrum was observed after 2004 October 30. Until this
date, the atmospheric dispersion compensation prisms of Hectospec
rotated in the wrong direction \citep{Fabricant08}. The spectra before
this date have improper spectrophotometric calibration. However, the
redshifts for these galaxies are not affected.
\item The model fit (Section~\ref{sec:modelfitting}) succeeds and the
  reduced $\chi^2$ of the fit is less than 10. This restriction mostly
  removes strong broad lined AGN, very strong emission-line galaxies,
  and spectra with bad sky-subtraction.
\item The median of the signal-to-noise ratio (S/N) per pixel for ten
  regions free of night-sky emission lines exceeds 2.
\item The redshift of the galaxy allows the determination of both the
  observed- and rest-frame magnitude.
\end{itemize}
After applying these selection criteria, we have samples of 11,707 and
5993 spectra to calculate the k-corrections for $g$ and $r$,
respectively.

\section{Method}
\label{sec:method}
Here, we describe the method for determining the k-corrections from our
spectra. We first outline the formulae used to derive the
magnitudes. We also describe the fitting of SSP models to our
spectra. Finally, we discuss the advantages of using \dn{} rather than
observed color as an identifying characteristic of a galaxy.

\subsection{Synthetic magnitudes}
\label{sec:magdet}

We use flux-calibrated spectra (and SSP model fits;
Section~\ref{sec:modelfitting}) to calculate the observed- and
rest-frame magnitudes required for the k-corrections. We compute the
effective flux-density $F_\lambda$ from the observed spectrum for a
particular bandpass
\begin{equation}
  F_\lambda = \frac{\int f(\lambda) \lambda T(\lambda) d\lambda}{\int \lambda T(\lambda) d\lambda},
\end{equation}
where $f(\lambda)$ is the flux density as a function of wavelength
$\lambda$ in \AA{}, i.e. the flux-calibrated spectrum in units of
\ergsPerAng{}, and $T(\lambda)$ is the filter transmission curve per
photon with wavelength.\footnote{\citet{Blanton07} note that many
authors give the transmission curve as the contribution to the
detector signal per unit of energy ($T'(\lambda)$), rather than per
photon ($T(\lambda)$). They relate to each other as $T(\lambda)
\propto T'(\lambda)/\lambda$. However, accidentally using
$T'(\lambda)$ as $T(\lambda)$ has a very small impact (at the very
worst 0.07\,mag, and typically 0.02\,mag or less for this survey).}

We convert $F_\lambda$ into the effective flux-density as a function
of frequency $F_\nu$:
\begin{equation}
  F_\nu = \frac{3.34\times\pow{-19} F_\lambda}{\lambda_\mathrm{eff}^2},
\end{equation}
where $F_\nu$ is in \ergsPerHz{}, $F_\lambda$ in \ergsPerAng{}, and
$\lambda_\mathrm{eff} = \int \lambda T(\lambda) d\lambda/\int
T(\lambda) d\lambda$ is in angstrom. We calculate the artificial magnitude
as
\begin{equation}
  m = -2.5 \log \frac{F_\nu}{3631\,\Jy},
\end{equation}
where 3631\,\Jy{} is the zeropoint of the AB-magnitude system
\citep{Oke83}.

Throughout this article, unless otherwise noted, we determine the
k-corrections from synthetic observer and rest-frame magnitudes.
We obtain the observed colors of the galaxies from the SDSS fiber
magnitudes; when these magnitudes are unavailable, we determine them
from our synthetic magnitudes.

\subsection{Model fitting}
\label{sec:modelfitting}

We use the method of \citet{Tremonti04} to fit a linear combination of
10 SSP models to each spectrum. The SSP models are from \cb{}. These
models include an improved treatment of the thermal pulse asymptotic
giant branch phase with respect to the models of \citet{Bruzual03}
\citep[see][for more details]{Bruzual07}. The 10 different ages for
the populations are 0.005, 0.025, 0.1, 0.25, 0.5, 1, 1.4, 2.5, 5, and
10\,\Gyr{}. The models have solar
metallicity. Figure~\ref{fig:cb07models} shows the individual model
spectra normalized to 1 at 5500\,\AA{} together with the SDSS $g$ and
$r$ bands. We use the \citet{Calzetti00} extinction law rather than
the \citet{Charlot00} extinction law used by \citet{Tremonti04},
but the difference is small (10\,\% at the extremes of the wavelength
range 4000\,\AA{} -- 10,000\,\AA{}).

We fit the models to the wavelength region between 4000 and
8500\,\AA{}; we use the fits to determine \dn{}
(Section~\ref{sec:d_vs_c}) and remove residuals of night-sky features
in our spectra (Section~\ref{sec:fitting}). The cut at the blue end of
the spectrum is due to the uncertain calibration of Hectospec in that
region. Beyond 8500\,\AA{} light emitted from the home sensors in the
fiber positioner of the spectrograph affects the spectrum. Because
this effect is additive, we cannot remove this problem.

Figure~\ref{fig:modeltracks} shows the predicted k-corrections as a
function of redshift for each model without attenuation. We include
two models with attenuation to show its effect on the k-correction. We
apply a typical reddening for star-forming galaxies to the model with
the youngest stellar population ($A_V = 1$) and a reasonably extreme
reddening for early-type galaxies to the model with the oldest stellar
population ($A_V = 0.5$). Figure~\ref{fig:modeltracks} shows
approximately the full range of expected k-corrections modulo the
effects of emission lines (see Section~\ref{sec:el}).

\subsection{\dn{} versus color}
\label{sec:d_vs_c}

We use \dn{} as a measure of galaxy type because it is an indicator of
the age of the stellar population \citep[see
Table~\ref{tab:d4000models};][]{Balogh99,Bruzual83}. \citet{Woods10}
examine the fraction of galaxies with emission lines as a function of
\dn{} (they remove AGNs from their sample). Galaxies with a low \dn{}
invariably show emission lines; the fraction of galaxies with emission
lines decreases rapidly with increasing \dn{}. \citet{Kauffmann03}
show a very clear relation between \dn{} and H$\delta_A$ (a measure
for the REW of \hd{}). \citeauthor{Kauffmann03} also show clear
evolution in both \dn{} and H$\delta_A$ after an instantaneous burst
of star formation as a function of time.

Figure~\ref{fig:woods} shows the distribution of \dn{} for our sample,
along with the fraction of galaxies with at least one emission
line. We classify a galaxy as having an emission line when the
absorption-corrected REW is at least 5\,\AA{} for any of the following
emission lines: \ha{}, \hb{}, \Oiiia{}, \Oiiib{}, or \Oii{}. Our
distribution is double-peaked with a local minimum around $\dn = 1.46$
({\it thick dashed line}), very close to the \citet{Woods10} value of
1.44 ({\it dotted line}).

Almost all galaxies with a low \dn{} ($\lesssim 1.2$) show emission
lines indicating an episode of active (or very recent, roughly less
than 10\,\Myr{} ago) star formation, yielding a spectrum dominated by
young, hot stars. The fraction of galaxies with emission lines
gradually decreases with increasing \dn{}. For galaxies with a large
4000\,\AA{}-break, an old stellar population dominates the
spectrum. In our sample, toward the highest \dn{} the number of
galaxies with emission lines increases slightly because of the
presence of AGNs.

\begin{figure}
  \centering
  \includeIDLfigP{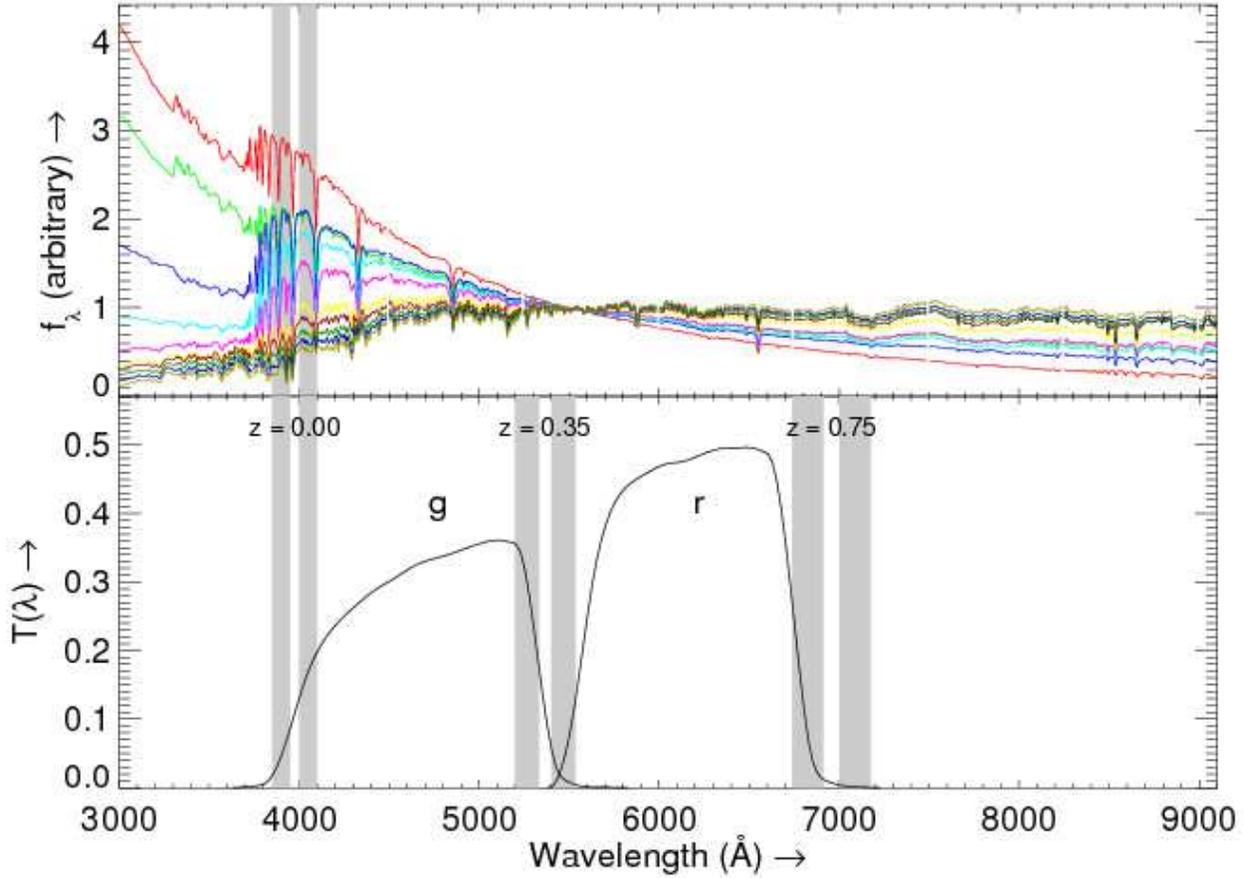}
  \caption{The 10 \cb{} SSP models scaled to unity at 5500\,\AA{}
  ({\it top}). The ages for the population in each model from top to
  bottom at the blue side of the spectrum are: 5\,\Myr, 25\,\Myr,
  100\,\Myr, 250\,\Myr, 500\,\Myr, 1.0\,\Gyr, 1.4\,\Gyr, 2.5\,\Gyr,
  5.0\,\Gyr, and 10.0\,\Gyr. The SDSS $g$ and $r$ bands and the
  location of the 4000\,\AA{}-break with respect to these filters for
  different redshifts ({\it bottom}). The shaded regions indicate the
  observed wavelength ranges where we determine \dn{} for three
  different redshifts. See eq.~\eqref{eq:d4000} for the definition of
  \dn.}
  \label{fig:cb07models}
\end{figure}

\begin{figure}
  \centering
  \includeIDLfigP{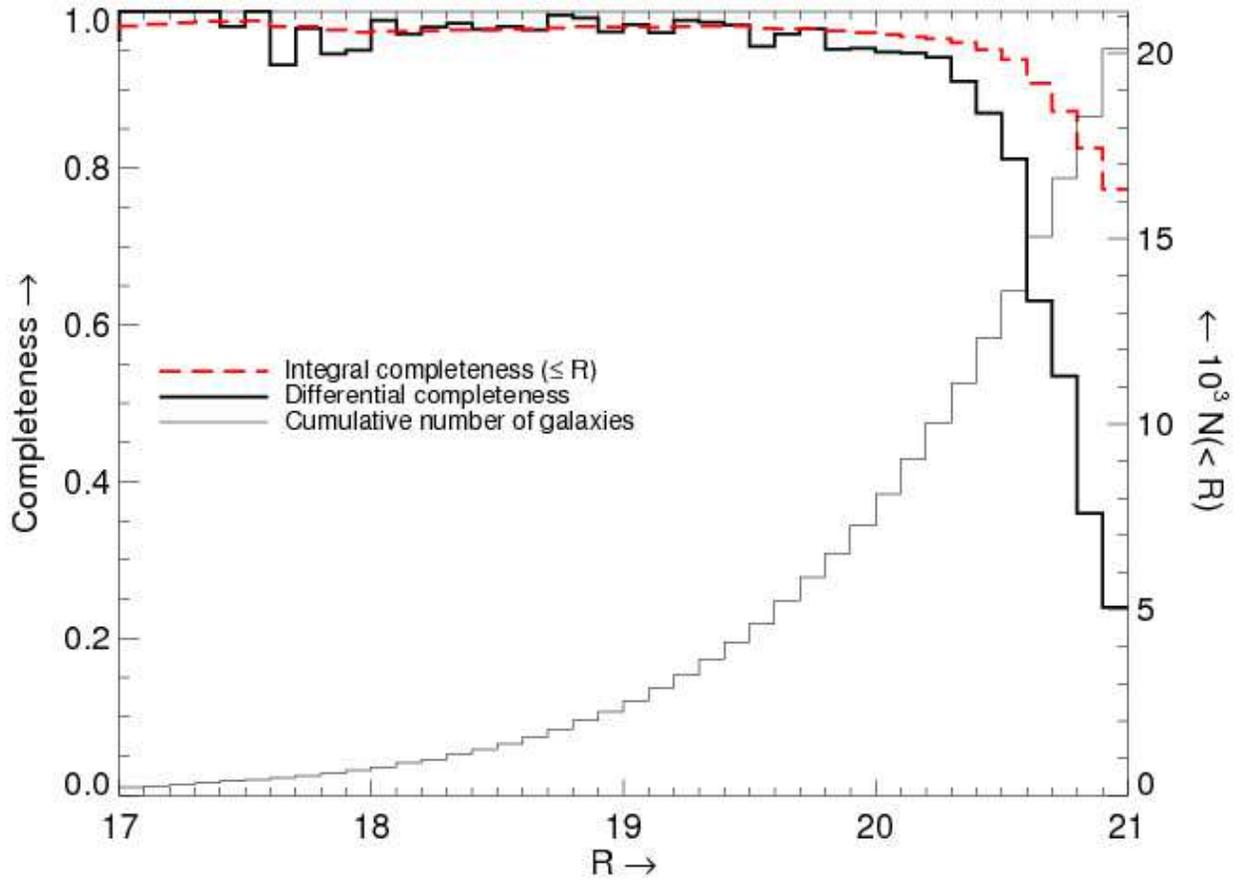}
  \caption{Integral and differential completeness of SHELS as a
  function of total $R$-band magnitude ({\it left axis}), and
  the cumulative number of galaxies with increasing magnitude ({\it
  right axis}).}
  \label{fig:completeness}
\end{figure}

\begin{figure}
  \centering
  \includeIDLfigP{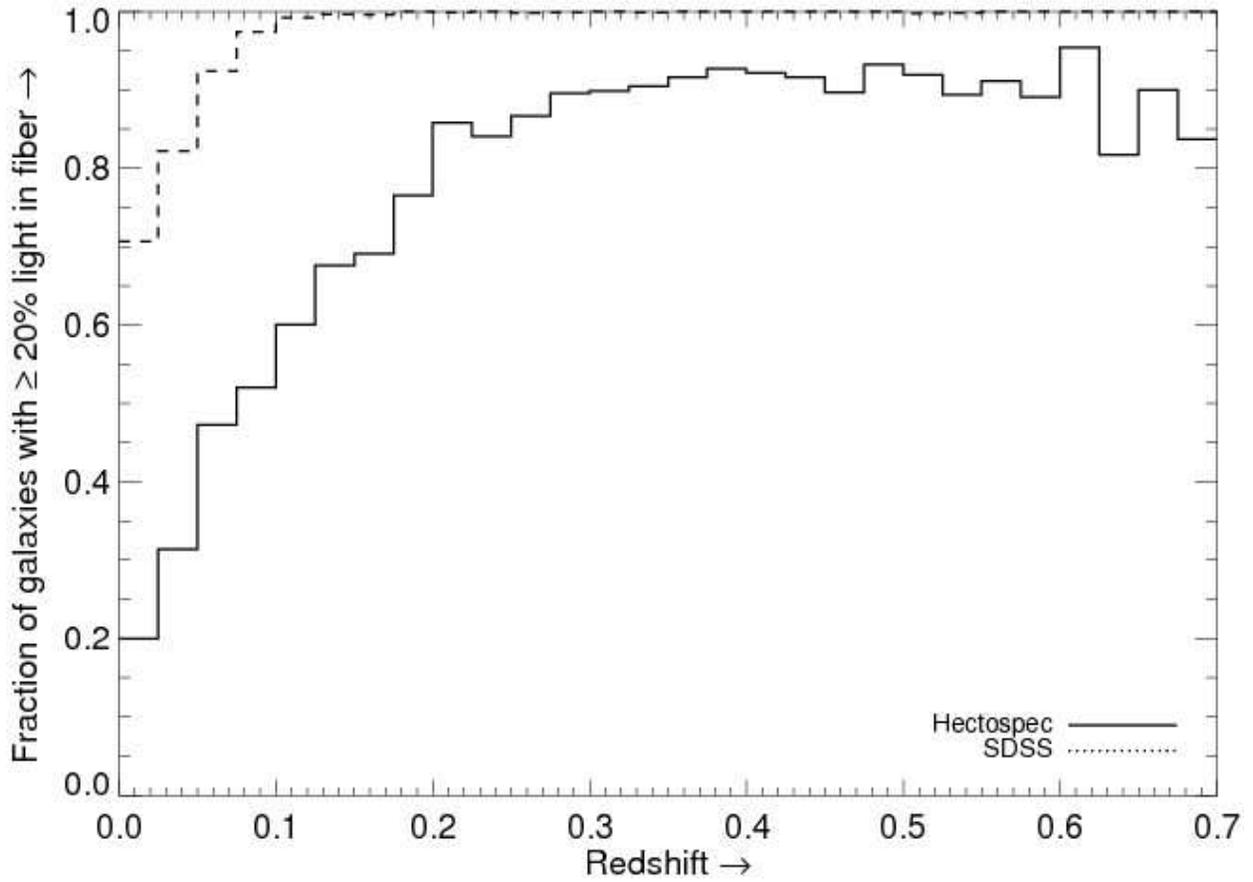}
  \caption{Fraction of galaxies where the apertures of the Hectospec
  fibers ($d = 1\farcs5$; {\it solid line}) and SDSS fibers ($d =
  3\arcsec$; {\it dashed line}) contain 20\,\% of the total light
  calculated from the $R$ band photometry.}
  \label{fig:lightfrac}
\end{figure}

\begin{figure}
  \centering
  \includeIDLfigPcustom{28pt}{12pt}{30pt}{25pt}{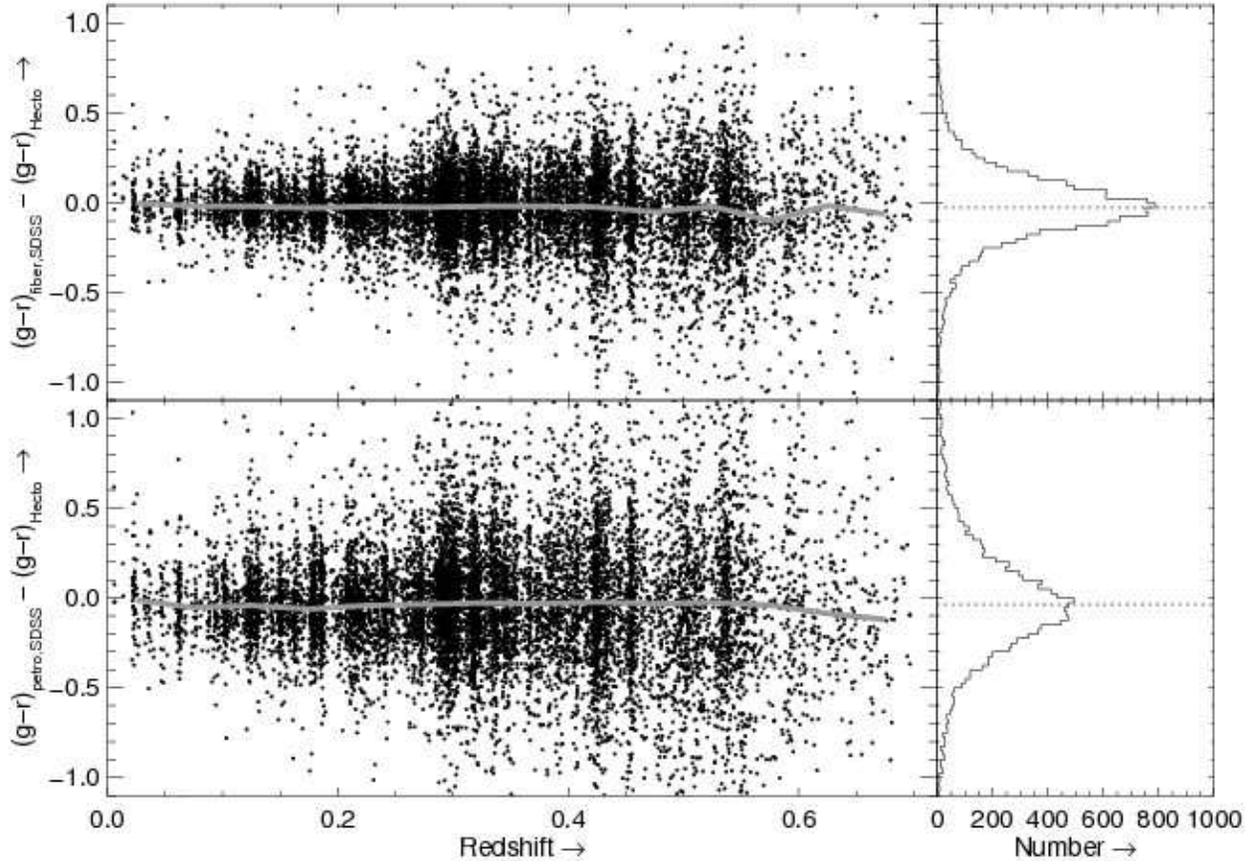}
  \caption{Comparison of \grCol{} for different sized apertures. ({\it
  Top}) SDSS 3\arcsec{} fiber \grCol{} and synthetic Hectospec
  1\farcs5 fiber \grCol{}. ({\it Bttom}) SDSS Petrosian \grCol{} and
  synthetic Hectospec 1\farcs5 fiber \grCol{}. The solid line traces
  the median of the differences as a function of redshift. A histogram
  shows the color difference ({\it right}). The dotted line indicates
  the median of the entire sample.}
  \label{fig:colcomp}
\end{figure}

\begin{figure}
  \centering
  \includeIDLfigP[0.495\textwidth]{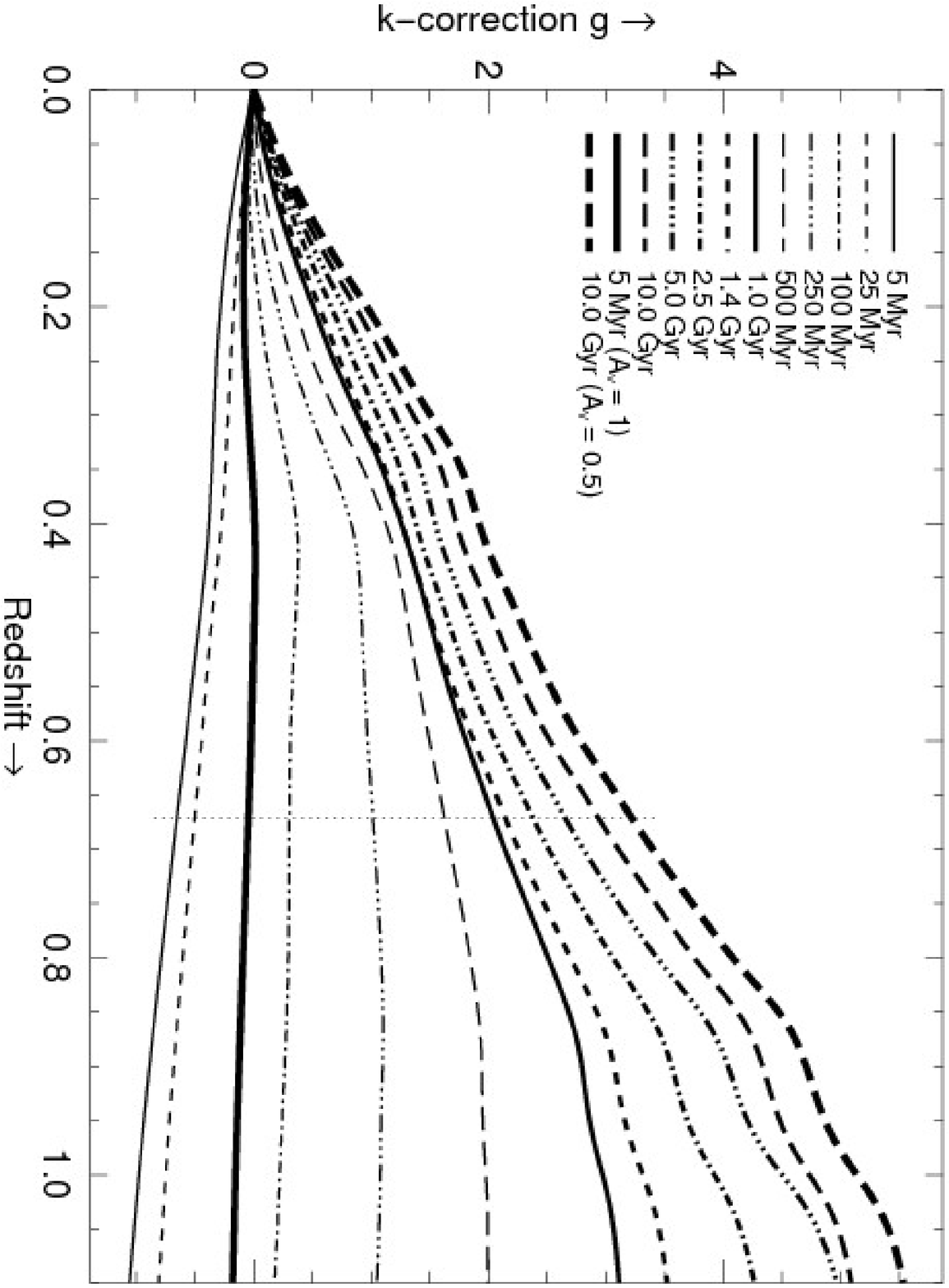}
  \includeIDLfigP[0.495\textwidth]{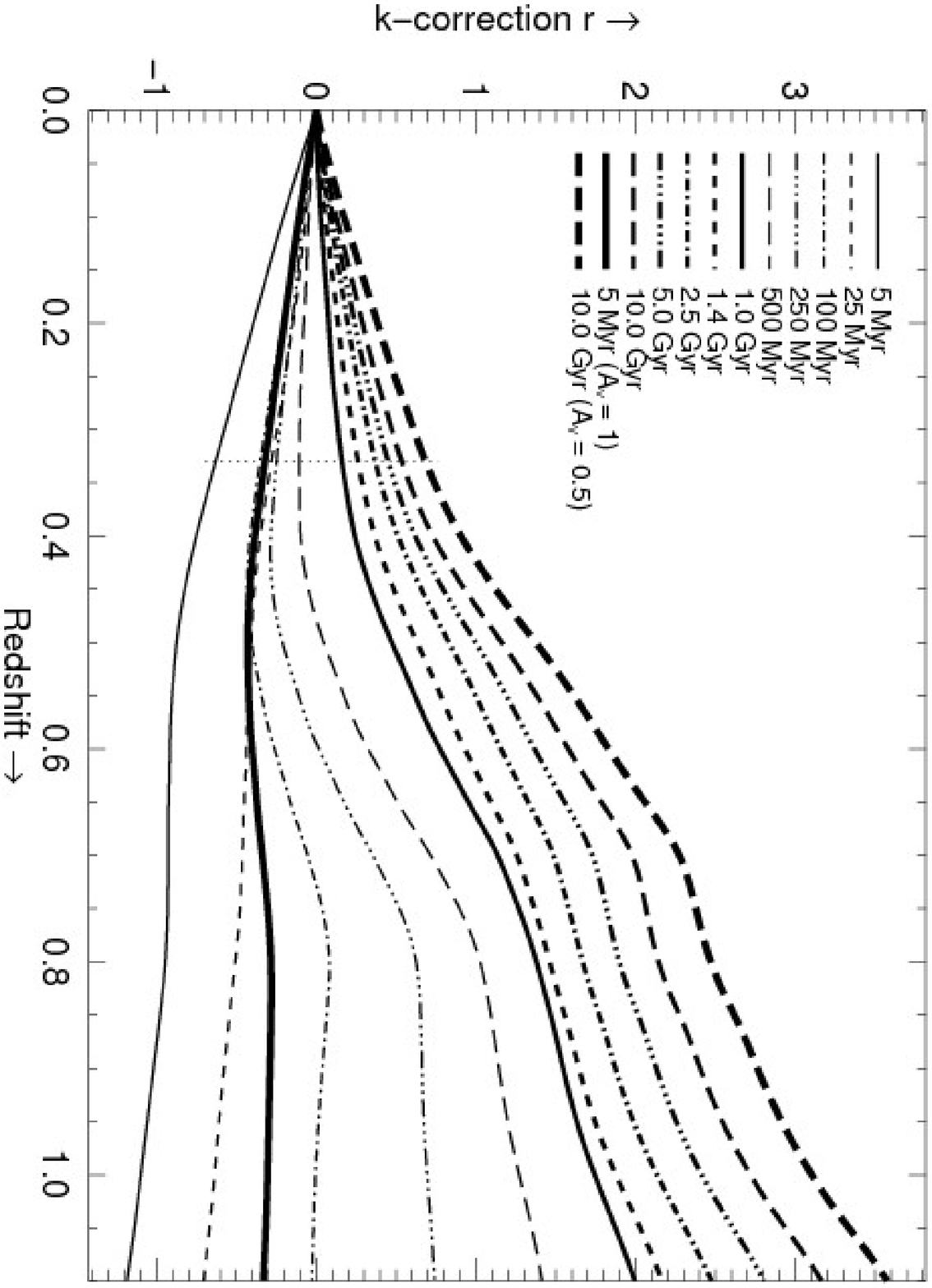}
  \caption{Predictions of the k-corrections for $g$ ({\it left}) and
  $r$ ({\it right}) as a function of redshift based on the 10 basic
  \cb{} SSP models and two of these models (those with the youngest
  and oldest population) with dust attenuation. The vertical dotted
  lines indicate the redshift up to which we can determine the
  k-correction from our spectra for these two filters.}
  \label{fig:modeltracks}
\end{figure}

\begin{figure}
  \centering
  \includeIDLfigP{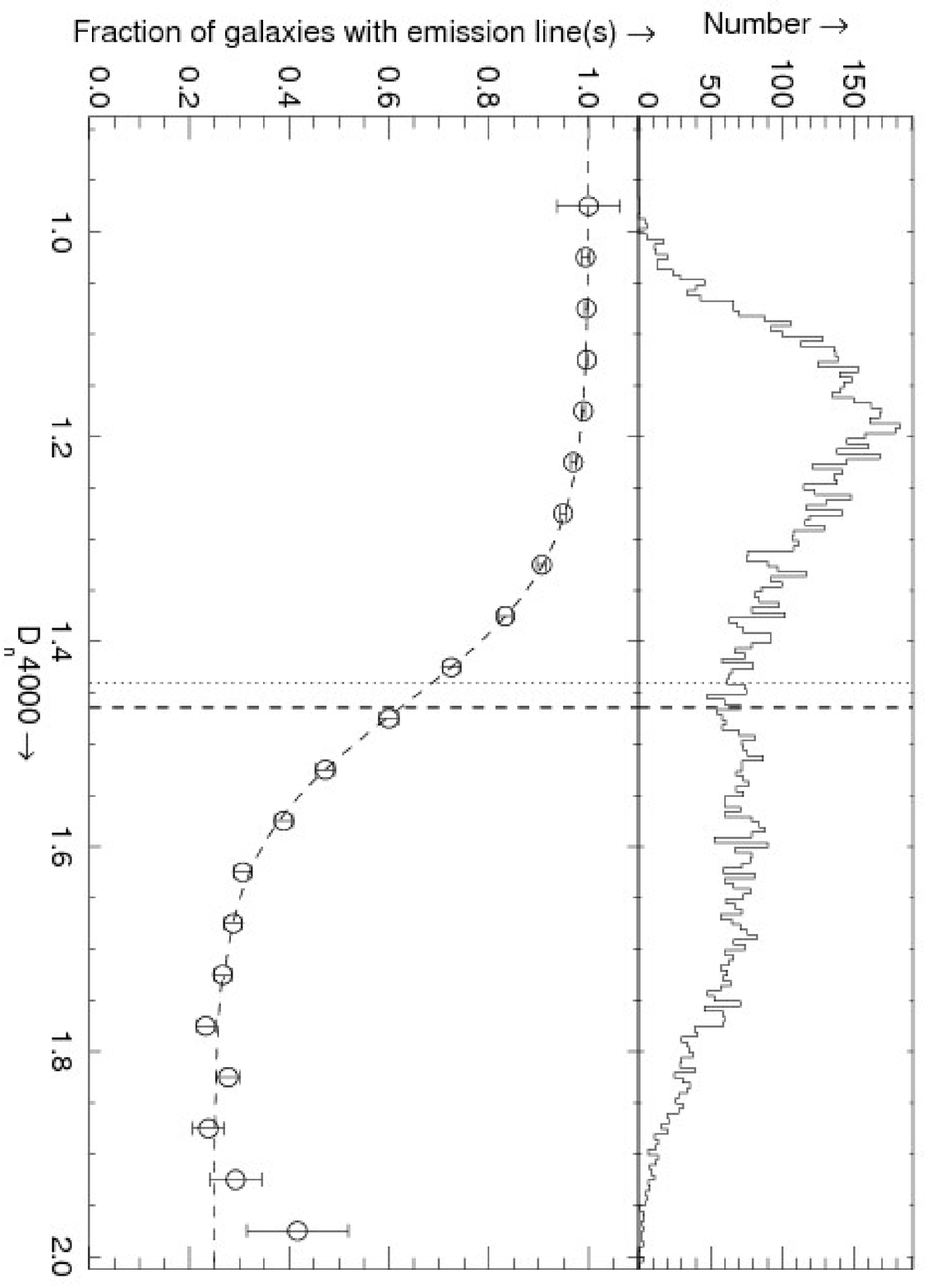}
  \caption{Distribution of \dn{} for SHELS galaxies ({\it top}) and
  the fraction of galaxies with one or more emission lines as a
  function of \dn{} ({\it bottom}). Emission-line galaxies have a REW
  of at least 5\,\AA{} for one or more of the following emission
  lines: \ha{}, \hb{}, \Oiiia{}, \Oiiib{}, or \Oii{}. The thick dashed
  line indicates the local minimum of the bimodal distribution in
  \dn{} at $\dn = 1.46$, very close to that of \citet{Woods10},
  indicated by vertical dotted line at $\dn = 1.44$. The dashed line
  is a simple $\tanh$ fit to the data points to guide the eye. The
  rise at large \dn{} reflects a contribution by AGN.}
  \label{fig:woods}
\end{figure}

An advantage of \dn{} over observed color is that \dn{} is insensitive
to reddening; \dn{} is measured over a small wavelength range (see
Fig.~\ref{fig:cb07models}). In addition to the \dn{} for the 10
SSP, we show the \dn{} for two of these models with some attenuation
in Table~\ref{tab:d4000models}. Attenuation affects \dn{} only
minimally. Furthermore, the small wavelength range needed to calculate
\dn{} allows the use of uncalibrated spectra, provided that the
sensitivity of the spectrograph does not vary significantly over the
relevant wavelengths.

Another advantage of \dn{} is---by its definition---independence of
redshift; the observed color of a galaxy changes dramatically with
redshift. At different redshifts, the bandpasses probe different parts
of the spectrum. Thus, the k-corrections are not (necessarily) equal
for different filters. Consider, for example, the 4000\,\AA{} break, a
large-scale feature (in wavelength) in the galaxy spectrum where the
flux can change significantly. The 4000\,\AA{} break is just in the
$g$ band at $z\sim0$, leaves it around $z\sim0.35$ and enters the
$r$ band, and leaves the $r$ band at $z\sim0.75$ (see
Fig.~\ref{fig:cb07models}).\footnote{Similarly, the Lyman break causes
drastic changes in galaxy colors as a function of redshift, which is a
property employed in selecting higher redshift galaxies with the
drop-out technique \citep{Steidel95}.}

\begin{figure*}
  \centering
  \includeIDLfigPcustom[0.49\textwidth]{10pt}{12pt}{5pt}{5pt}{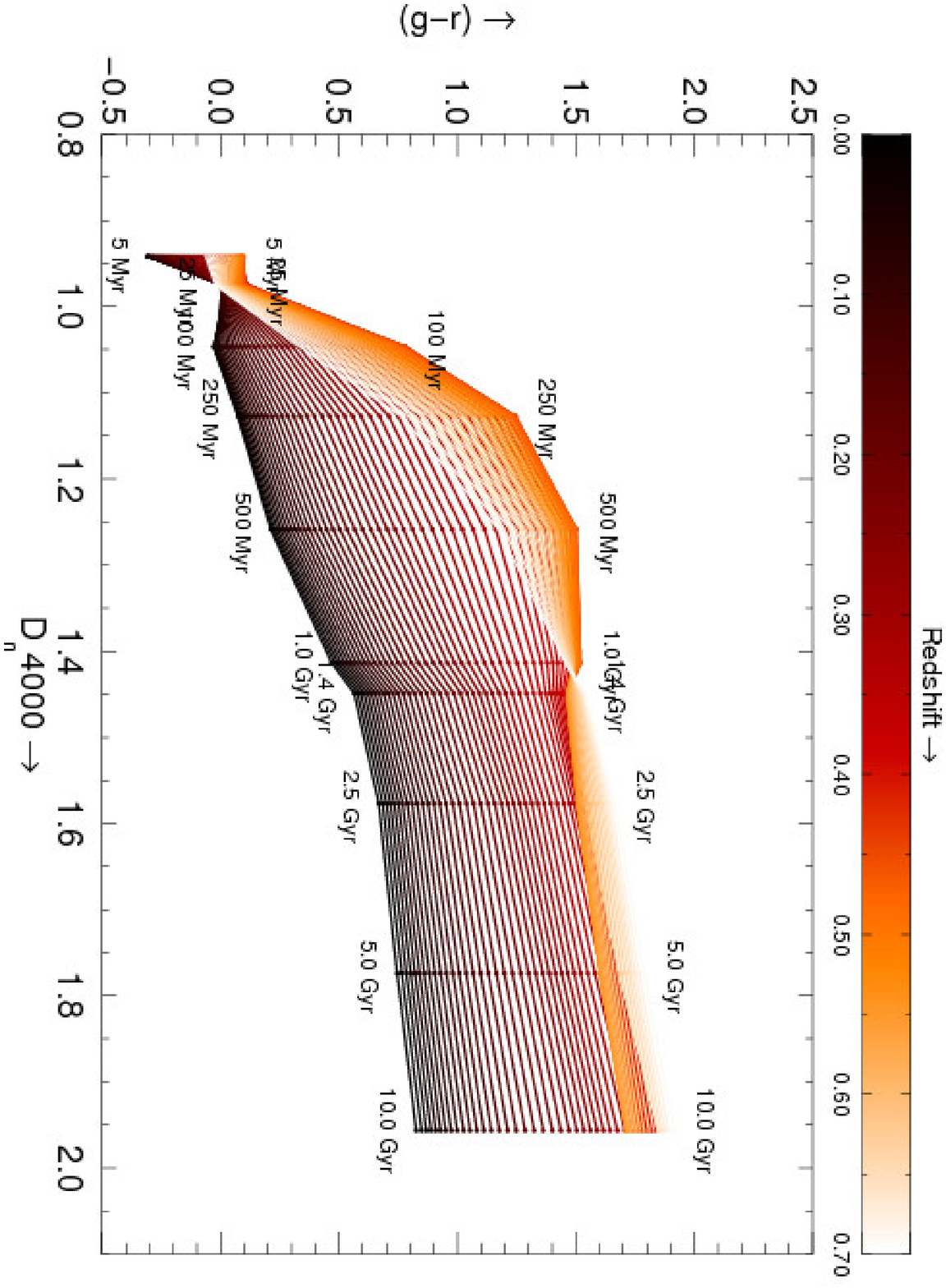}
  \includeIDLfigPcustom[0.49\textwidth]{10pt}{12pt}{5pt}{5pt}{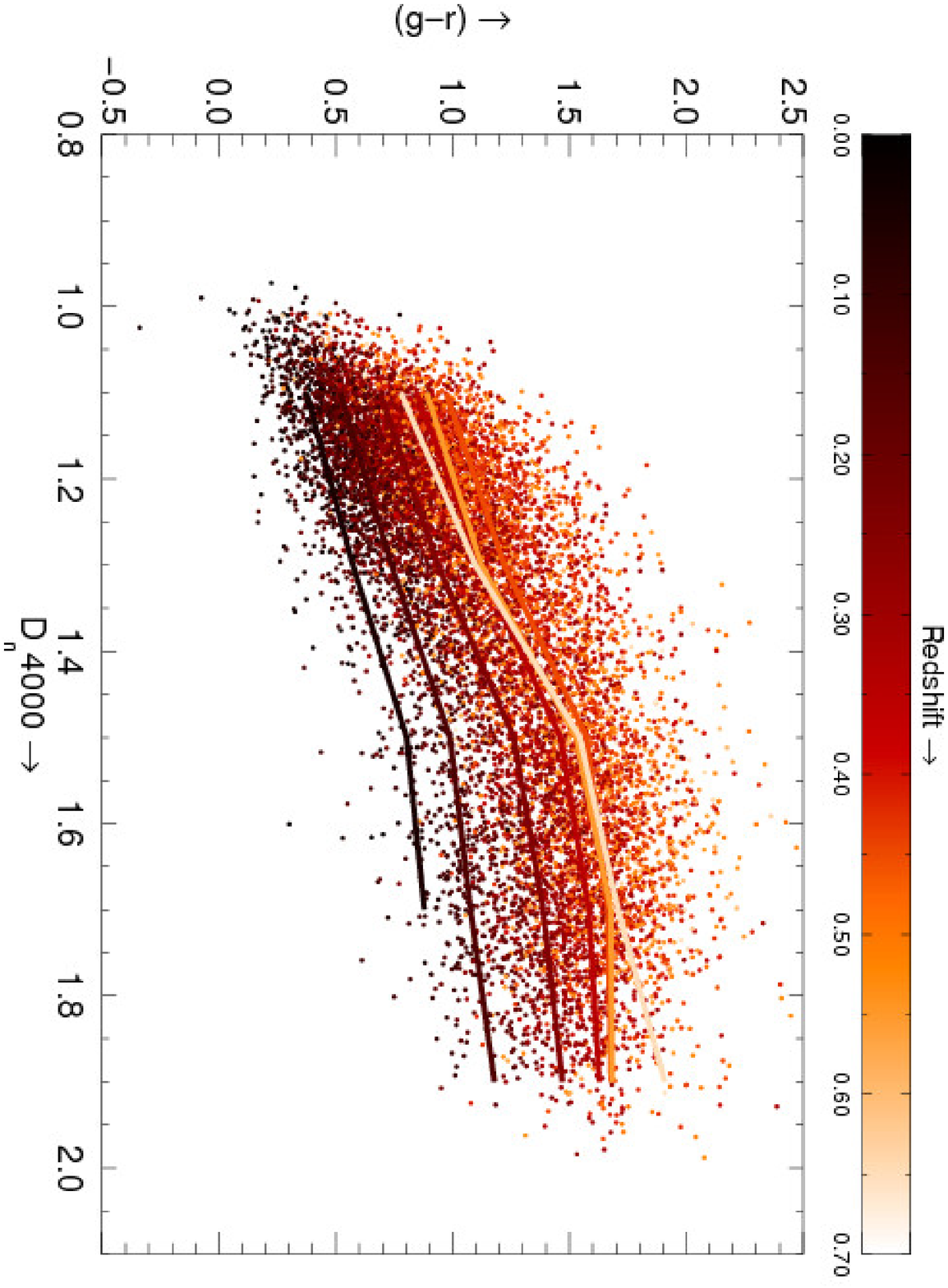}
  \caption{Observer frame \grCol{} color as a function of \dn{} for
  the 10 \cb{} SSP models without attenuation ({\it left}). The solid
  lines connect the individual points at equal
  redshifts. Observer-frame \grCol{} color as a function of \dn{} for
  the SHELS galaxies ({\it right}). The points are color-coded by the
  redshift of the galaxy. The colored solid lines indicate the median
  of \grCol{} binned by redshift as a function of \dn{}. Both panels
  show that \grCol{} becomes less sensitive to the age of the stellar
  population of a galaxy toward higher \dn{}.}
  \label{fig:d4000gr}
\end{figure*}

Figure~\ref{fig:d4000gr} shows observed \grCol{} as function of
\dn{} for both the separate SSP models and the individual galaxies. We
indicate the median color of the galaxies binned by redshift as
function of \dn{}. Toward higher \dn{}, \grCol{} becomes less
sensitive than \dn{} to the age of the stellar population of a
galaxy. Furthermore, the models show that \grCol{} is not a monotonic
function of redshift; \dn{} is redshift-independent.

A spectrum of a galaxy adequate to determine its redshift (a necessary
quantity to calculate the k-correction) usually straightforwardly
provides the \dn{} of that galaxy. The \dn{} is more robust for
galaxies where older stellar populations dominate the spectrum; the
redshift is then usually determined from the Ca H+K lines.

We use the \dn{} determined from our model fits. This approach allows
us to determine the \dn{} more accurately for lower S/N spectra,
because we use almost the entire wavelength range of the spectrum to
constrain the model fit; the fits reflect the noise over the entire
spectrum rather than the noise in a small portion of the spectrum
around 4000\,\AA{}.

\section{Empirically determined k-corrections}
\label{sec:fitting}

Here, we derive the empirically determined k-corrections from our
spectra and from the model fits to the spectra as a function of
redshift and \dn{}. We also provide an online calculator.\footnote{See
\url{http://tdc-www.cfa.harvard.edu/instruments/hectospec/progs/EOK/}.}

\begin{figure*}
  \centering
  \includeIDLfigPcustom[0.49\textwidth]{13pt}{12pt}{5pt}{5pt}{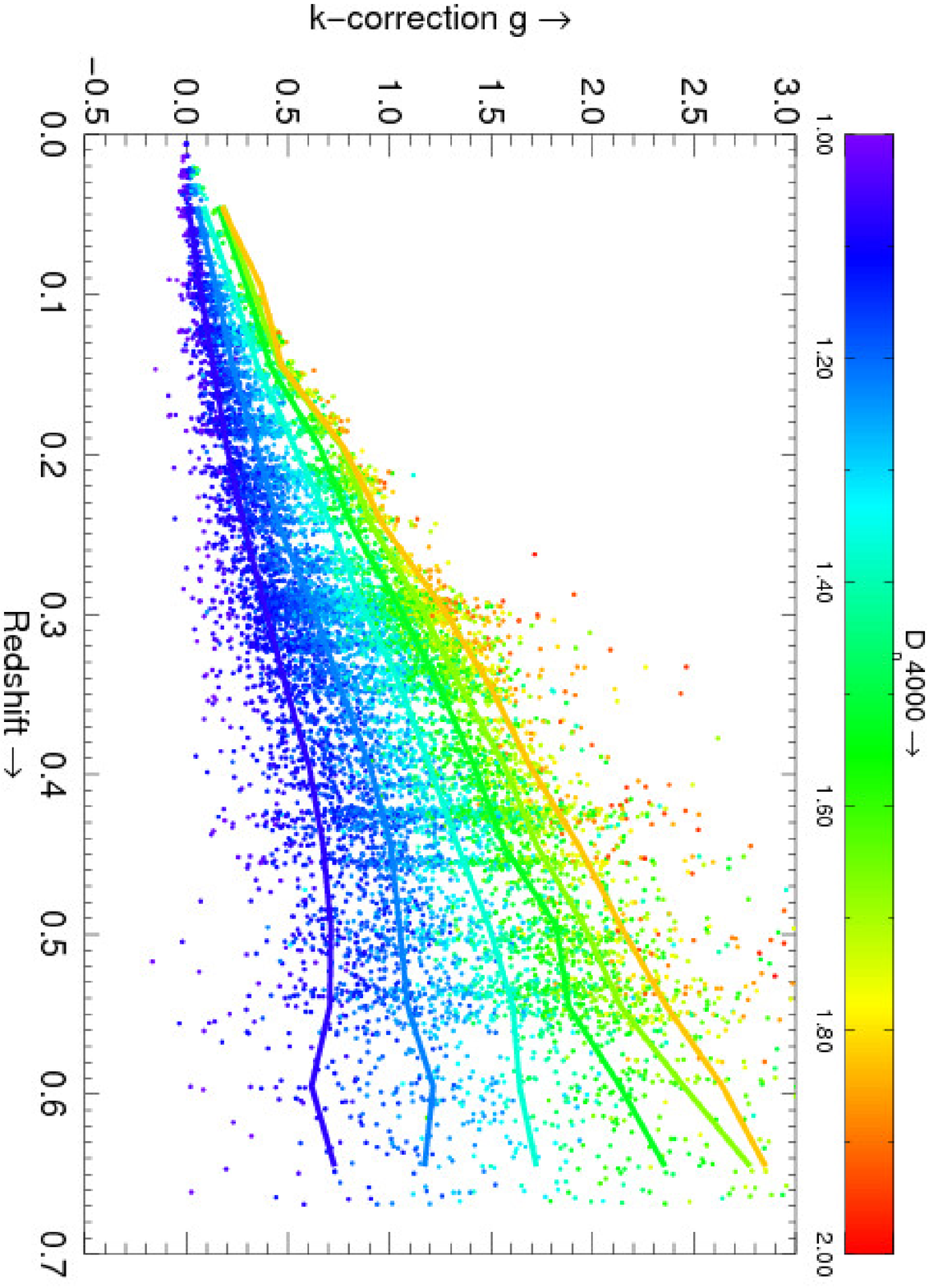}
  \includeIDLfigPcustom[0.49\textwidth]{13pt}{12pt}{5pt}{5pt}{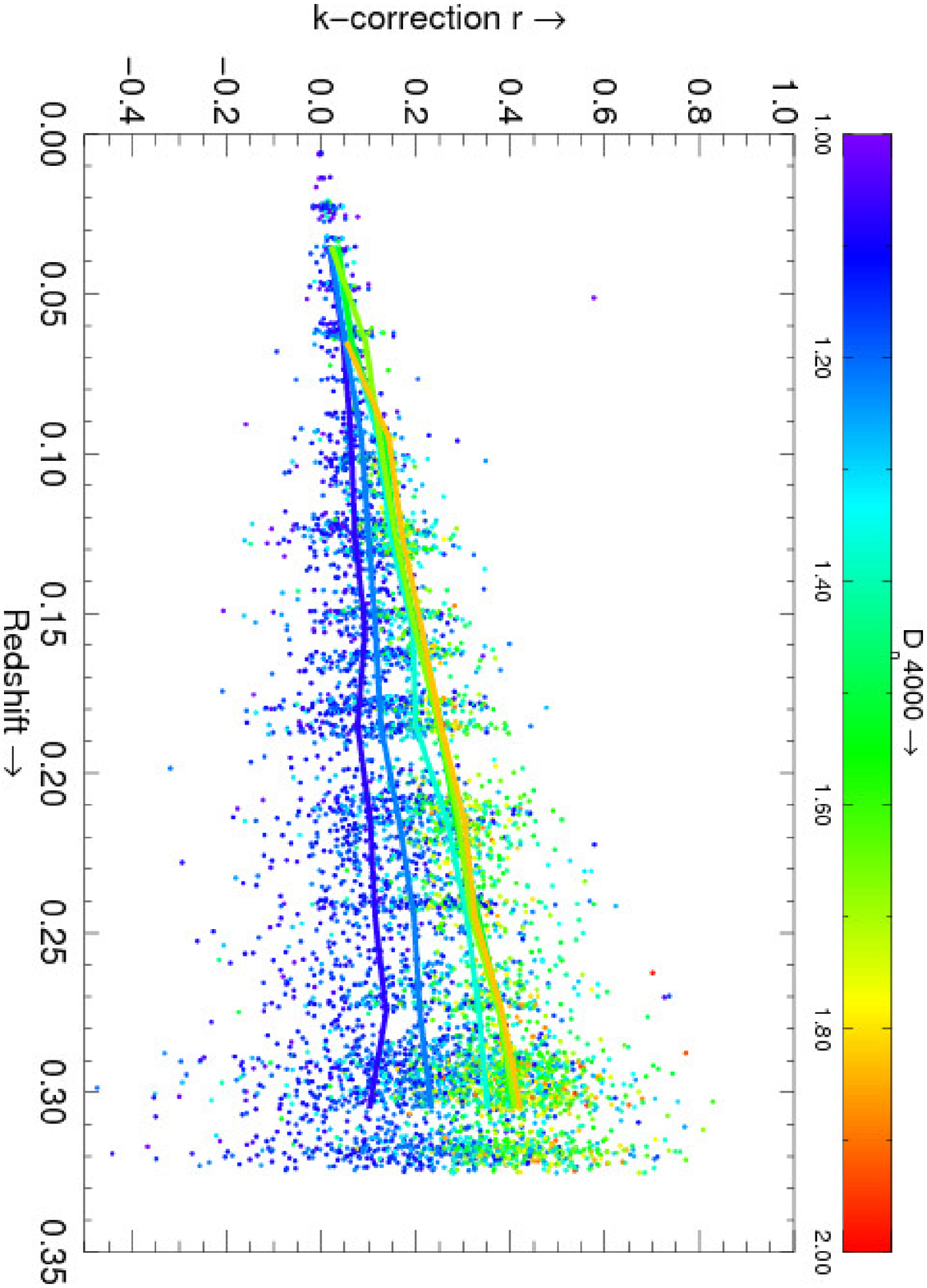}
  \includeIDLfigPcustom[0.49\textwidth]{13pt}{12pt}{5pt}{5pt}{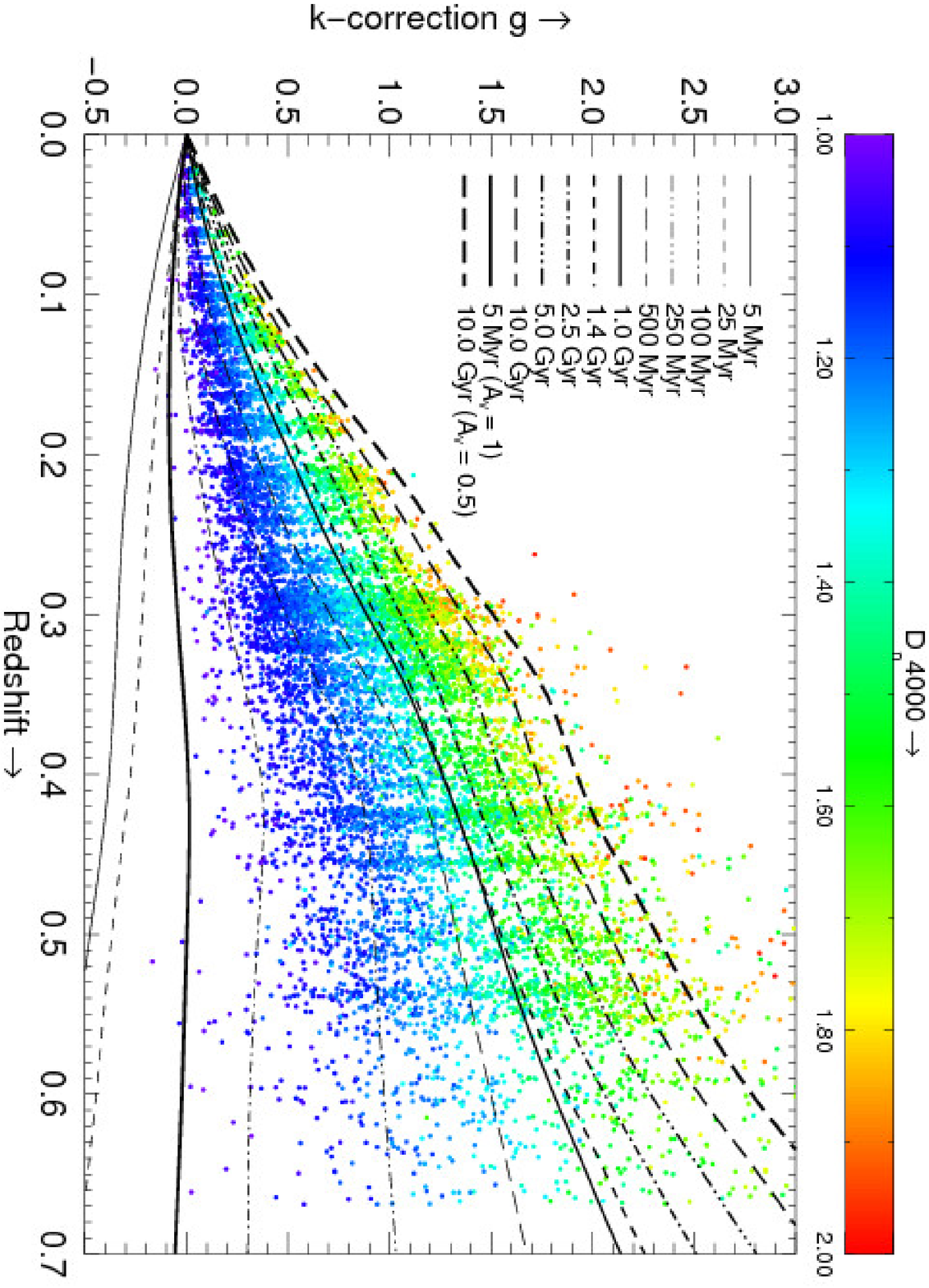}
  \includeIDLfigPcustom[0.49\textwidth]{13pt}{12pt}{5pt}{5pt}{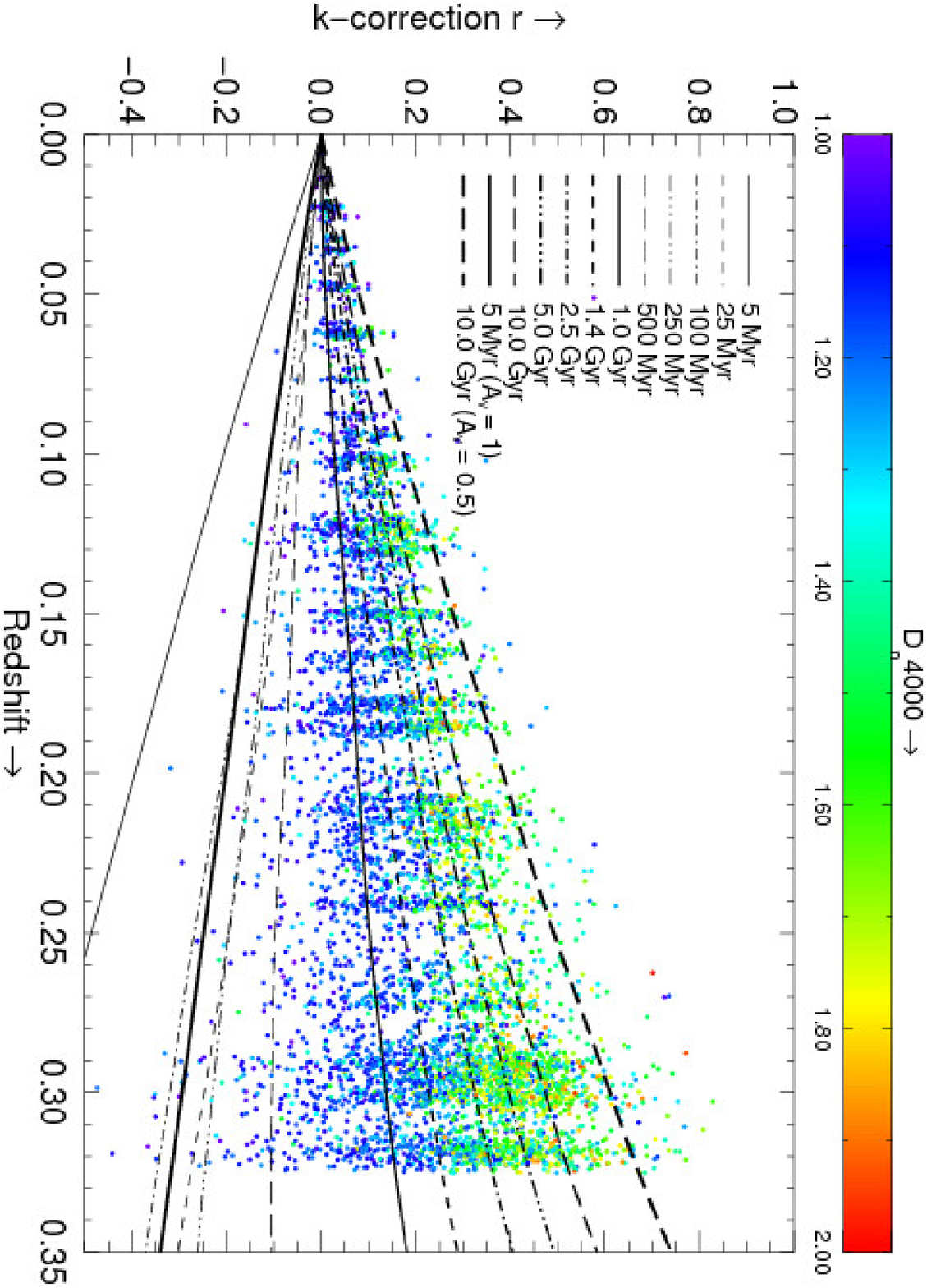}
  \includeIDLfigPcustom[0.49\textwidth]{13pt}{12pt}{5pt}{5pt}{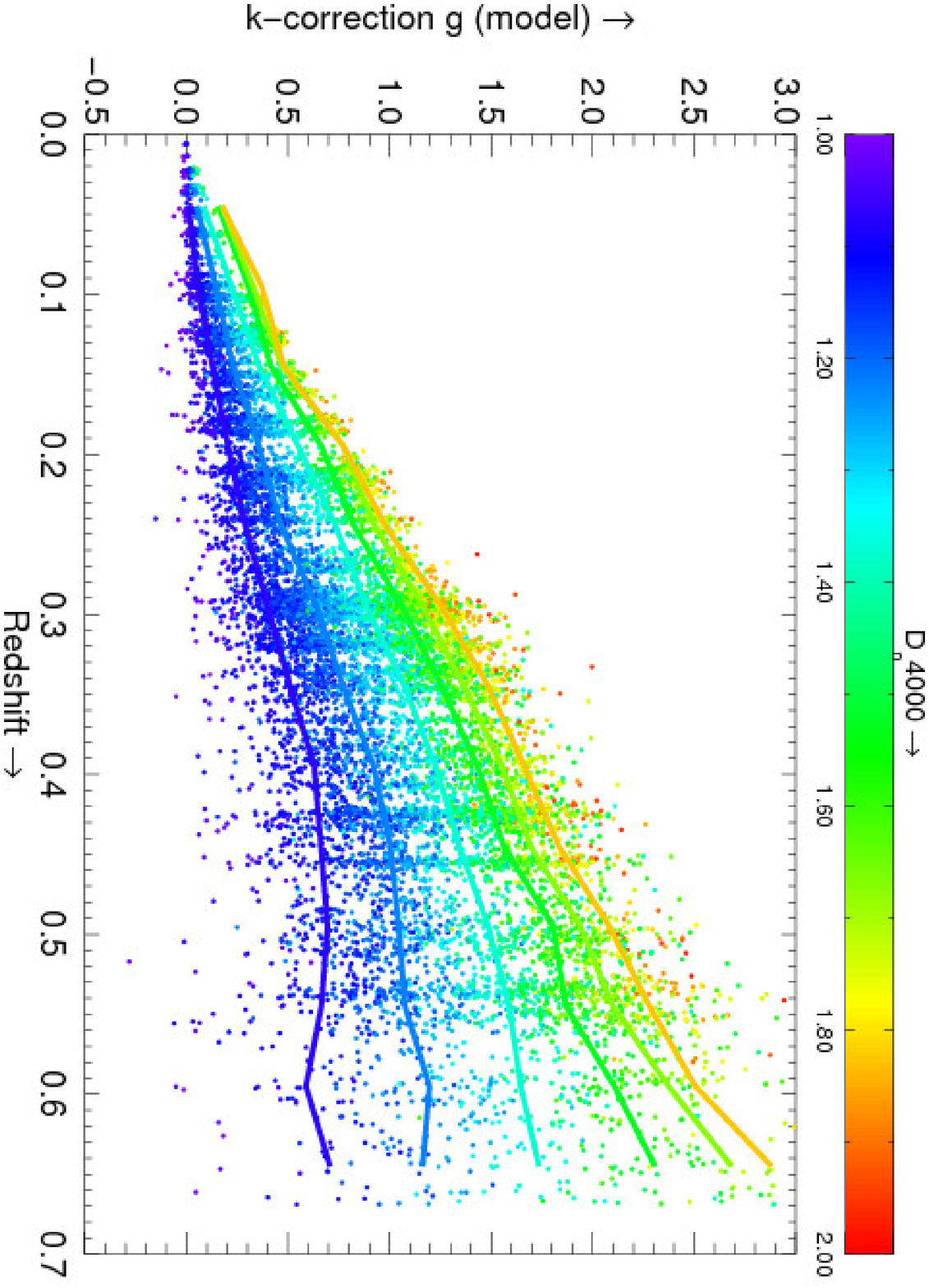}
  \includeIDLfigPcustom[0.49\textwidth]{13pt}{12pt}{5pt}{5pt}{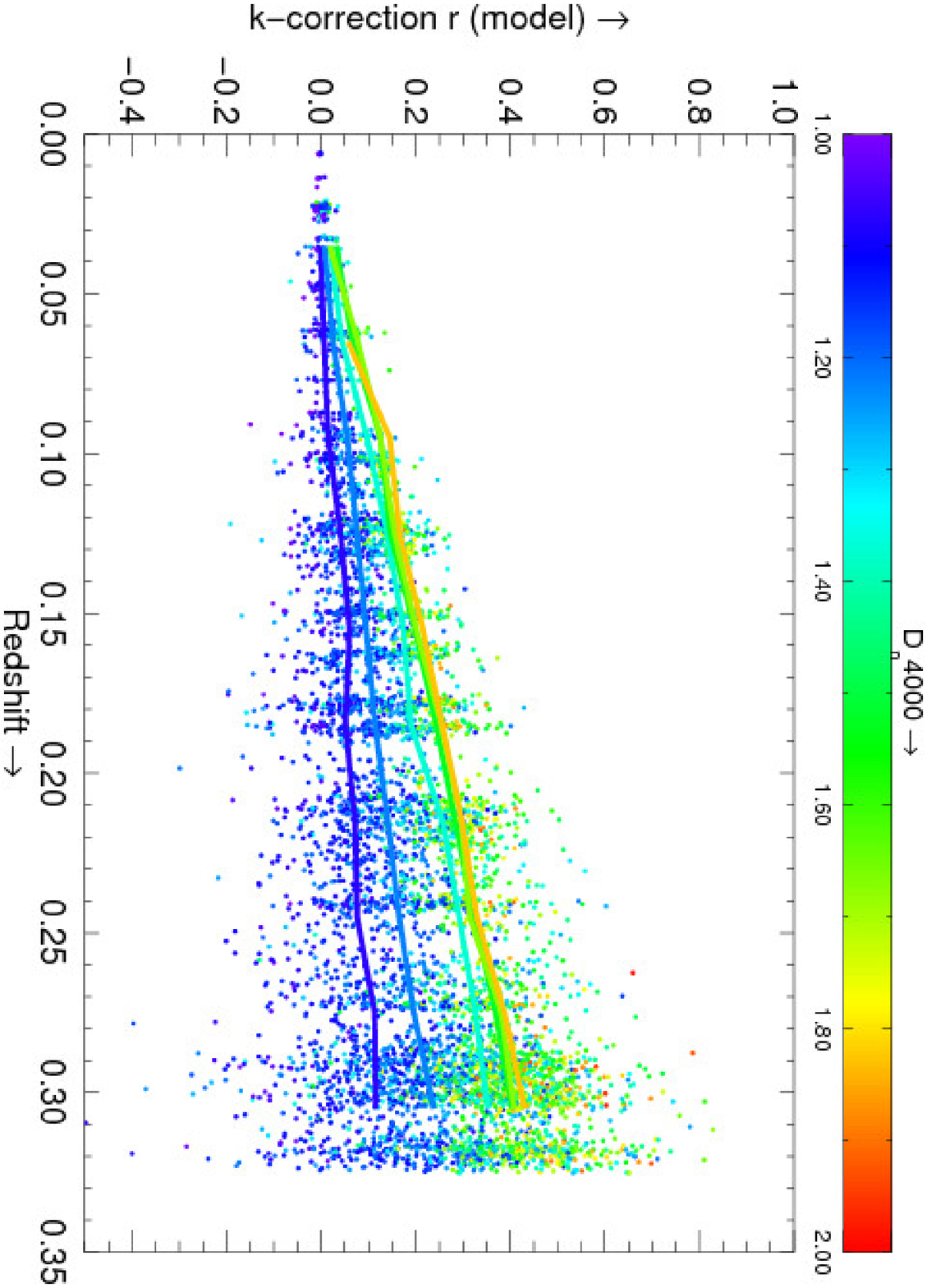}
  \caption{K-correction derived from the spectra ({\it top and
  middle}) and the models ({\it bottom}) as a function of redshift
  where each point is color-coded by the \dn{} of the galaxy for $g$
  ({\it left}) and $r$ ({\it right}). The colored solid lines indicate
  the median of the k-correction binned by \dn{} as a function of
  redshift in the {\it top} and {\it bottom} panels. The black lines
  in the {\it middle} panel compare the k-corrections predicted from
  the model tracks from Fig.~\ref{fig:modeltracks} with our
  empirically determined k-corrections.}
  \label{fig:kcor_d4000}
\end{figure*}

To minimize the impact of large residuals of night-sky features in our
spectra, we replace the observed spectrum around these wavelengths
with the model fit. We follow the same procedure for observed
wavelengths blueward of 4000\,\AA{} and redward of 8500\,\AA{}
because of the uncertainties in the flux calibration of our
spectra. Using the definition of the k-correction
(eq.~\eqref{eq:kcor}) and the prescription for determining artificial
magnitudes from our spectra (Section~\ref{sec:magdet}), we determine
the k-corrections from both our spectra and the model
fits. Figure~\ref{fig:kcor_d4000} gives the resulting k-corrections
for the $g$ and $r$ band. Each galaxy is color-coded by its \dn{}.

To use these results for other observed galaxies, we approximate these
results with a two-dimensional surface in a fashion similar to that of
\citetalias{Chilingarian10}. We use a $\chi^2$ minimization to fit a
two-dimensional surface to our data. We describe the fitting method in
detail in the Appendix. We give a short summary here. To
derive the fit, we use the redshift, \dn{}, and determined
k-correction (respectively, $x_k$, $y_k$, and $z_k$ in
eq.~\eqref{eq:chi2} and eq.~\eqref{eq:3}) for all galaxies satisfying
the sample selection of Section~\ref{sec:selection} to derive the
fit. We choose the maximum polynomial orders for the redshift and
\dn{} ($N_x$ and $N_y$ in eq.~\eqref{eq:sum}, respectively) to be the
lowest orders that do not show an obvious pattern in the residuals
after fitting. If we use too many orders, the fit provides inaccurate
results for the highest redshifts or extreme \dn{} values, because the
number of galaxies constraining the fit is limited. At $z=0$ the fit
is constrained by the definition that the k-correction is zero
(Section~\ref{sec:intro}). For our fits, $N_x = 3$ and $N_y = 3$ are
appropriate.

Tables with the coefficients for the approximation are in the Appendix
(Tables~\ref{app:coefffirst_d}-\ref{app:coefflast_d}).
Figure~\ref{fig:residuals_d4000_g} shows the residuals as a function
of \dn{} and redshift for $g$ and $r$. The thick colored lines
indicate the median of the residuals binned by \dn{} to show the
absence of structure in the residuals as a function of
redshift. Table~\ref{tab:stats} contains statistics ($\sigma$ of the
Gaussian fit, rms, and the 68.3\,\% range around
the mean) on the residuals for the fit.

We also determine the coefficients for our fit as a function of \grCol{}
(now $y_k$ in eqs.~\eqref{eq:chi2} and ~\eqref{eq:3}) and redshift
in order to compare our results with \citetalias{Chilingarian10} (see
Section~\ref{sec:chilcomp}). We show the residuals in
Figure~\ref{fig:residuals_gr_g} for $g$ and $r$. The coefficients for
the approximation are in the Appendix
(Tables~\ref{app:coefffirst_c}-\ref{app:coefflast_c}). Table~\ref{tab:stats}
contains statistics on the residuals. We note that the analytic
approximations using on \dn{} show the same, or lower, residuals for
the $r$ and $g$ bands, respectively, compared with those approximations of the
same polynomial order based on \grCol{}.

\section{Comparison with previous work}
\label{sec:chilcomp}

The advantage of using k-corrections determined directly from spectra
relative to those from models or model fits is that we make no
assumptions about galaxy evolution, attenuation laws, metallicities,
orientation, etc. Using the spectra directly also eliminates the need
for constructing a set of models that---in linear
combination---represent fundamental galaxy types. However, models
do not have the (photon)noise that the observations contain. Here, we
compare our empirical k-corrections with the models of
\citetalias{Chilingarian10}.

Figure~\ref{fig:kcor_d4000} shows the model tracks of pure SSP models
from Figure~\ref{fig:modeltracks} plotted over our data. The range of
spectroscopically determined k-corrections matches the range of the
models at the high \dn{} end extremely well. At the low \dn{} end,
there are few galaxies with k-corrections as low as predicted by the
youngest SSP models without attenuation. This difference implies that
these low \dn{} galaxies are always (significantly) attenuated and/or
these galaxies rarely have a stellar population that consists purely
of young stars.

We next compare our k-corrections with those from the
literature. \citetalias{Chilingarian10} provides k-corrections as a
function of redshift and just one other parameter, the observed galaxy
color. They use photometry in up to nine optical and near-infrared
bands for galaxies with known redshift to fit a grid of SSP models
from {\sc pegase.2} \citep{Fioc97}. In Figure~\ref{fig:chilkcor} we
show the k-corrections for our galaxies determined with the
prescription of
\citetalias{Chilingarian10}.\footnote{\citet{Blanton07} also provide a
method to determine the k-corrections. The method of
\citetalias{Chilingarian10} is similar to that of \citet{Blanton07} in
the sense that both use template spectra to fit the photometric data,
but the template set of \citetalias{Chilingarian10} differs from the
set of \citet{Blanton07}. Furthermore,
\citetalias{Chilingarian10} do a detailed comparison and conclude that
the results from both approaches are similar.}

Figure~\ref{fig:chilcomp} shows the difference between the
k-corrections derived from our data and those from
\citetalias{Chilingarian10} for SHELS galaxies as a function of both
\grCol{} and \dn{}. The k-corrections for the $r$ band agree extremely
well. The k-corrections for the $g$ band also agree well up to $z
\lesssim 0.3$ (68.3\,\% range around the median difference is 0.15 and
0.12 for $g$ and $r$, respectively at $z=0.3$). For larger $z$, they
start to disagree for galaxies with large \dn{}. The cause for this
difference is twofold.

At a redshift of $\sim$ 0.3, the \citetalias{Chilingarian10} sample
does not contain as many green-to-blue galaxies as red galaxies. They
drew their sample of galaxies from SDSS DR7 \citep{SDSS7} and UKIRT
Infrared Deep Sky Survey DR5 \citep{UKIDSS5} with redshifts between
0.03 and 0.6. The galaxies in SDSS DR7 have spectroscopy for at least
two distinct samples: the main galaxy sample ($r < 17.77$) with a
median redshift around 0.10 and the luminous red galaxy sample ($r
\lesssim 19.5$) selected with the majority of the galaxies at
redshifts higher than 0.15.

In contrast, SHELS reaches much fainter than SDSS ($R < 21$) and
selects galaxies based only on their total magnitude. Thus, SHELS is
not biased toward or against a particular type of galaxy (except
against AGN due to their stellarlike morphology). SHELS
contains more (in number) intrinsically bluer galaxies at higher
redshifts than \citetalias{Chilingarian10}.

\citetalias{Chilingarian10} only selects galaxies with redshifts up to
0.6. Thus, their fit is only valid to approximately that redshift;
SHELS contains galaxies to $z = 0.7$.

However, the differences with \citetalias{Chilingarian10} for the
galaxies with the largest \dn{} require further
discussion. Figure~\ref{fig:discrepancy} shows the differences between
the magnitudes derived from the spectra and the models in the observed
and rest frames. Modulo the presence of emission lines
(Section~\ref{sec:el}) and noise, these magnitudes should be the
same. Apart from the panel with observed $g$-band magnitudes, all
other panels show that the magnitudes derived from the spectra and the
models are consistent with each other.

We investigate the overestimation in flux indicated by
Figure~\ref{fig:discrepancy} of the high-\dn{} galaxies from the
models with respect to the spectra by combining the spectra of these
galaxies. We note that for the galaxies with the oldest stellar
populations there is less flexibility in fitting the models than for
galaxies with younger populations because only old SSP models can fit;
for younger populations conceivably different combinations of young
{\it and} old SSP models (and an increased/decreased amount of
attenuation) can yield a good fit.

Figure~\ref{fig:summed} shows the averaged luminosity-weighted
rest-frame spectrum---of both the actual spectra and the model
fits---of 84 galaxies with $0.4 < z < 0.6$, $\dn > 1.7$, and a
difference between the magnitude derived from the spectrum and the fit
that is larger than 0.2. We also show the difference between the
averaged spectrum and averaged model in the middle panel and the
relative difference between the averaged spectrum and averaged model
with respect to the averaged model.

The models deviate from the spectra at the bluest rest-frame
wavelengths. The relative difference there is large. Because the
features in the average fit also appear in the spectrum, the flux of
the average spectrum goes only marginally below 0. These rest-frame
wavelengths correspond to the bluest part of the spectrum (where the
spectrograph is not very sensitive) and it appears that the sky is
somewhat oversubtracted. We suggest the use of the k-corrections
determined from the model fits for the galaxies with high \dn{} at the
highest redshifts.

Overall, we find very good agreement between our k-corrections using
the full spectrum and the prescription of \citetalias{Chilingarian10}
based on SSP fits to photometry. Similar to
\citetalias{Chilingarian10} (who use redshift and observed color), we
can approximate the k-correction with only two parameters: redshift
and \dn{}. There are no systematics remaining in the residuals of our
approximation.

\section{Uncertainties}
\label{sec:unc}

In Section~\ref{sec:d_vs_c} we determined \dn{} from the model fits to
the spectra. Not every spectrum is suitable for fitting SSP models; in
those cases one can determine \dn{} from the spectrum. In this section
we consider the effects of using the spectrum-derived \dn{} on the
determination of the k-correction instead of using the model-derived
\dn{}. We compare the accuracy of the model-derived \dn{} with the
spectrum-derived \dn{}.

The accuracy of \dn{} is obviously a function of the S/N of the
spectrum. The S/N for each of our spectra is the median of the S/N per
Hectospec pixel measured in 10 wavelength ranges in the observed
spectrum that are free of night-sky emission and sample the range over
which we fit our models to the spectrum (4100--4300, 5000--5200,
5470--5570, 5700--5850, 6000--6200, 6400--6600, 6650--6700,
7000--7200, 7600--7700, and
8050--8250\,\AA{}). Figure~\ref{fig:sntest} shows the relative
difference of the spectrum- and model-derived \dn{} as a function of
S/N. The accuracy of \dn{} increases, as one might expect, with
increasing S/N.

Figure~\ref{fig:sntest} also shows the distribution of the relative
difference between the model- and spectrum-derived \dn{} binned by
S/N. The 68.3\,\% range around the median increases with decreasing
S/N (0.21, 0.13, 0.11, and 0.07, respectively, for the four S/N ranges
indicated). The values of the higher S/N spectra are consistent with
the comparison between SDSS and Hectospec spectra of
\citet[][$\mathrm{rms} = 0.09$]{Fabricant08}.

Figure~\ref{fig:sntest} also shows the distribution of model-derived
\dn{} (because we use this throughout this article) in the same S/N
bins. Galaxies with the most poorly determined \dn{} are typically the
low surface brightness---also intrinsically bluer---galaxies.

Table~\ref{tab:exampleunc} shows the k-corrections and their
uncertainties for a few example galaxies given the uncertainties in
\dn{} and \grCol{}. The three uncertainties for \dn{} correspond to
the 68.3\,\% range around the median for very low, moderate, and high
S/N spectra. The uncertainties in the corresponding color are those in
the SDSS Petrosian and fiber colors. The column $\sigma_k$ indicates
the corresponding range in the calculated k-correction. We choose the
redshifts of the galaxies, $z = 0.30$ and $0.55$, close to the maximum
redshifts where we can determine k-corrections for the $r$ and $g$
bands, respectively. We choose \dn{} values that are typical for the
two populations of galaxies (Fig.~\ref{fig:woods}; $\dn = 1.25$ and
$1.65$).

The \dn{} from a spectrum with $S/N > 2$ ($\sigma_\dn \lesssim 0.11$)
yields an uncertainty in the k-correction, similar to the uncertainty
in the k-correction determined from \grCol{} given the typical
uncertainties in the galaxy color. For the \dn{} from a spectrum with
$S/N < 2$ ($\sigma_\dn > 0.11$), the uncertainty in the k-correction
is somewhat larger than the uncertainty in the k-correction determined
from \grCol{}. In most of the examples, the uncertainty in the SDSS
magnitude dominates the uncertainty in the final absolute magnitude,
not the uncertainty in the k-correction.

\section{Emission lines}
\label{sec:el}

Emission lines with large REW can significantly affect the flux
density measured through a filter.\footnote{Narrowband surveys employ
exactly this feature to increase the contrast between observations
done with a very narrow filter and a broadband filter; the narrower
the filter becomes, the larger the difference in magnitude between the
two filters \citep[e.g.,][]{Waller90,Pascual07}.}
Figure~\ref{fig:discrepancy} shows the influence of the emission lines
on the magnitudes. The galaxies with low \dn{} and $|m_\mathrm{spec} -
m_\mathrm{mod}| \gg 0$ are almost all emission-line galaxies.

As \citet{Woods10} show (see also Section~\ref{sec:d_vs_c} and
Figure~\ref{fig:woods}), nearly all galaxies with low \dn{} have
emission lines. These galaxies make up a large fraction of our
sample. We thus investigate the influence of the emission lines on the
derived k-corrections.

We examine five galaxies with differing emission-line REWs.
Figure~\ref{fig:ewspectra} shows the spectra of these galaxies. We
calculate the magnitudes and k-corrections as a function of redshift
from the rest-frame spectra and the model fits of these galaxies. We
assume that the difference between the spectrum and the model results
only from the emission lines. We plot the difference in derived
magnitudes and k-corrections for both bands in
Figure~\ref{fig:elinfluence}.\footnote{The spectra of these galaxies
normally evolve over time. The point of this figure, however, is to
demonstrate the influence of emission lines, not galaxy evolution.}

The entrance and departure of emission lines from the filter as a
function of increasing redshift dominate the curve that traces the
difference in magnitudes. At $z=0$ the \hb{} and \Oiii{} complex are
in the $g$ filter and leave it with increasing redshift. Around
$z\sim0.1$ the \Oii{} doublet enters the filter, causing the
difference in the magnitude to increase only slightly. For the
$r$ band, the rippling shape of the curve results from \Sii{}, \Nii{},
and (predominantly) \ha{} leaving the filter with increasing redshift
and having left around $z\sim0.1$, while the \hb{} and \oiii{} complex
starts to enter the filter at the same time. Around $z\sim0.35$ the
complex starts to leave, while \hc{} enters. At $z\sim0.55$ \hc{}
leaves and \hd{} enters.

Figure~\ref{fig:elinfluence} shows that (1) the influence of
emission lines is clearly correlated with their equivalent width, and
(2) apart from galaxies with extremely large REW emission lines
($\mathrm{REW} \gg 100$\AA{}), the influence of emission lines on the
broadband $g$ and $r$ magnitudes---and, consequently,
k-corrections---is reasonably small. The significance depends on the
accuracy of the photometry and/or the required precision of the
k-corrections. For an accuracy $\lesssim 0.10$\,mag k-corrections need
to take emission-line flux into account. For example, \citet{Brown08}
use the influence of strong emission lines on the observed galaxy
colors to select extremely metal poor galaxies.

\section{Additional k-corrections}
\label{sec:modkcor}

In Section~\ref{sec:fitting} we determined the k-corrections for the
SDSS $g$ and $r$ bands; here we consider k-corrections in other
photometric bands. Because emission lines are important for only the
strongest emission lines in a galaxy spectrum and our model fit
extends beyond the wavelength range of the spectrum, we can calculate
the k-corrections for any optical bandpass over the redshift range
where we sample enough galaxies: i.e., $z \le 0.7$.

Here, we derive the k-corrections as a function of redshift and \dn{}
for the SDSS bandpasses $ugriz$ and the Johnson-Cousins $UBVRI$ given
in Table~2 of \citet{Bessell90}. Figure~\ref{fig:allfilters} shows the
10 normalized bandpasses we used to calculate the
k-corrections. Figure~\ref{fig:sdss_u} showcases the results for the
SDSS $u$ and the Johnson-Cousins $V$ band. The figures for these and
the remaining filters are also available online.\footnote{See
\url{http://tdc-www.cfa.harvard.edu/instruments/hectospec/progs/EOK/figures.shtml}.}
Tables~\ref{tab:mod_kx_u}-\ref{tab:mod_kx_I} contain the coefficients
for the analytic approximation using redshift and \dn{},
Tables~\ref{tab:mod_kx_u_col}-\ref{tab:mod_kx_z_col} contain the
coefficients for the approximations using redshift and color (SDSS
filters only; we do not have imaging available in the Johnson-Cousins
bandpasses). We give the statistics of the residuals in
Table~\ref{tab:sigmas} to indicate the quality of the fits.

For the $ugriz$ bandpasses, our k-corrections match those of
\citetalias{Chilingarian10} at $z \lesssim 0.6$ for galaxies with high
\dn{} and at $z \lesssim 0.3$ for galaxies with low
\dn{}. \citetalias{Chilingarian10} selects galaxies with redshifts up
to 0.6 and their number of green galaxies decreases steeply beyond
redshift 0.3. Furthermore, toward redder filters the differences
between our k-corrections and those of \citetalias{Chilingarian10}
decrease; because the 4000\,\AA{} break goes through the redder
filters at higher redshifts, the k-corrections for the redder filters
are also smaller than for the bluest filters.

The residuals from the analytic approximation using \dn{} and redshift
are about the same as or smaller than those using observed galaxy color
and redshift. This result is, however, not surprising; the optical
observed galaxy colors used by \citetalias{Chilingarian10} coarsely
probe \dn{} for the particular filter where they calculate the
k-correction.

\section{Summary}
\label{sec:summary}

We use the 15,513 SHELS spectra to determine the k-corrections for the
SDSS $g$ and $r$ bands {\it empirically}. These k-corrections have
the major advantage that they are model-independent. We express the
analytic approximations for the k-correction as a function of two
parameters: redshift and \dn{}. These parameters have some advantages
over those from \citetalias{Chilingarian10}: redshift and galaxy
color. The advantages include (1) \dn{} is redshift-independent,
(2) \dn{} is an indicator of the age of the stellar population
of the galaxy, (3) \dn{} barely suffers from attenuation,
(4) the spectrum used to determine the redshift of a galaxy can
also yield \dn{}, and (5) the spectrum does not necessarily have
to be flux-calibrated to determine \dn{}. Furthermore, the analytic
approximations based on \dn{} have about the same ($r$ band) or less
scatter ($g$ band), compared with the approximations using \grCol{}. The
$\sigma$ of the scatter in the residuals is $\sim 0.08$\,mag, the rms
is $\sim 0.15$; the residuals also show no systematic bias.

From \cb{} SSP models we indicate the range of k-corrections that we
determine from observed galaxies. The models with the youngest and
oldest stellar populations with a small amount of attenuation span the
entire range of determined k-corrections. We conclude that galaxies
are always (significantly) attenuated and/or rarely have a stellar
population that consists purely of young stars.

We compare our results to the analytic approximations given by
\citetalias{Chilingarian10}. Up to $z \lesssim 0.3$ our results agree
well for both bands. Differences between the two surveys arise due to
the fact that SHELS goes to significantly fainter magnitudes than SDSS
(the survey from which \citetalias{Chilingarian10} draws). Furthermore, our
total magnitude selection criterion allows us to sample the entire
range of galaxies up to $z\sim0.7$, rather than selecting for luminous
red galaxies (like SDSS). We make the analytic approximations
available in an online calculator (see footnote 7).

We check the influence of emission lines on the k-corrections. The
influence of the emission lines depends on the REW of the line; the
larger the REW of the line, the larger the influence on the
k-correction is. Depending on the accuracy of the k-corrections
required, the influence of emission lines can typically be ignored,
except those few cases in which the REW is very large ($\mathrm{REW}
\gg 100$\,\AA{} gives $\Delta m \gg 0.10\,$mag).

Finally, we provide k-corrections as a function of redshift and \dn{}
using our model fits for SDSS $ugriz$ and Johnson-Cousins $UBVRI$. We
determine these k-corrections over the redshift range where we sample
enough galaxies: i.e., $z \le 0.7$. We compare our results to the
analytical approximations given by \citetalias{Chilingarian10} and
show that we cover a larger redshift range, where we sample more (in
number) ``green'' galaxies than \citetalias{Chilingarian10} do. For
spectra with S/N $>$ 2 ($\sigma_\dn \lesssim 0.11$), the k-corrections
based on \dn{} are as good as those based on color.

Despite the fact that the scatter in our k-corrections is typically
$\sim 0.1$\,mag, the uncertainties in the observed magnitude are of a
similar size or larger. Thus, uncertainties in the magnitudes dominate
the uncertainty in the final, k-corrected magnitude.

This empirical approach to determine k-corrections can be extended to
the near- and midinfrared to include spectroscopic features, such as the
Paschen and Brackett hydrogen lines (and breaks) and polycyclic
aromatic hydrocarbon features. The technique can also be
extended to higher redshifts where the 4000\,\AA{} break moves through
the reddest optical filters and into the near-infrared.

\section*{Acknowledgments}

EW acknowledges the Smithsonian Institution for the support of his
postdoctoral fellowship.

Observations reported here were obtained at the MMT Observatory, a
joint facility of the Smithsonian Institution and the University of
Arizona. The SDSS is managed by the Astrophysical Research Consortium
for the Participating Institutions.

We thank Susan Tokarz and Bill Wyatt for reducing and redshifting the
Hectospec data, John Roll for actively maintaining and updating {\sc
Starbase}, Bill Wyatt for maintaining a local copy of the SDSS
database. We acknowledge the work of the Hectospec instrument
assistants Perry Berlind and Mike Calkins, and the MMT telescope
operators Mike Alegria, John McAfee, and Ale Milone. We thank all the
staff at the Telescope Data Center and Fred Lawrence Whipple
Observatory that make the observations possible. We have benefited
from discussions with Warren Brown, Nelson Caldwell, and Scott Kenyon.

\appendix

\section{K-correction approximations: surface fitting}
\label{app:surfit}
In Section~\ref{sec:fitting} we describe the empirical k-correction as
a result of redshift and either \dn{} or color. Here, we describe the
method used for the fitting and give the coefficients for the
approximations.

We use $\chi^2$ minimization to determine the coefficients for our
analytic approximations. We describe our method in the context of
fitting a two-dimensional surface. The surface $z$ as a function of
coordinates $x$ and $y$ is given by
\begin{equation}
z(x,y) = \sum _{i=0,j=0} ^{i=N_x,j=N_y} a_{ij} x^i y^j,
\label{eq:sum}
\end{equation}
where $a_{ij}$ are the coefficients and $N_x$ and $N_y$ are the
maximum polynomial terms for $x$ and $y$, respectively.

The $\chi^2$ for this surface for $N_g$ data points is defined as
\begin{equation}
\chi^2 \equiv \sum _{k=1} ^{N_g} \left (z_k - z(x_k,y_k) \right )^2.
\label{eq:chi2}
\end{equation}
The minimization $\partial \chi^2/ \partial a_{ij} = -2 \sum _k (z_k -
z(x_k,y_k)) x_k^i y_k^j = 0$ for each coefficient $a_{ij}$ yields
\begin{equation}
\label{eq:3}
\sum _k z_k x_k^i y_k^j = \sum _k z(x_k,y_k) x_k^i y_k^j.
\end{equation}

We solve these equations using lower/upper (LU) decomposition. For LU
decomposition we need to write all equations from eq.~\eqref{eq:3} as
a linear system $\mathbf{A}\mathbf{x}=\mathbf{B}$, where $\mathbf{x}$
is a vector containing the coefficients. Let $S_{ij} \equiv \sum _k
x_k^i y_k^j$ and $Z_{ij} \equiv \sum _k z_k x_k^i y_k^j$, then the
linear system we solve looks like
\begin{equation}
\newcommand{\cvdots}{\multicolumn{1}{c}{\vdots}}
\newcommand{\cldots}{\multicolumn{1}{c}{\ldots}}
\newcommand{\cddots}{\multicolumn{1}{c}{\ddots}}
\begin{array}{cccc}
  \left [ \begin{array}{llllllll}
      S_{00} & S_{10} & S_{20} & \cldots & S_{i0} & S_{01} & \cldots & S_{ij}\\
      S_{10} & S_{20} & S_{30} & \cldots & S_{(i+1)0} & S_{11} & \cldots & S_{(i+1)j}\\
      S_{20} & S_{30} & S_{40} & \cldots & S_{(i+2)0} & S_{21} & \cldots & S_{(i+2)j}\\
      \cvdots & \cvdots & \cvdots & \cddots & \cvdots & \cvdots & \cvdots & \cvdots\\
      S_{i0} & S_{(i+1)0} & S_{(i+2)0} & \cldots & S_{(i+i)0} & S_{i1} & \cldots & S_{(i+i)j}\\
      S_{01} & S_{11} & S_{21} & \cldots & S_{i1} & S_{02} & \cldots & S_{i(j+1)}\\
      \cvdots & \cvdots & \cvdots & \cvdots & \cvdots & \cvdots & \cddots & \cvdots\\
      S_{ij} & S_{(i+1)j} & S_{(i+2)j} & \cldots & S_{(i+i)j} & S_{i(j+1)} & \cldots & S_{(i+i)(j+j)}\\
    \end{array} \right ]
  \!\!\!\!&\!\!\!\!
  \left [ \begin{array}{c}
      a_{00}\\
      a_{10}\\
      a_{20}\\
      \cvdots\\
      a_{i0}\\
      a_{01}\\
      \cvdots\\
      a_{ij}\\
    \end{array} \right ]
  & \!\!\!\!=\!\!\!\! &
  \left [ \begin{array}{c}
      Z_{00}\\
      Z_{10}\\
      Z_{20}\\
      \cvdots\\
      Z_{i0}\\
      Z_{01}\\
      \cvdots\\
      Z_{ij}\\
    \end{array} \right ].\\
  \mathbf{A} & \mathbf{x} & \!\!\!\!=\!\!\!\! & \mathbf{B}\\
\end{array}
\end{equation}
We use the IDL routine {\sc la\_linear\_equation} based on LAPACK
\citep{lapack} to solve for $\mathbf{x}$.

\begin{figure*}
  \centering
  \includeIDLfigPcustom[0.95\textwidth]{0pt}{70pt}{10pt}{80pt}{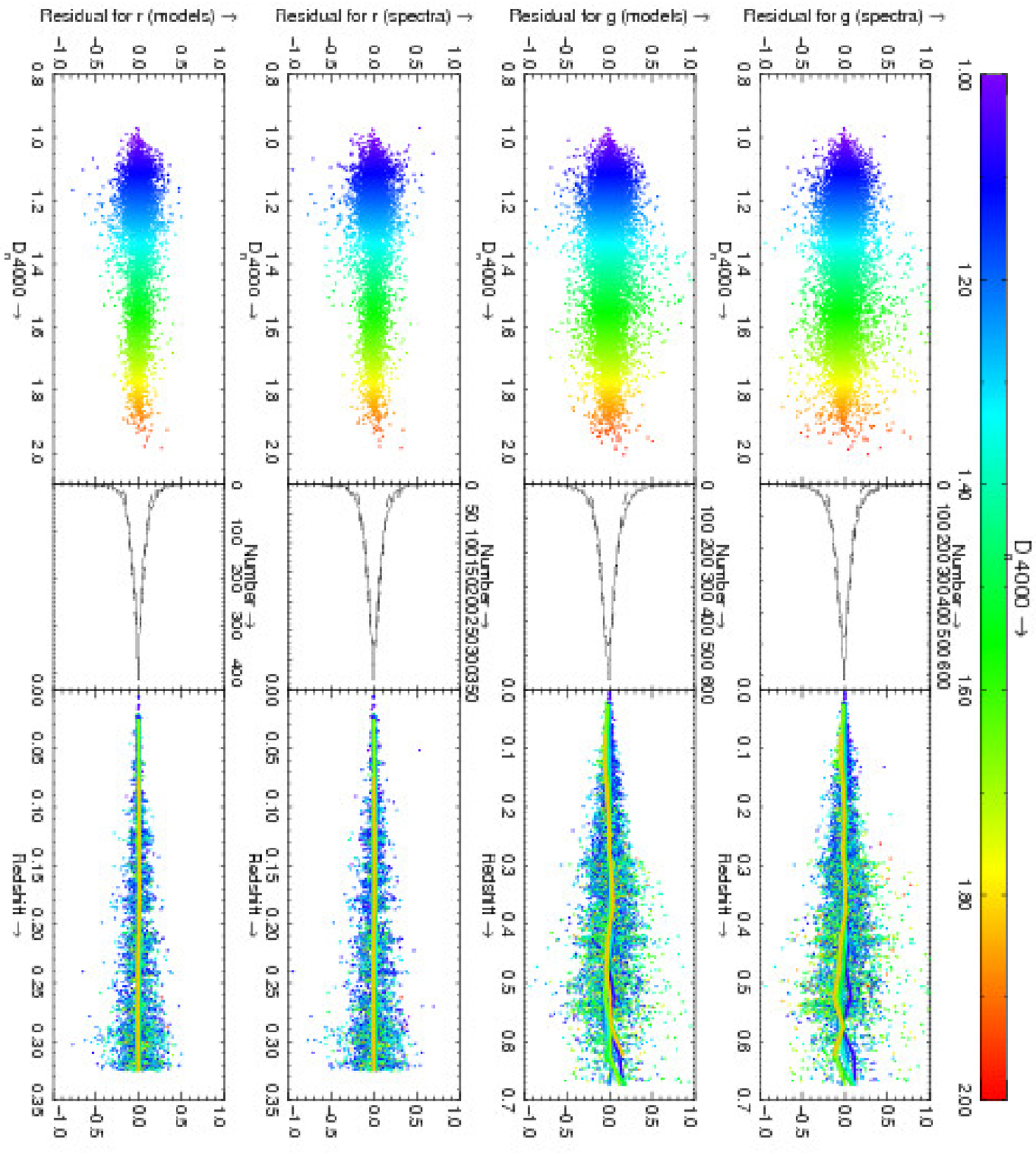}
  \caption{Assessment of the accuracy of the analytic approximations
  for the k-corrections in the $g$ band as a function of \dn{} and
  redshift. From left to right, the residuals from the surface fitting
  as a function of \dn{}, the distribution of residuals ({\it solid
  line}) and overplotted a Gaussian fit to the distribution ({\it
  dashed line}), and the residuals as a function of redshift, with the
  solid lines indicating the median of the difference binned by \dn{}
  as a function of redshift. We plot these quantities for the
  k-corrections in the SDSS $g$ band derived from the spectra ({\it
  top row}) and the SDSS $g$ band derived from the models ({\it second
  row}). The points are color-coded by the \dn{} of the galaxy. The
  bottom two rows are the same as the top two rows but for the SDSS
  $r$ band.}
  \label{fig:residuals_d4000_g}
\end{figure*}

\begin{figure*}
  \centering
  \includeIDLfigPcustom[0.95\textwidth]{0pt}{70pt}{10pt}{80pt}{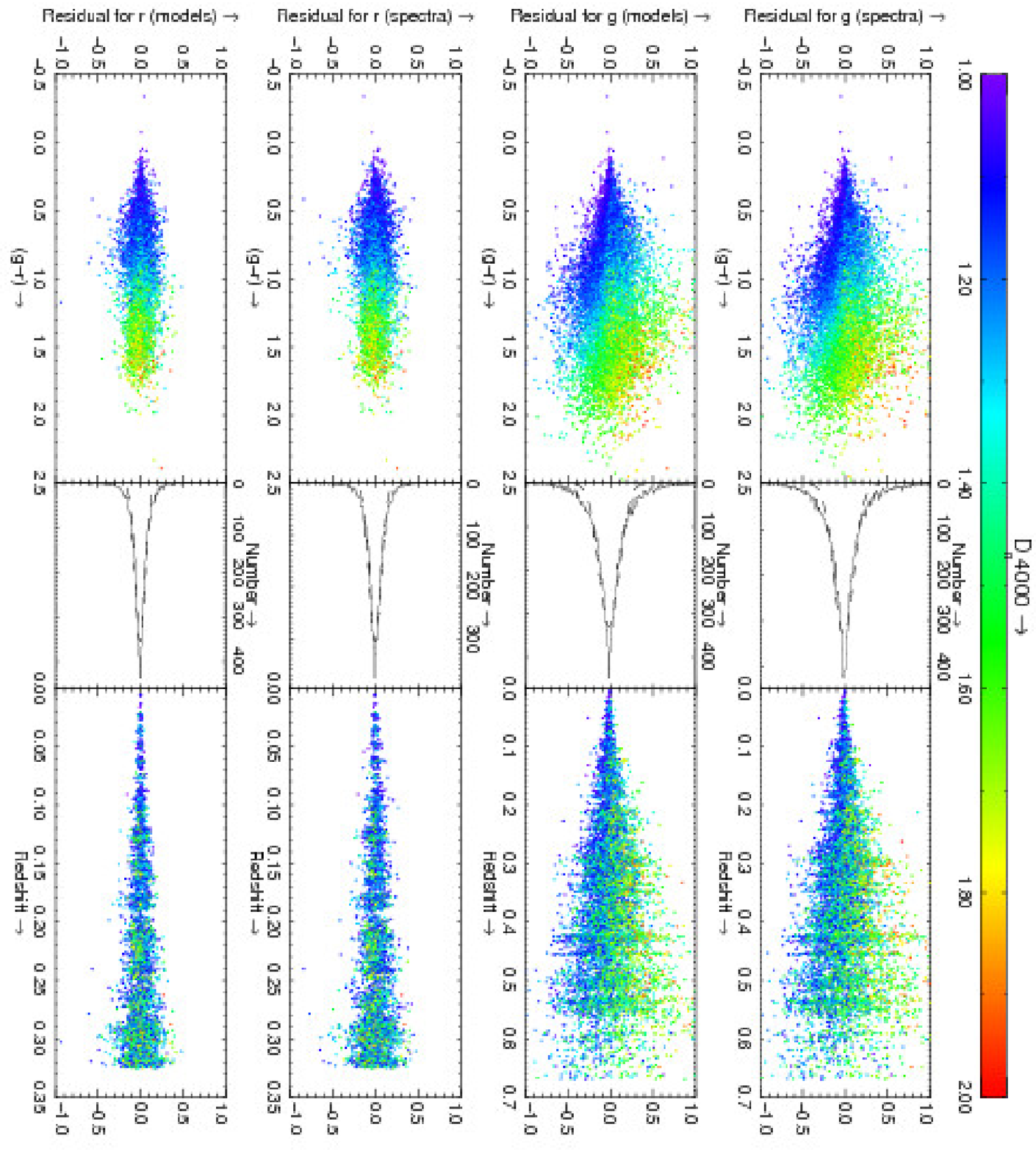}
  \caption{Assessment of the accuracy of the analytic approximations
  for the k-corrections in the $g$-band as a function of \grCol{} and
  redshift. From left to right, the residuals from the surface fitting
  as function of \grCol{}, the distribution of residuals ({\it solid
  line}) and overplotted a Gaussian fit to the distribution ({\it
  dashed line}), and the residuals as a function of redshift with the
  solid lines indicating the median of the difference binned by \dn{}
  as a function of redshift. We plot these quantities for the
  k-corrections in the SDSS $g$ band derived from the spectra ({\it
  top row}) and the SDSS $g$ band derived from the models ({\it second
  row}). The points are color-coded by the \dn{} of the galaxy. The
  bottom two rows are the same as the top two rows but for the SDSS
  $r$ band.}
  \label{fig:residuals_gr_g}
\end{figure*}

\begin{figure*}
  \centering
  \includeIDLfigPcustom[0.495\textwidth]{13pt}{12pt}{5pt}{5pt}{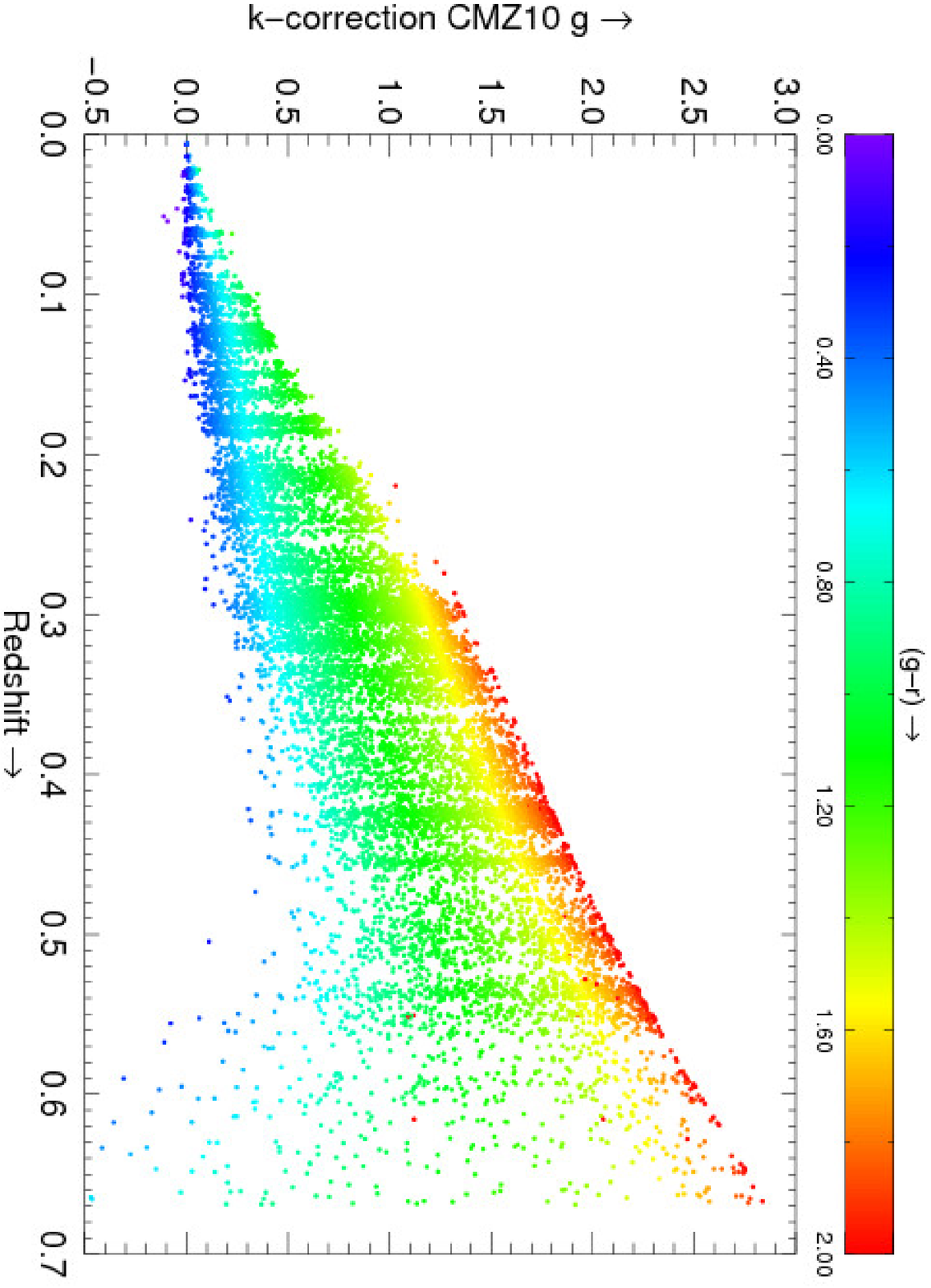}
  \includeIDLfigPcustom[0.495\textwidth]{13pt}{12pt}{5pt}{5pt}{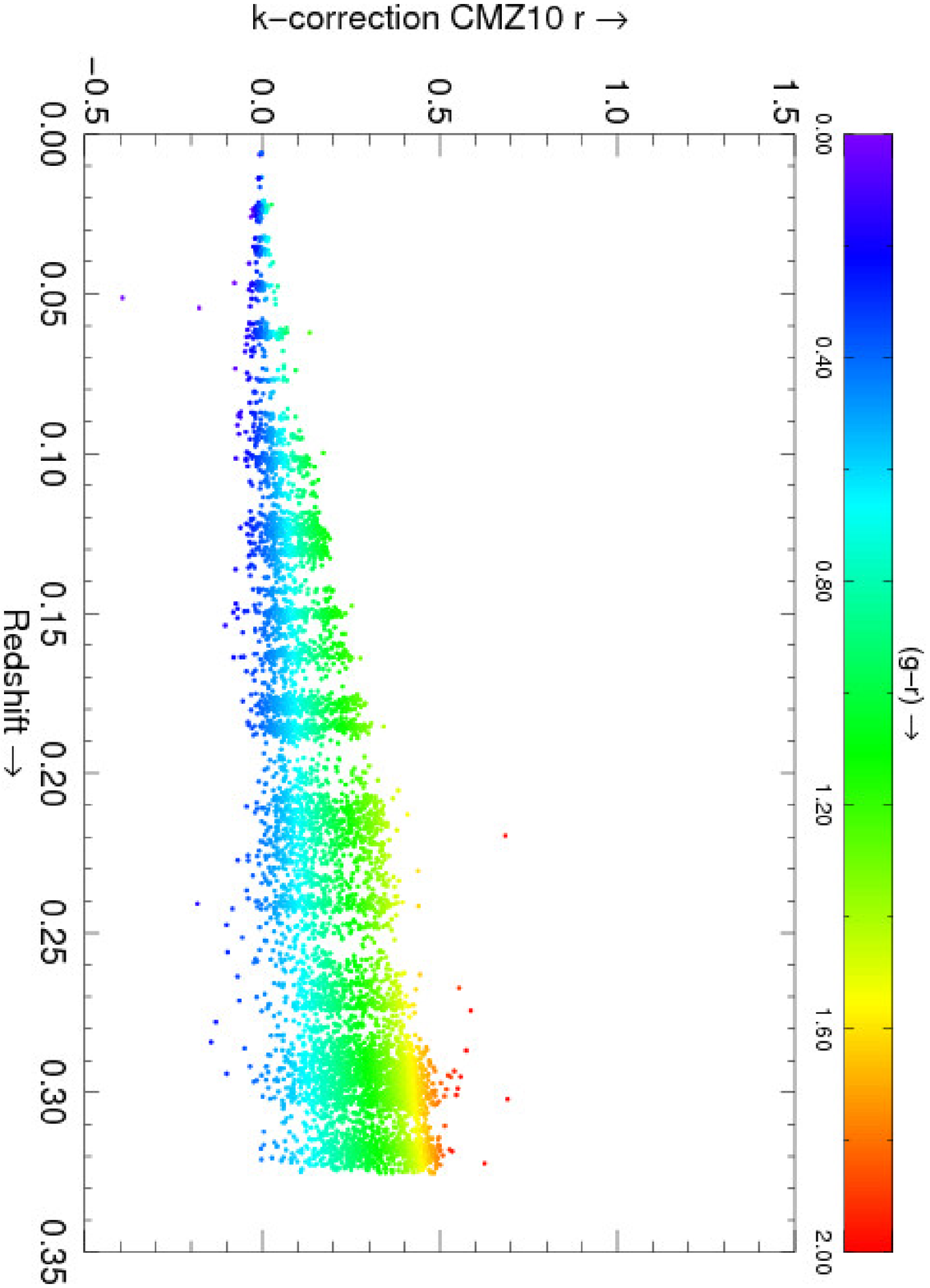}
  \caption{K-corrections for our galaxies derived from the
  prescription given by \citetalias{Chilingarian10} based on their
  {\sc pegase.2} models for the $g$ ({\it left}) and $r$ band ({\it
  right}). Each point is color-coded by the \grCol{} of the galaxy.}
  \label{fig:chilkcor}
\end{figure*}

\begin{figure*}
  \centering
  \includeIDLfigPcustom[0.495\textwidth]{13pt}{12pt}{5pt}{5pt}{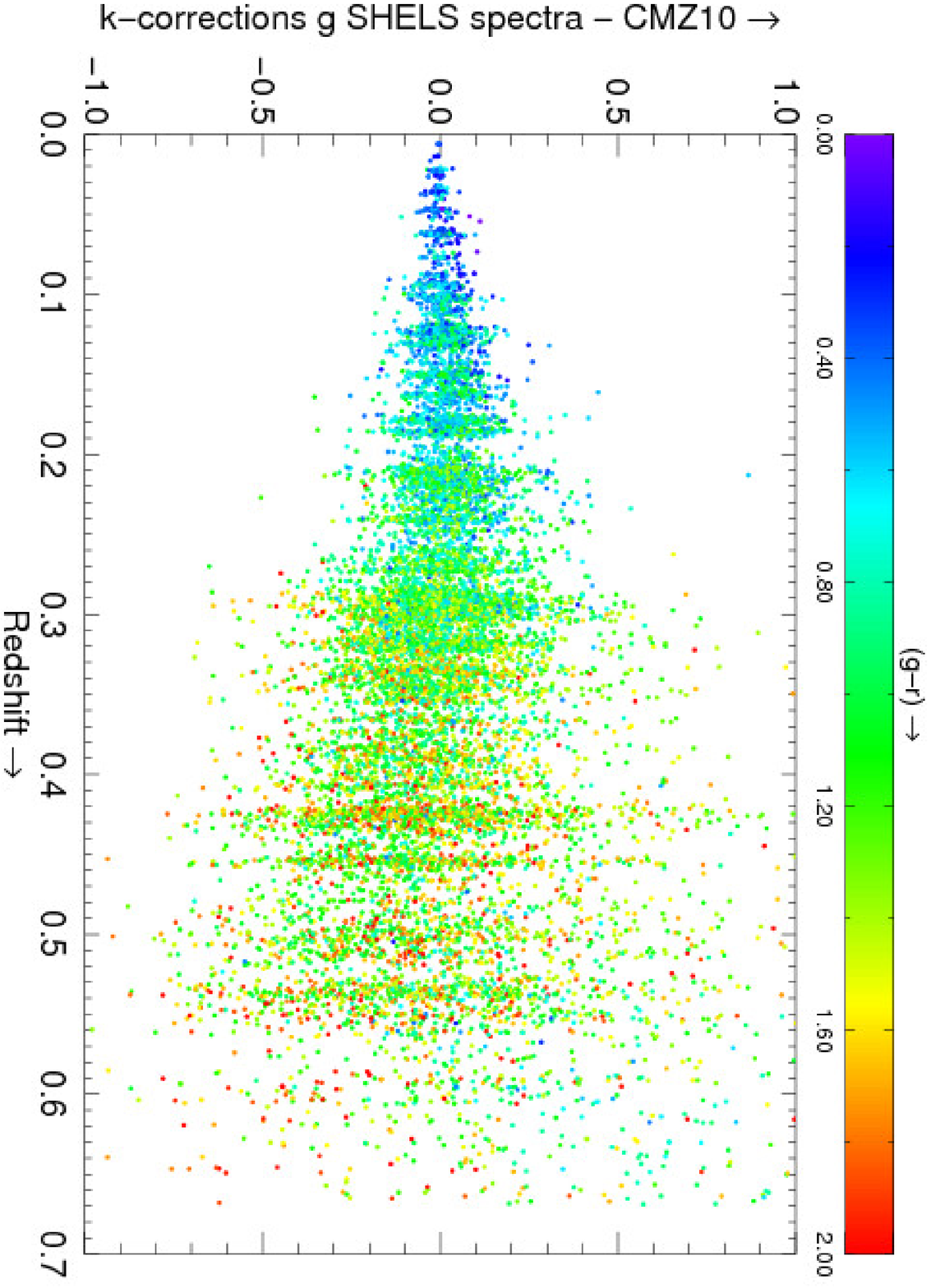}
  \includeIDLfigPcustom[0.495\textwidth]{13pt}{12pt}{5pt}{5pt}{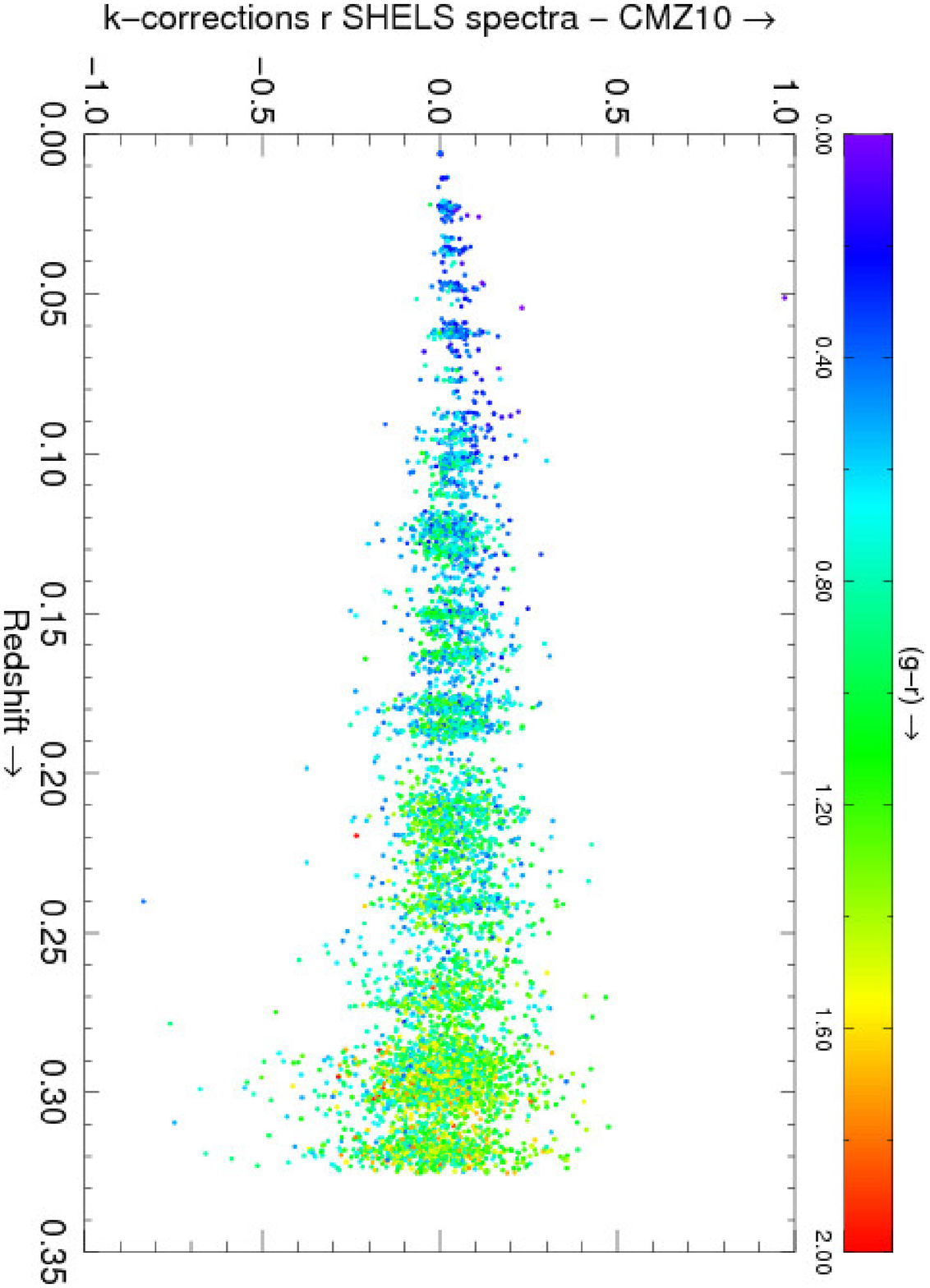}
  \includeIDLfigPcustom[0.495\textwidth]{13pt}{12pt}{5pt}{5pt}{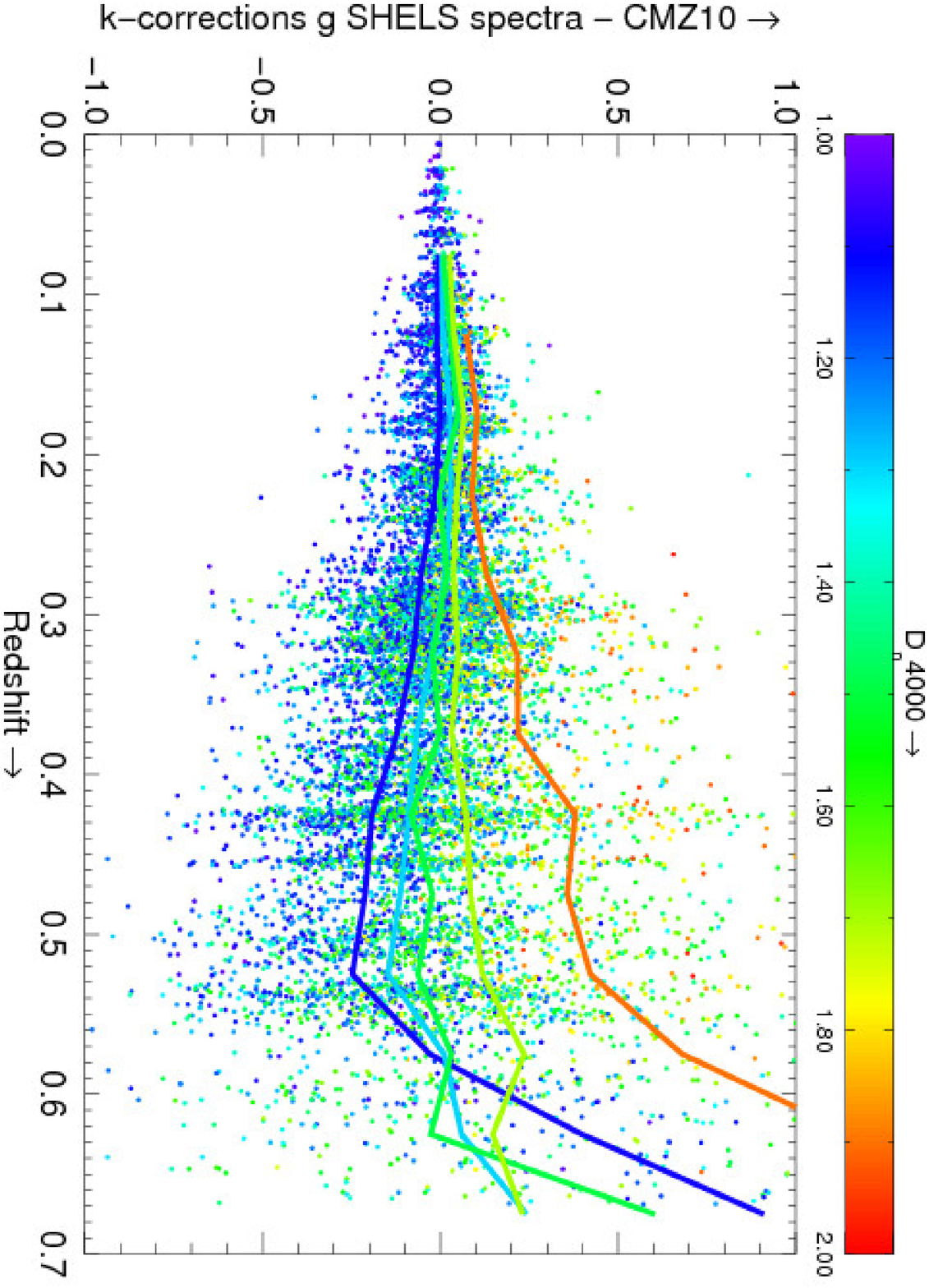}
  \includeIDLfigPcustom[0.495\textwidth]{13pt}{12pt}{5pt}{5pt}{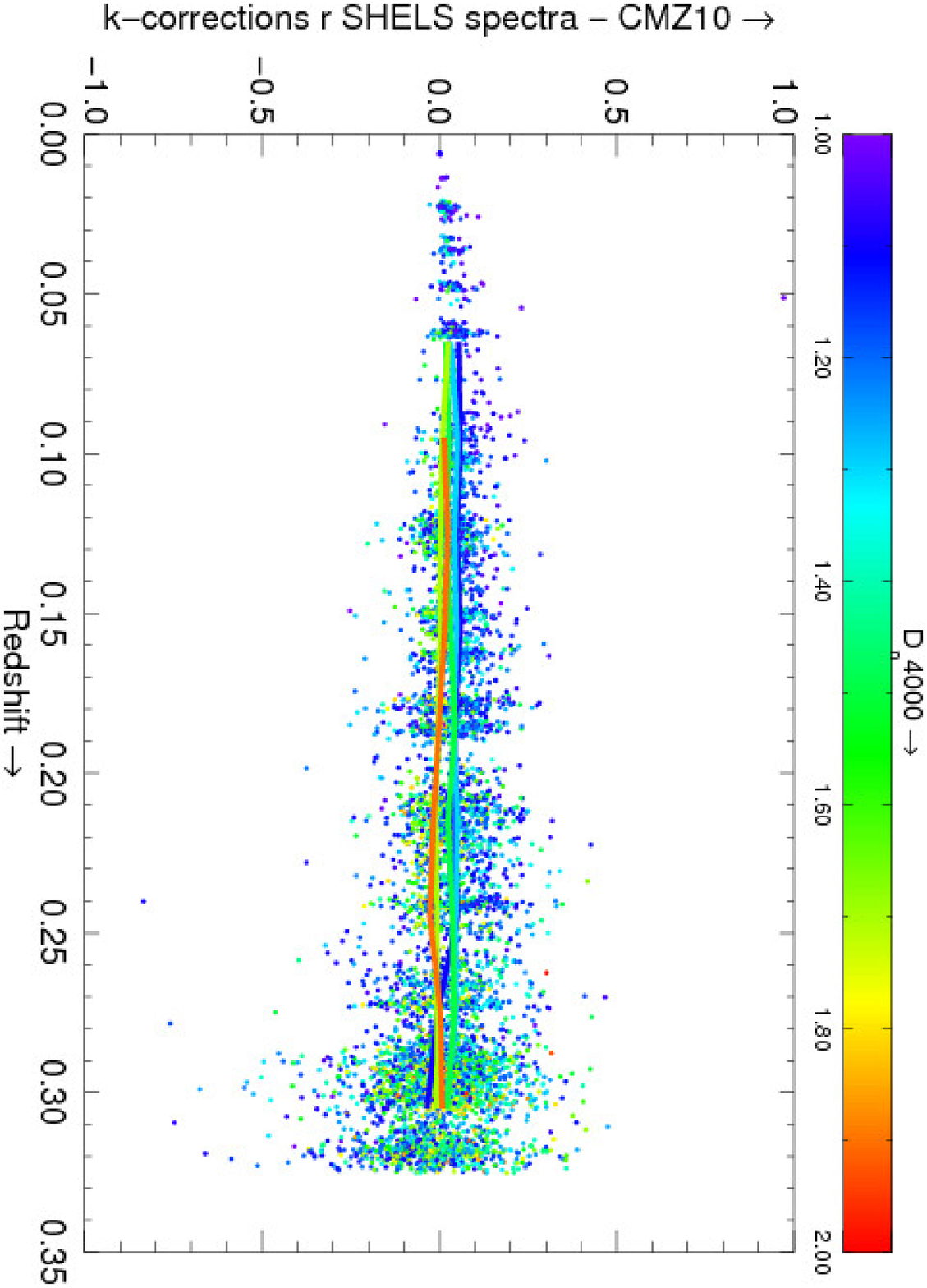}
  \includeIDLfigPcustom[0.495\textwidth]{13pt}{12pt}{5pt}{5pt}{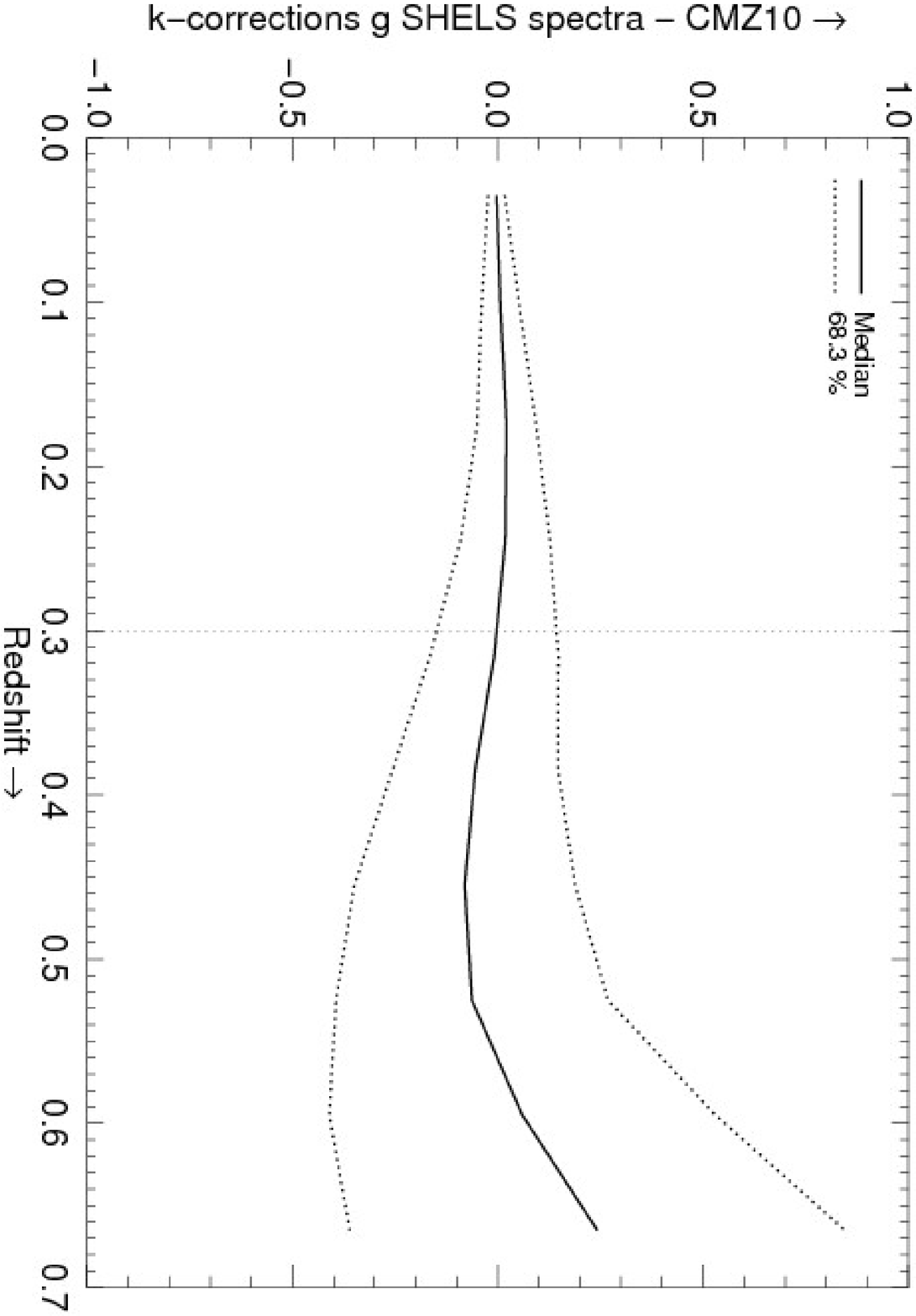}
  \includeIDLfigPcustom[0.495\textwidth]{13pt}{12pt}{5pt}{5pt}{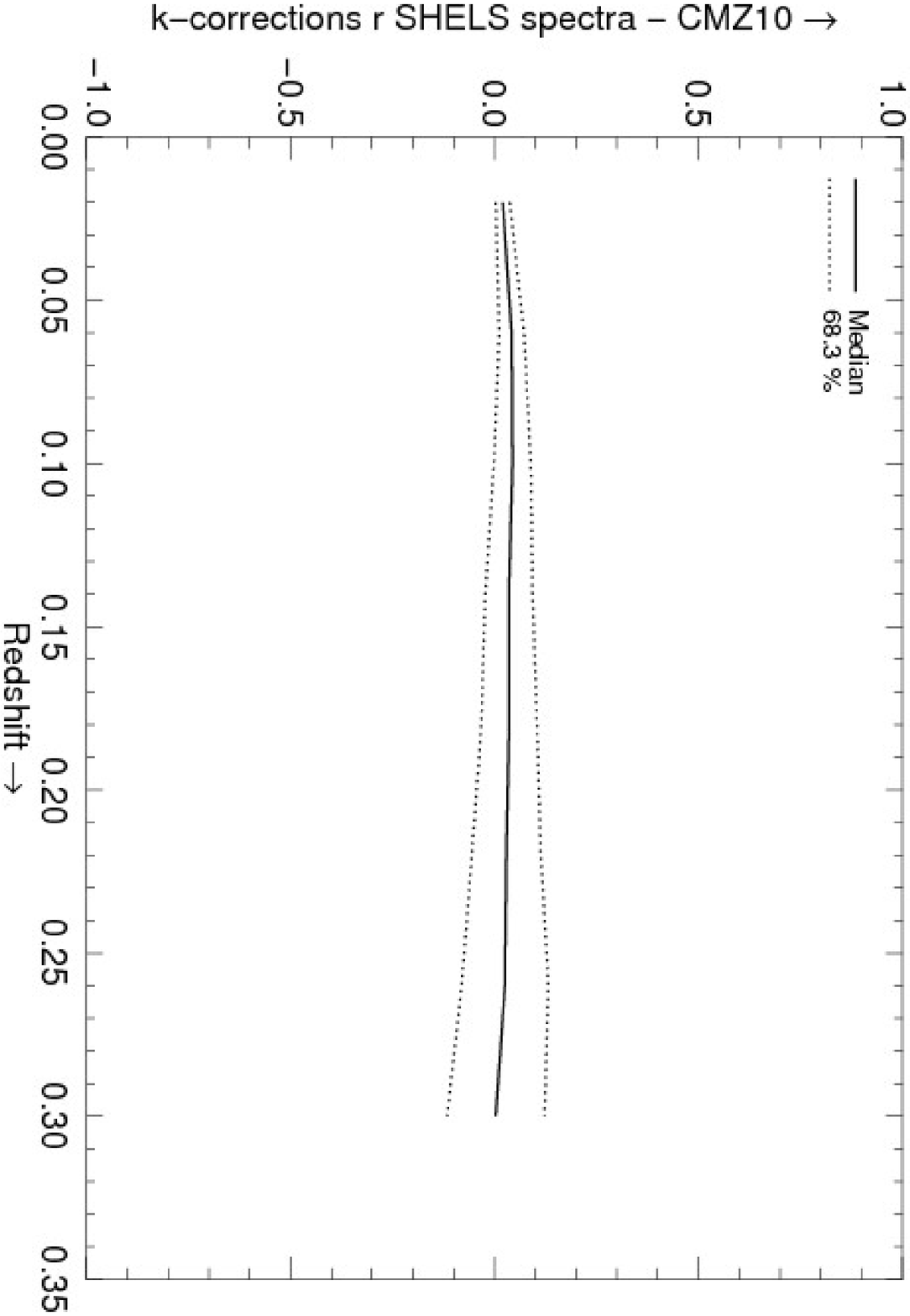}
  \caption{Comparison of the k-correction derived from our spectra and
  the prescription given by \citetalias{Chilingarian10} based on their
  {\sc pegase.2} models for the $g$ ({\it left}) and $r$ band ({\it
  right}). Each point is color-coded by the \grCol{} ({\it top}) and
  the \dn{} ({\it middle}) of the galaxy. The colored solid lines
  indicate the median of the difference binned by \dn{} as a function
  of redshift. The median ({\it solid line}) and the 68.3\,\% range
  ({\it dotted lines}) around the median for graphs in the top and
  middle panel ({\it bottom}). The vertical dotted line in the left
  panel indicates the maximum redshift where
  \citetalias{Chilingarian10} have enough green and blue galaxies
  to constrain their prescription.}
  \label{fig:chilcomp}
\end{figure*}

\begin{figure*}
  \centering
  \includeIDLfigPcustom[0.495\textwidth]{13pt}{12pt}{5pt}{5pt}{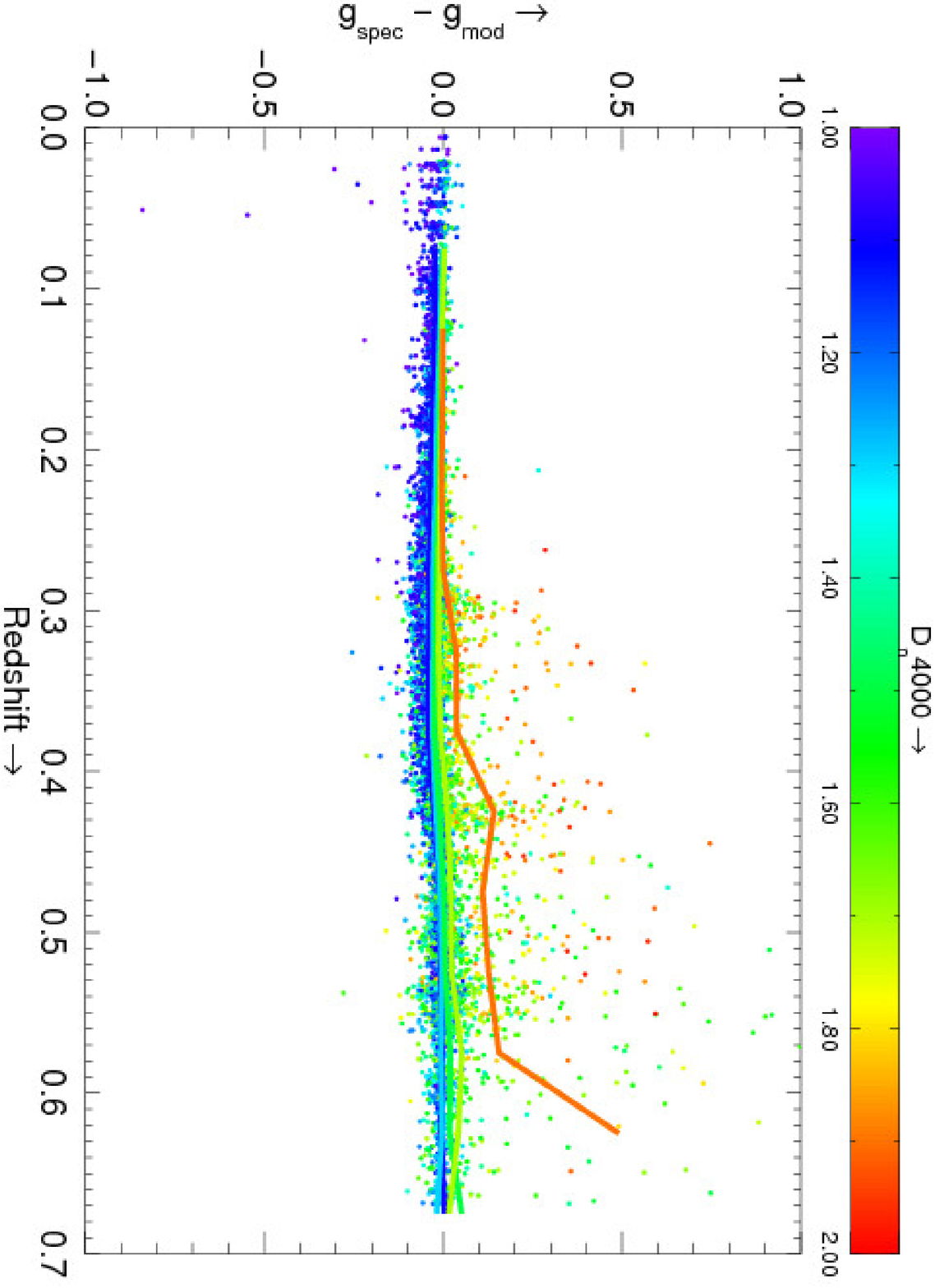}
  \includeIDLfigPcustom[0.495\textwidth]{13pt}{12pt}{5pt}{5pt}{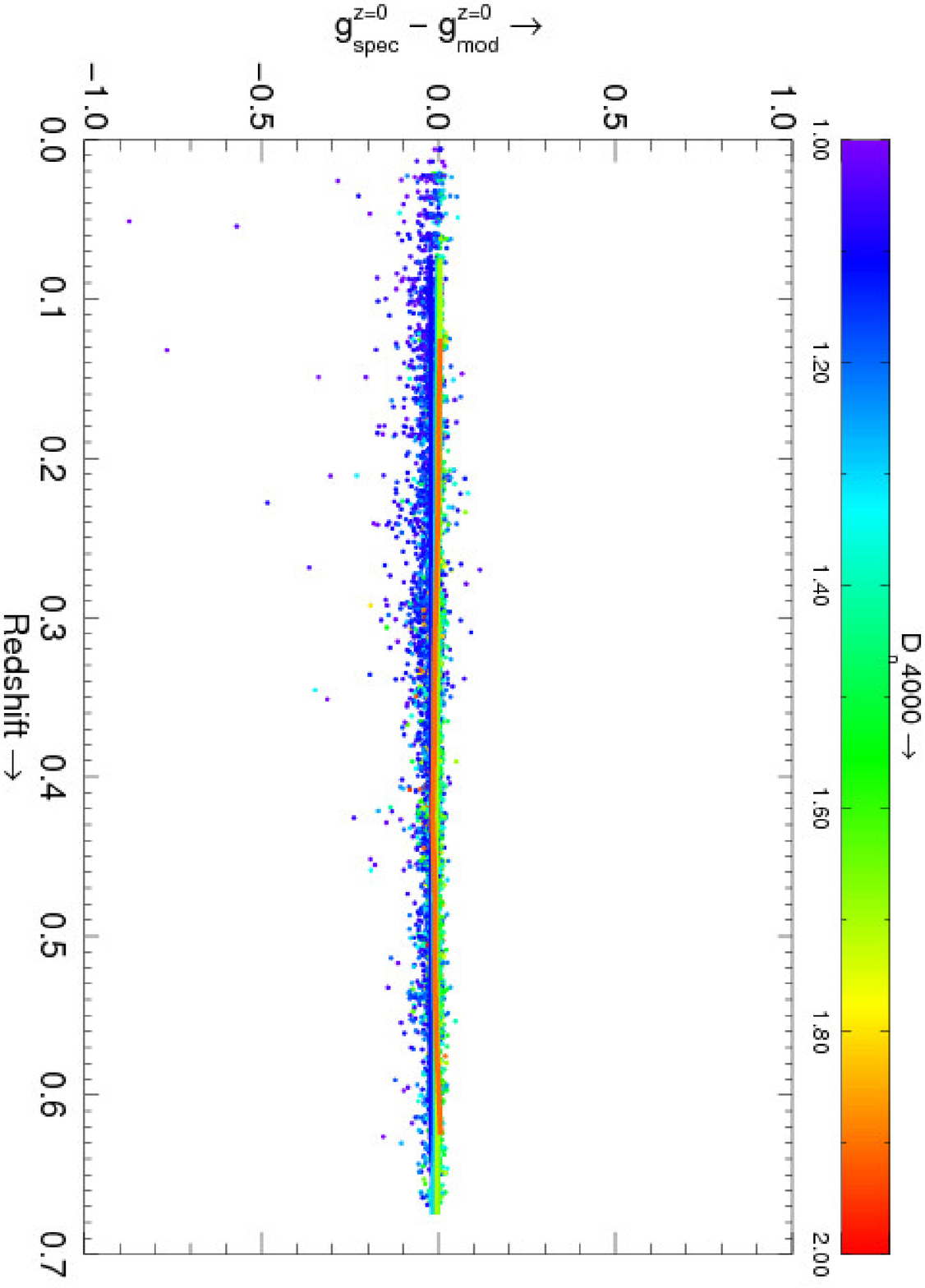}
  \includeIDLfigPcustom[0.495\textwidth]{13pt}{12pt}{5pt}{5pt}{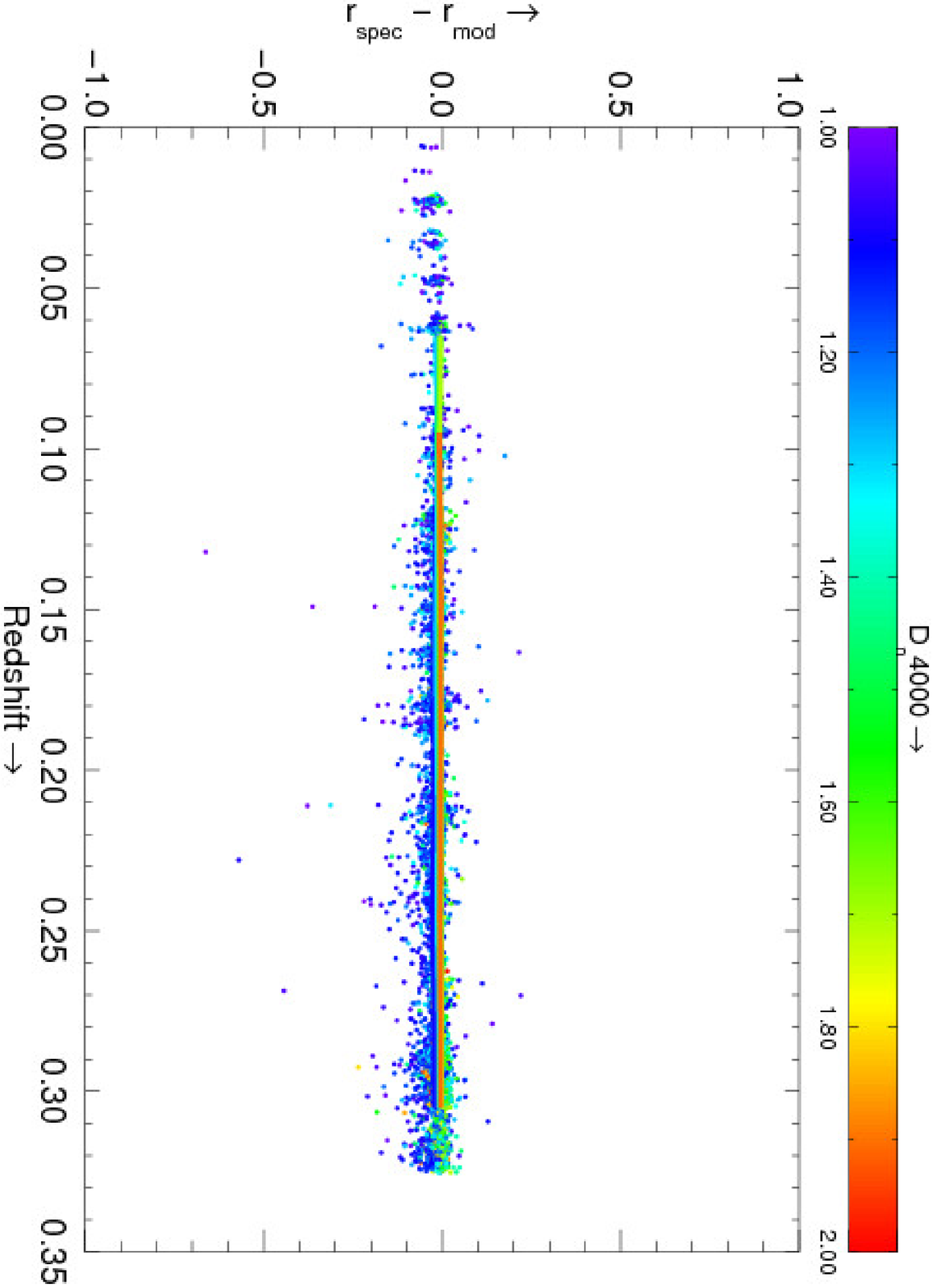}
  \includeIDLfigPcustom[0.495\textwidth]{13pt}{12pt}{5pt}{5pt}{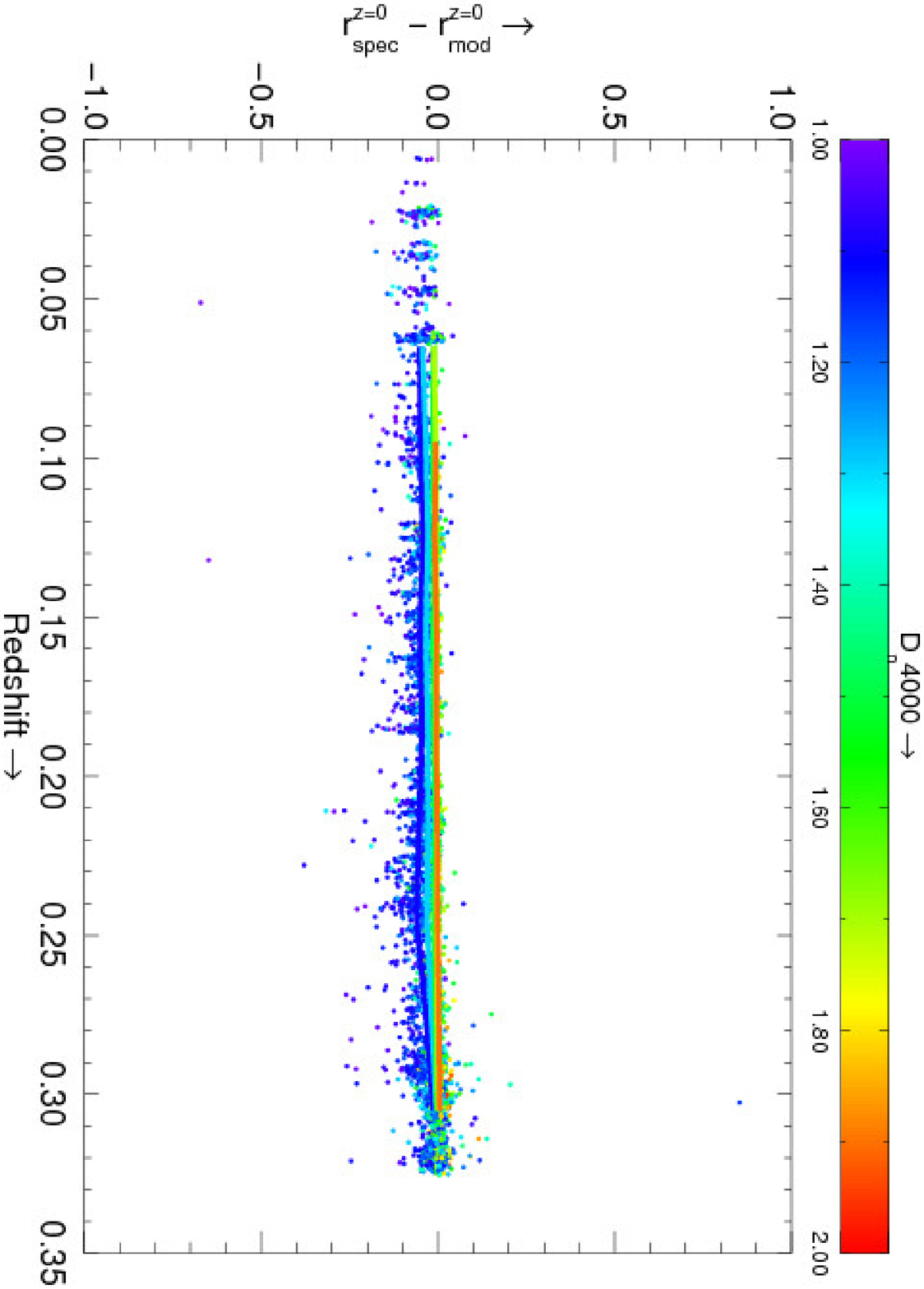}
  \caption{Test to determine the discrepancy between our k-corrections
    and those from \citetalias{Chilingarian10} for high \dn{}
    galaxies; difference between the magnitude derived from the
    spectrum and of the model fit for the $g$ ({\it top}) and
    $r$ band ({\it bottom}) for the observed ({\it left}) and rest
    frame ({\it right}). Each point is color-coded by the \dn{} of the
    galaxy. The solid lines indicate the median of the difference
    binned by \dn{} as a function of redshift. The Figure also shows
    the influence of emission lines on magnitudes; almost all galaxies
    with low \dn{} and away from the median of the difference have
    emission lines.}
  \label{fig:discrepancy}
\end{figure*}

\begin{figure}
  \centering
  \includeIDLfigPcustom{15pt}{14pt}{25pt}{25pt}{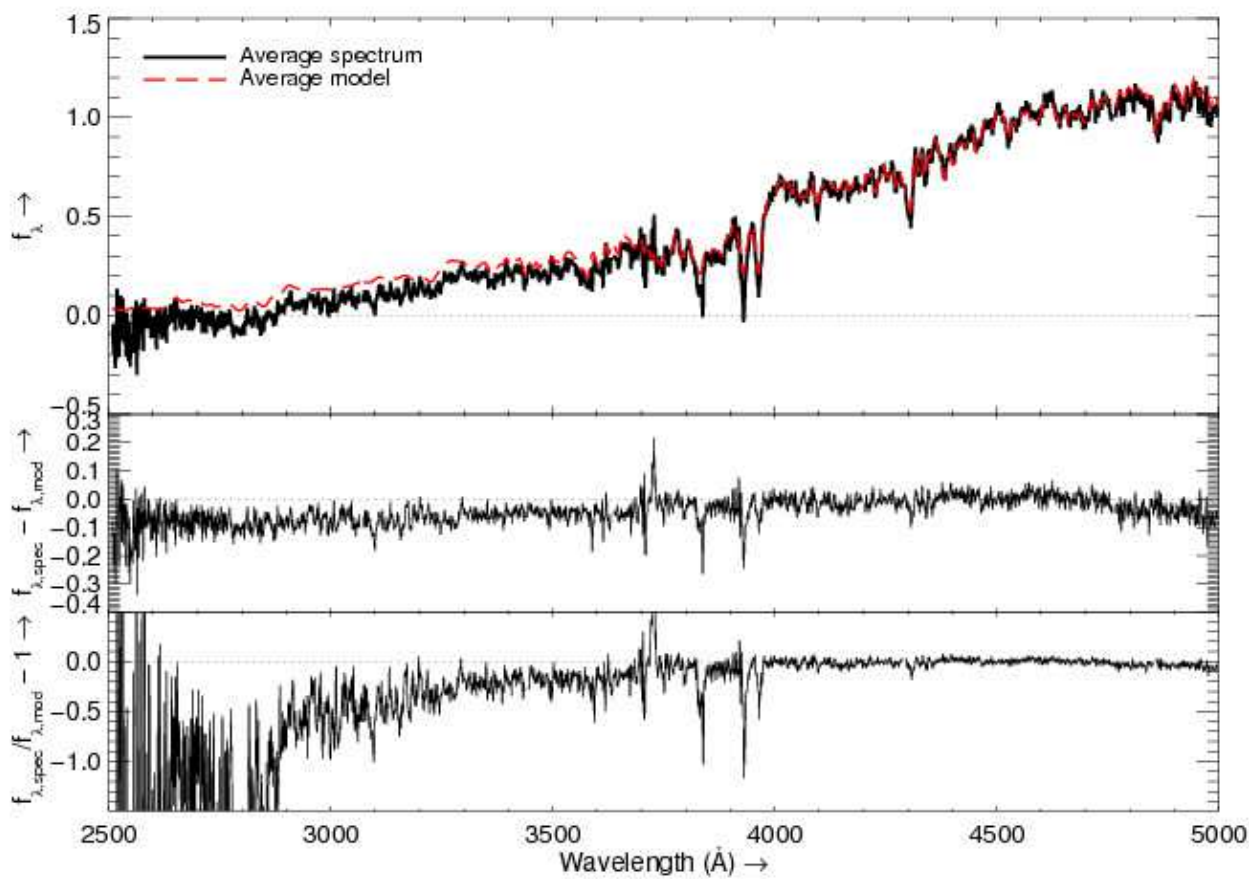}
  \caption{The averaged luminosity-weighted rest-frame spectrum of 84
  galaxies with $0.4 < z < 0.6$, $\dn > 1.7$ and a difference between
  the magnitude derived from the spectrum and the fit larger than 0.2
  ({\it top}; see Figure~\ref{fig:discrepancy}), the difference
  between the average spectrum and fit ({\it middle}), and the relative
  difference with respect to the fit ({\it bottom}). The dotted line
  in each panel indicates the zero-level.}
  \label{fig:summed}
\end{figure}

\begin{figure*}
  \centering
  \includeIDLfigPcustom[\textwidth]{130pt}{12pt}{130pt}{20pt}{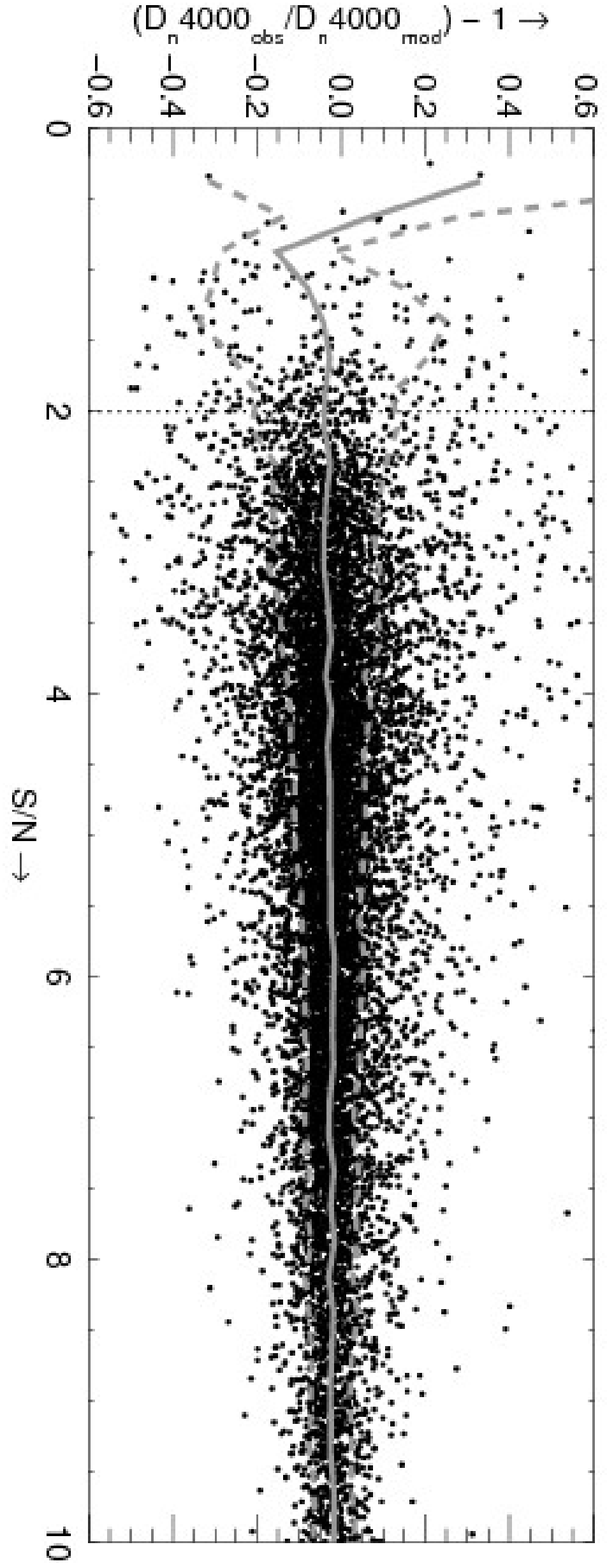}
  \includeIDLfigPcustom[0.49\textwidth]{10pt}{12pt}{5pt}{5pt}{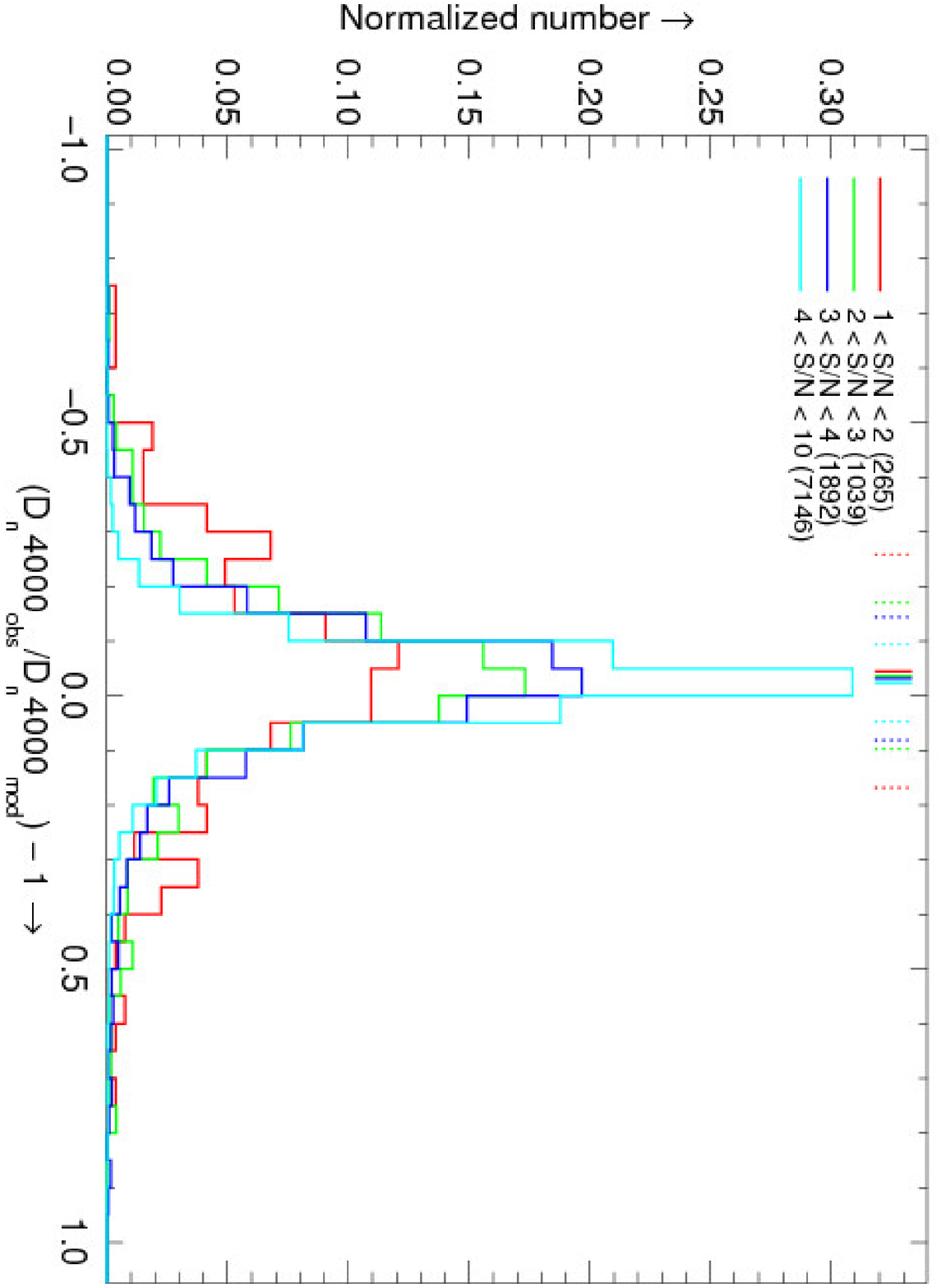}
  \includeIDLfigPcustom[0.49\textwidth]{10pt}{12pt}{5pt}{5pt}{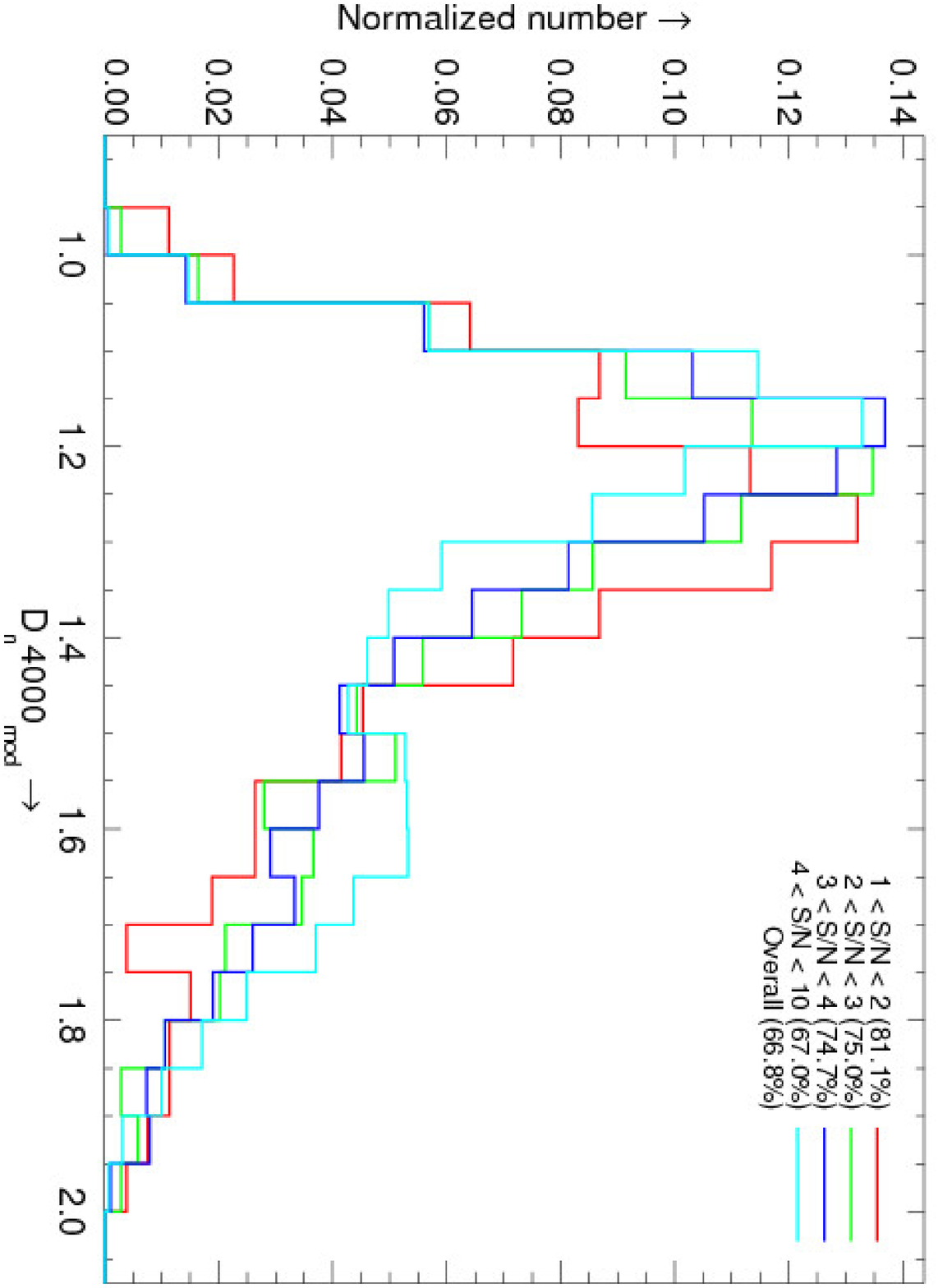}
  \caption{Test of the accuracy of the model-derived \dn{}. Relative
  difference between the spectrum- and model-derived \dn{} as a
  function of S/N of the spectrum ({\it top}). The {\it solid line}
  indicates the median and the {\it dashed line} the 68.3\,\% range
  around the median. Normalized histograms of relative difference
  between the spectrum- and model-derived \dn{} in different S/N
  ranges ({\it bottom left}). The vertical {\it solid} and {\it
  dashed} lines indicate the median and 68.3\,\% range around the
  median for each sample. The number in brackets is the number of
  galaxies in each S/N range. Normalized distribution of model-derived
  \dn{} for each of the S/N range ({\it bottom right}). The percentage
  in the brackets is the fraction of galaxies with $\dn \le 1.46$ for
  each sample.}
  \label{fig:sntest}
\end{figure*}

\begin{figure*}
  \centering
  \includeIDLfigPcustom[\textwidth]{25pt}{25pt}{25pt}{20pt}{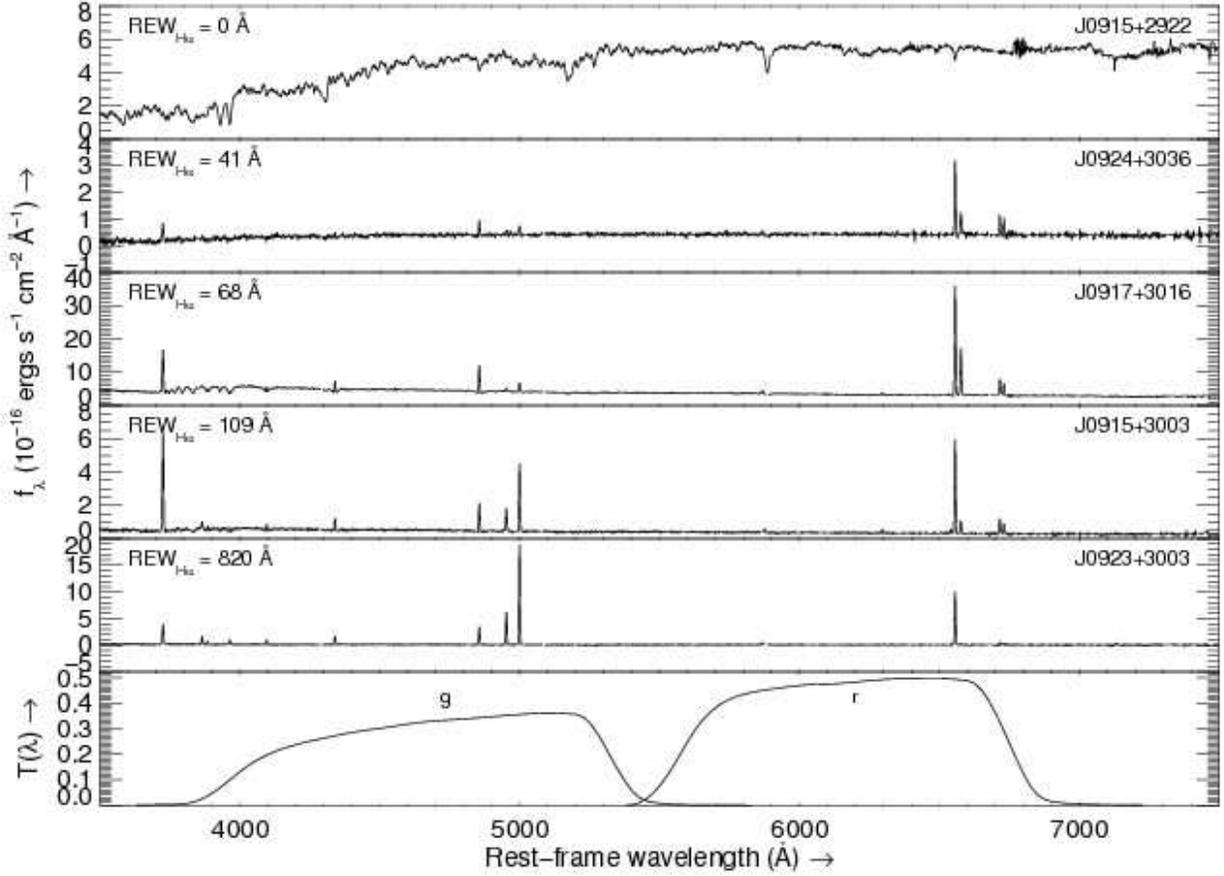}
  \caption{Rest frame spectra of five galaxies with different amounts
  of line emission and (detectable) emission lines to study the effect
  of the emission lines on the k-correction. The REW of \ha{} quoted
  in each panel is corrected for underlying stellar absorption. Each
  of these galaxies has a redshift of $\sim 0.13-0.14$. The bottom
  panel shows the transmission curves of the $g$ and $r$ band for
  reference to the rest-frame spectra.}
  \label{fig:ewspectra}
\end{figure*}

\begin{figure*}
  \centering
  \includeIDLfigPcustom[0.495\textwidth]{18pt}{17pt}{25pt}{10pt}{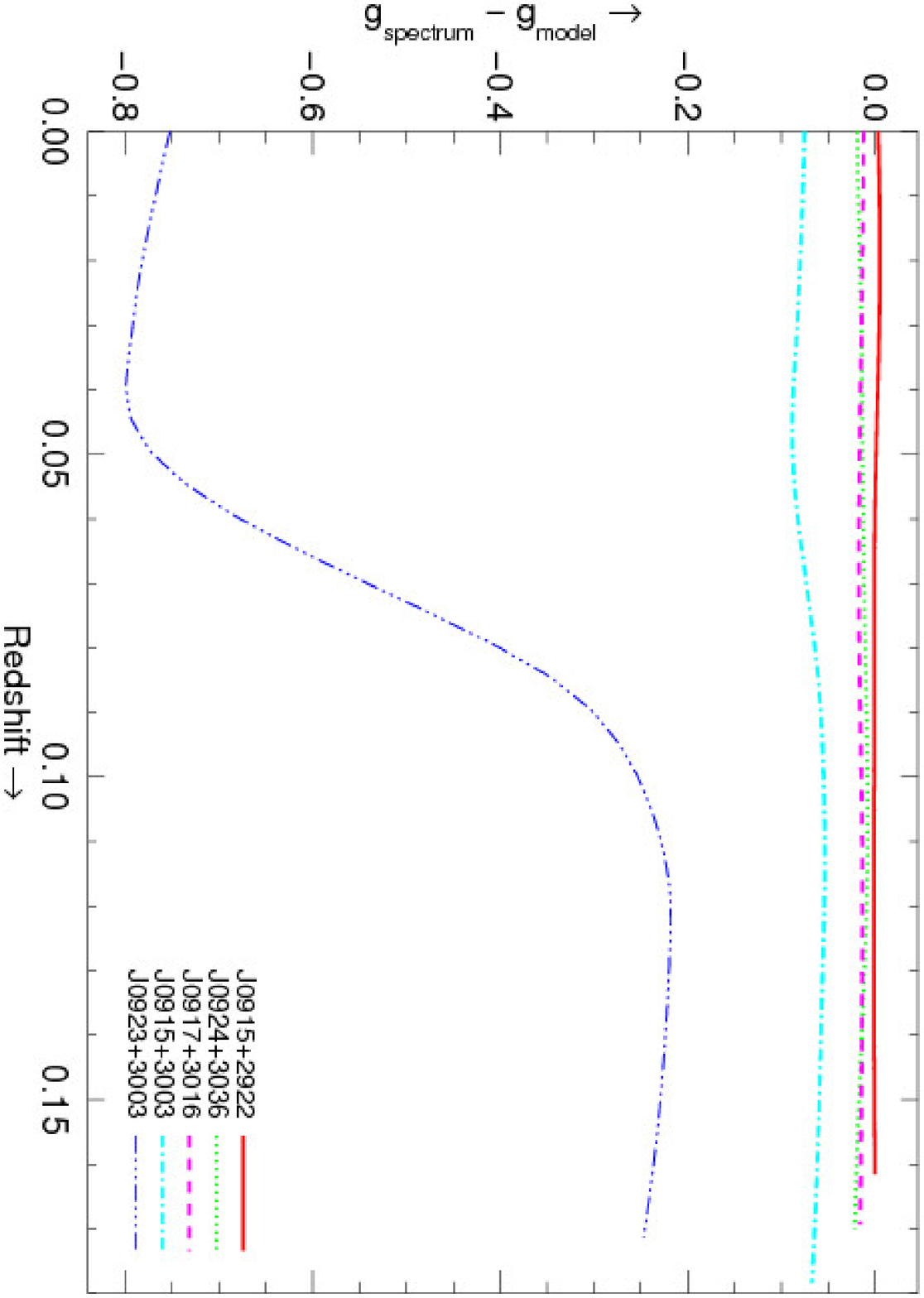}
  \includeIDLfigPcustom[0.495\textwidth]{18pt}{17pt}{25pt}{10pt}{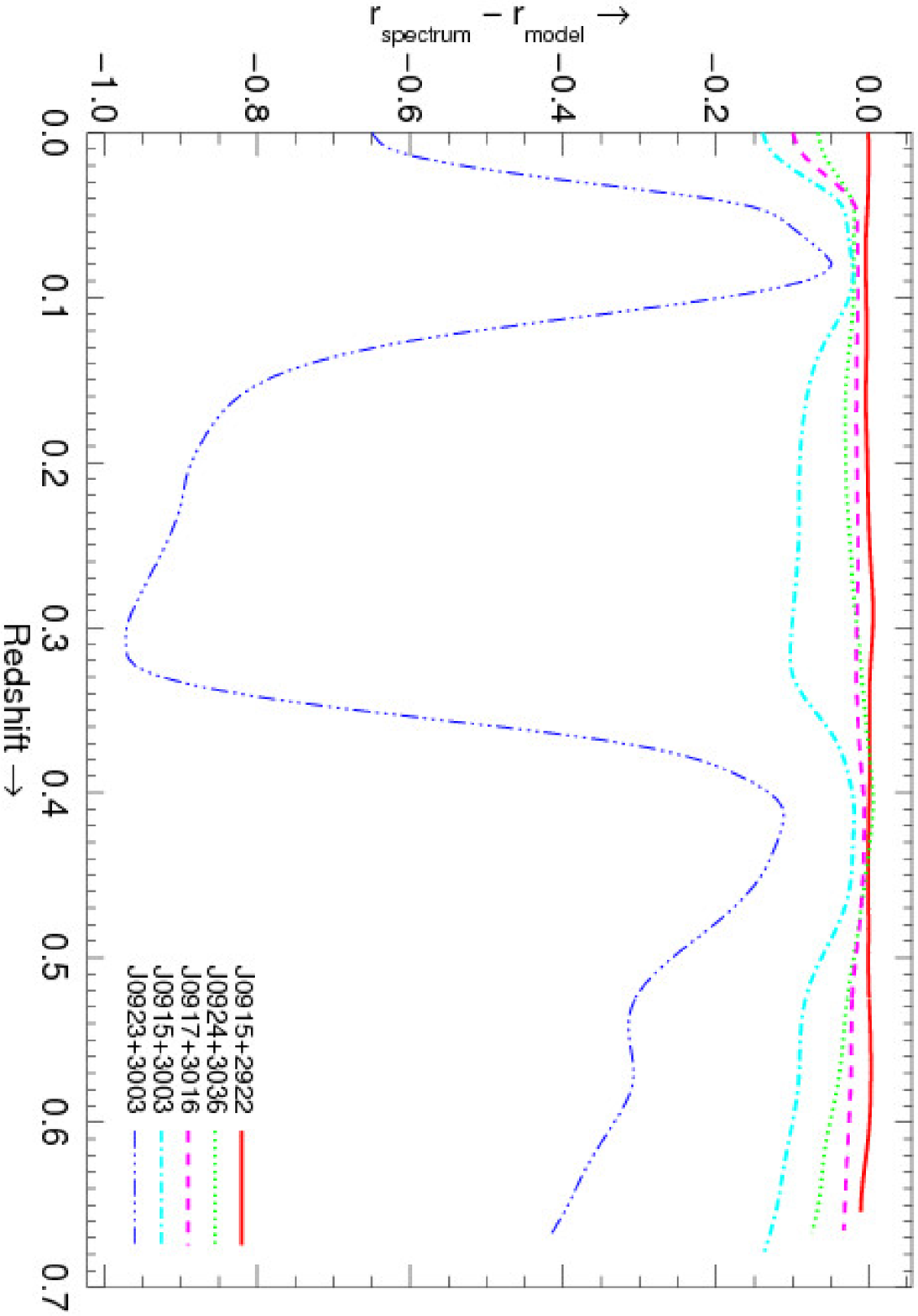}
  \includeIDLfigPcustom[0.495\textwidth]{18pt}{17pt}{25pt}{10pt}{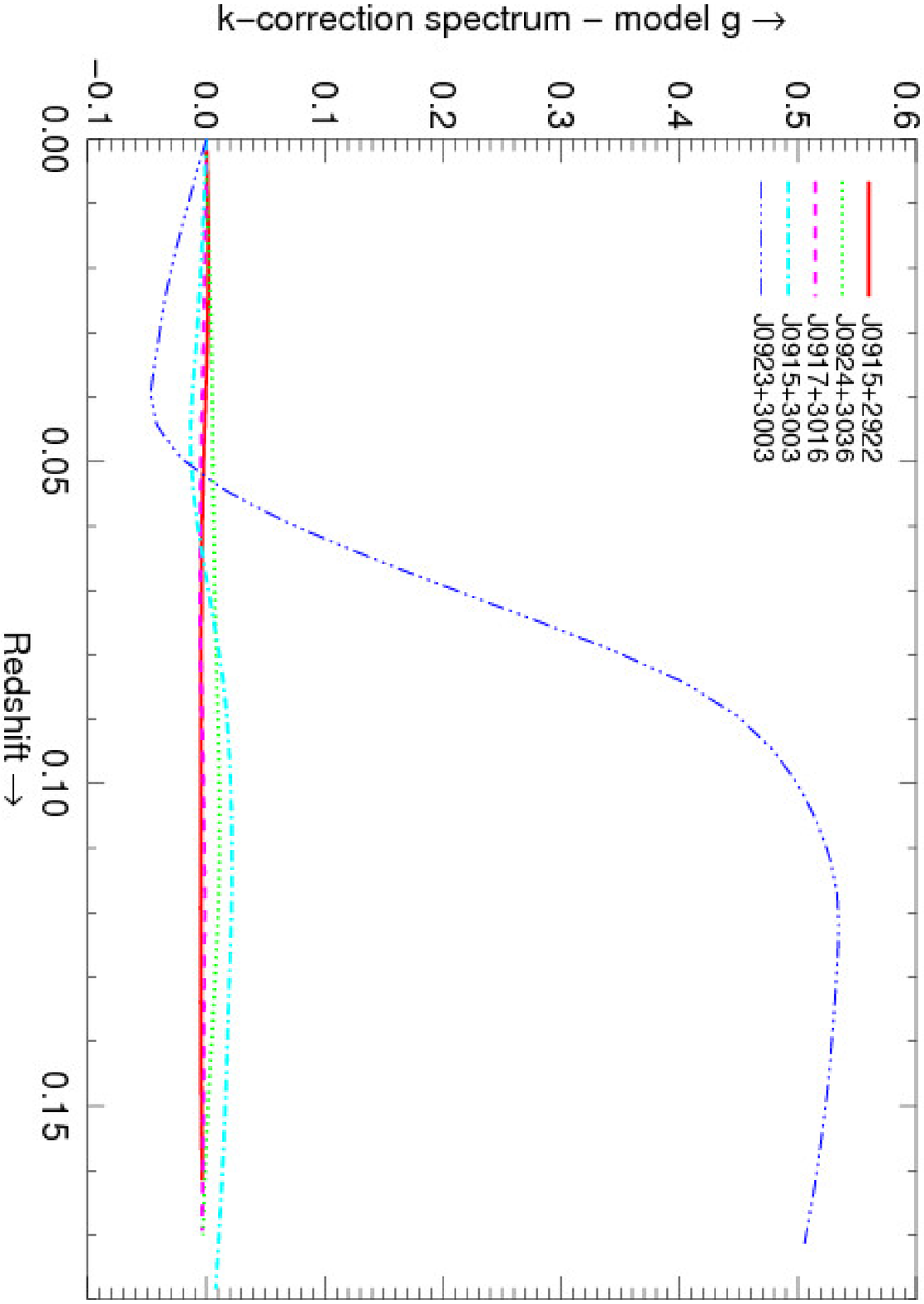}
  \includeIDLfigPcustom[0.495\textwidth]{18pt}{17pt}{25pt}{10pt}{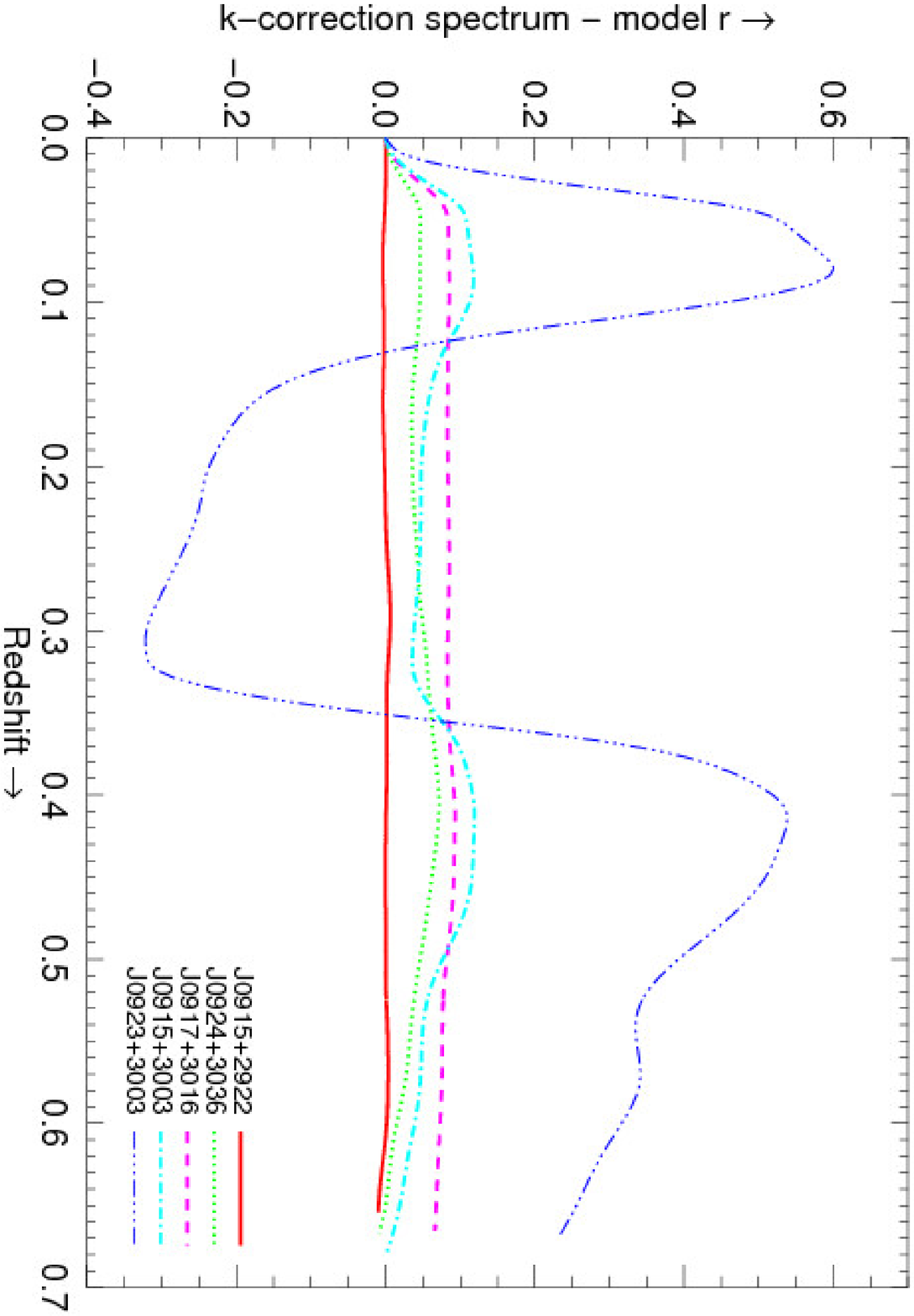}
  \caption{Influence of the presence of emission lines on the
  magnitude ({\it top}) and k-correction ({\it bottom}) for the $g$
  ({\it left}) and $r$ band ({\it right}). For each panel the quantity
  derived from the model is subtracted from that derived from the
  model fit to the data.}
  \label{fig:elinfluence}
\end{figure*}

\begin{figure}
  \centering
  \includeIDLfigP{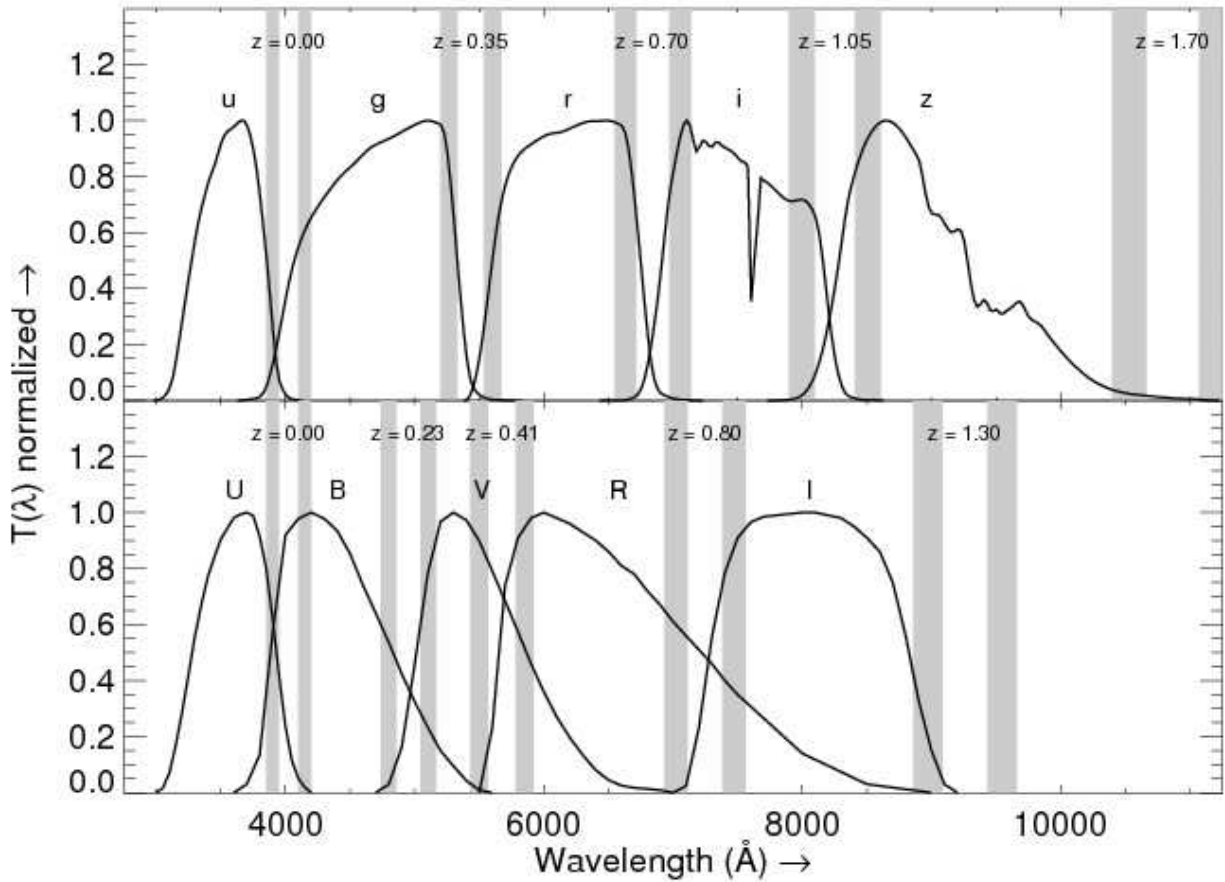}
  \caption{Normalized bandpasses for SDSS $ugriz$ bands ({\it top})
  and Johnson-Cousins $UBVRI$ bands ({\it bottom}) as a function of
  wavelength. The shaded regions indicate the observed wavelength
  ranges where \dn{} would (roughly) be between filters.}
  \label{fig:allfilters}
\end{figure}

\begin{figure*}
  \centering
  \includeIDLfigPcustom[0.35\textwidth]{13pt}{12pt}{5pt}{5pt}{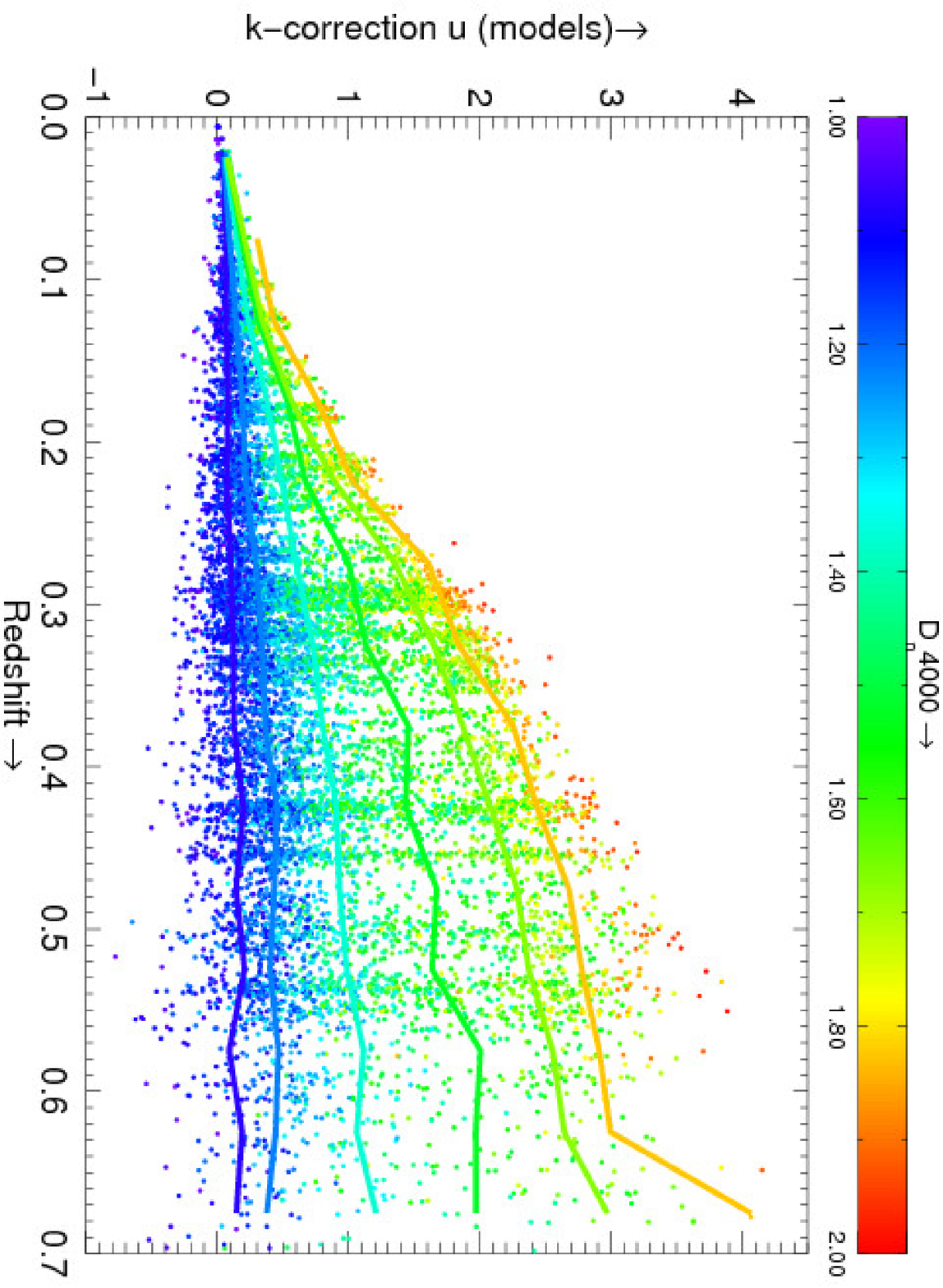}
  \includeIDLfigPcustom[0.35\textwidth]{13pt}{12pt}{5pt}{5pt}{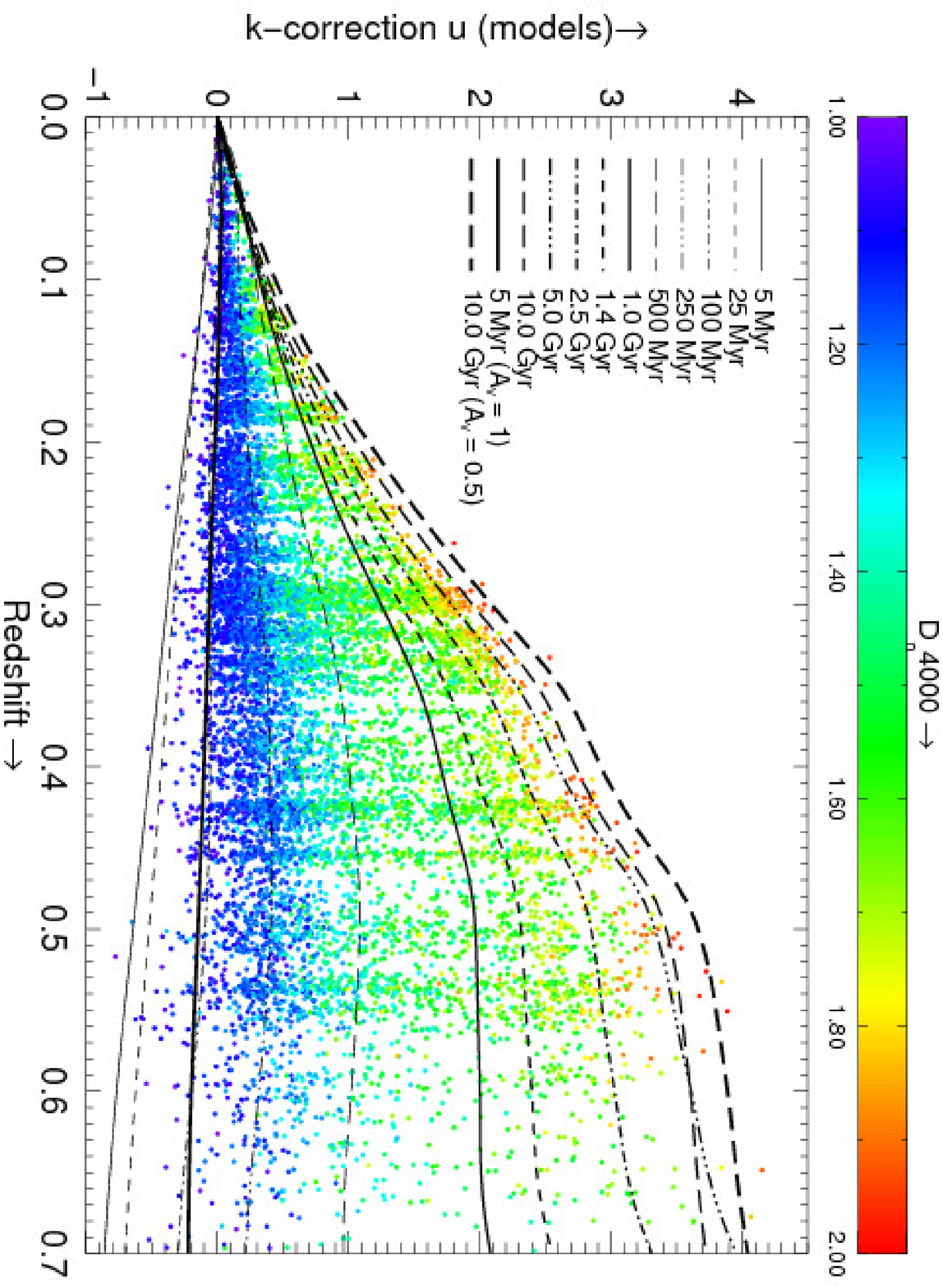}
  \includeIDLfigPcustom[0.70\textwidth]{130pt}{0pt}{158pt}{25pt}{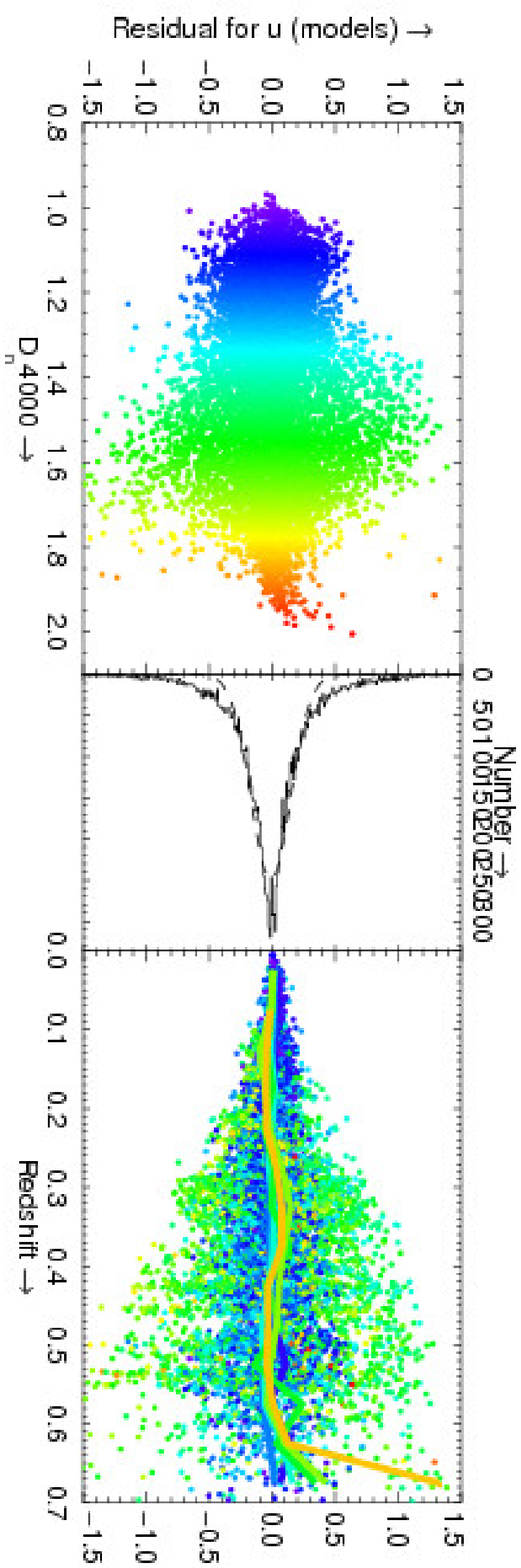}
  \includeIDLfigPcustom[0.35\textwidth]{13pt}{12pt}{5pt}{5pt}{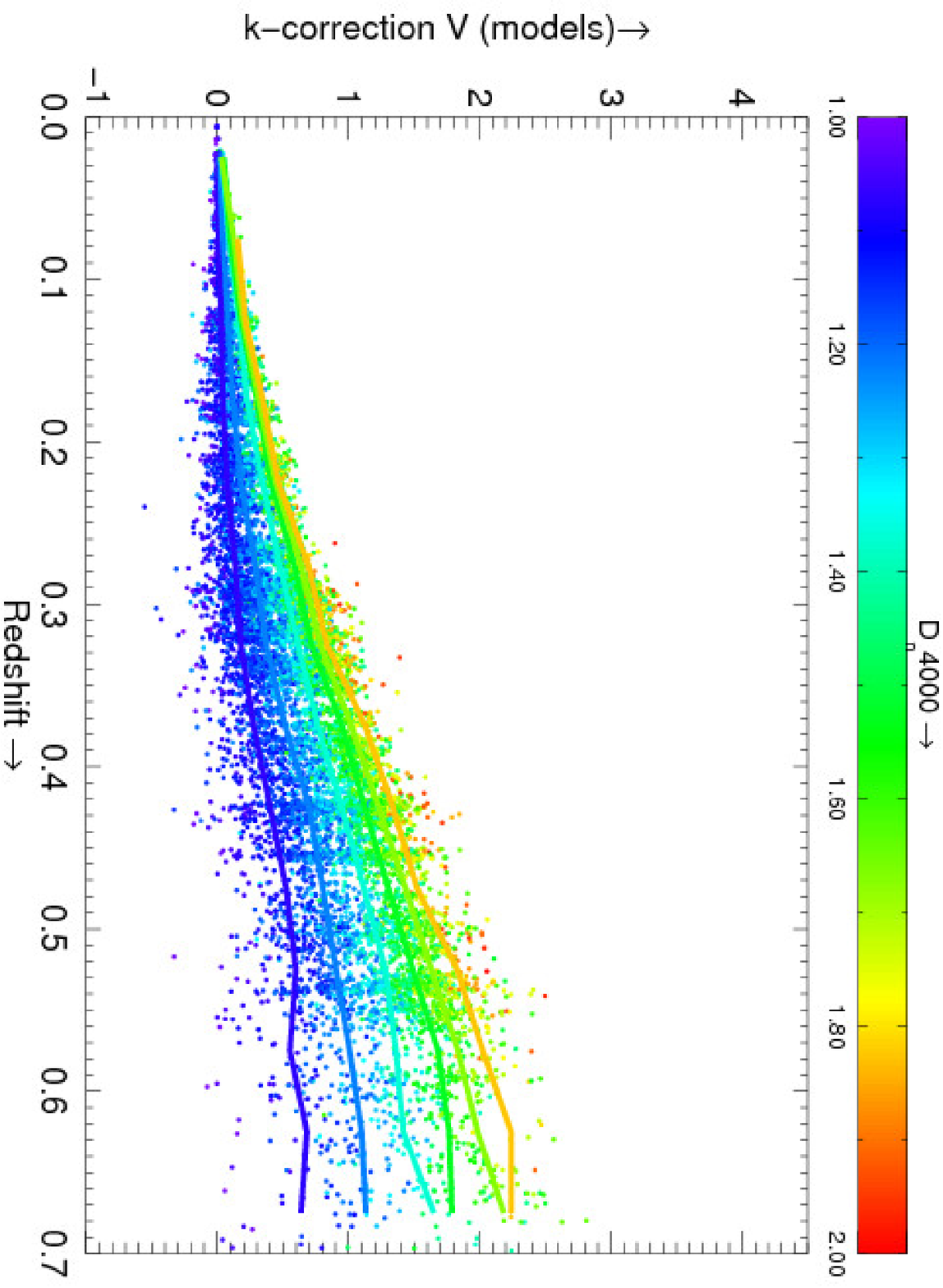}
  \includeIDLfigPcustom[0.35\textwidth]{13pt}{12pt}{5pt}{5pt}{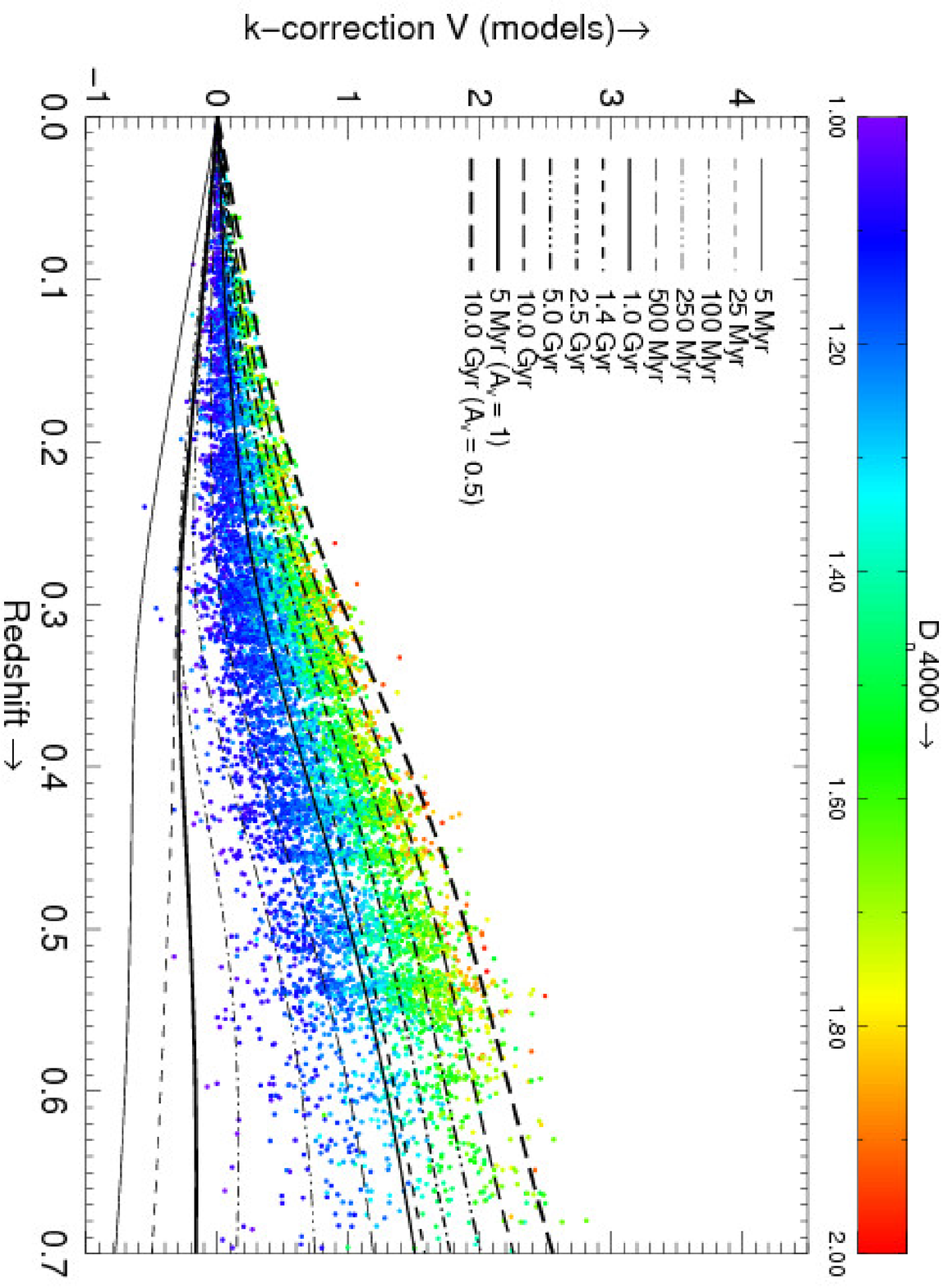}
  \includeIDLfigPcustom[0.70\textwidth]{130pt}{0pt}{158pt}{25pt}{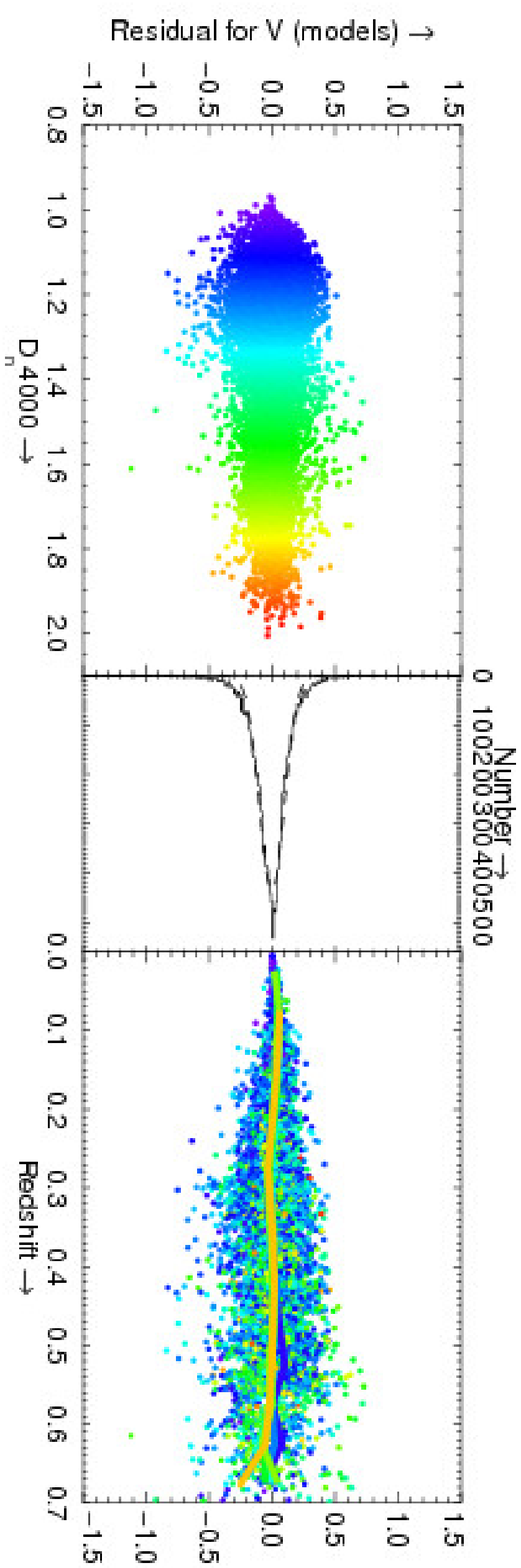}
  \caption{The k-corrections derived from the model fits ({\it top})
  for the SDSS $u$ band. The colored solid lines indicate the median
  of the k-correction binned by \dn{} as a function of redshift ({\it
  left}) and the black lines indicate the k-corrections from models shown in
  Figure~\ref{fig:cb07models} ({\it right}). Assessment of the
  accuracy of the analytic approximations for the k-corrections ({\it
  bottom}); from left to right: the residuals from the surface fitting
  as a function of \dn{}, the distribution of residuals ({\it solid
  line}) and overplotted a Gaussian fit to the distribution ({\it
  dashed line}), and the residuals as a function of redshift where the
  colored solid lines indicate the median of the residual binned by
  \dn{}. The bottom two rows are the same as the top two rows but for
  the Johnson-Cousins $V$ band.}
  \label{fig:sdss_u}
\end{figure*}

We apply this method to the data presented in this article. The variables
$x_k$, $y_k$, and $z_k$ represent the redshift, \dn{} (or \grCol{},
depending on which of the two we fit), and the k-correction,
respectively, for each individual galaxy. $N_g$ is the number of
galaxies. We show the residuals of our fits in
Figures~\ref{fig:residuals_d4000_g}, \ref{fig:residuals_gr_g},
and \ref{fig:sdss_u} and online (see footnote 11).

Tables~\ref{app:coefffirst_d}--\ref{tab:mod_kx_z_col} contain the
coefficients for the fits described in this article \citepalias[similar
to][]{Chilingarian10}. One calculates the k-correction by taking the
index of the row multiplied with the index of the column (i.e., $z^i
(\dn{})^j$, or $z^i \grCol^j$), and multiplying that with the coefficient
($a_{ij}$) in the table; one needs to do this for all coefficients in
the table. The sum of those values is the k-correction for the given
redshift and \dn{} (or \grCol{} where appropriate). Note that no table
contains a row with $z^0$ because we demand, by definition, that at
redshift $z=0$ the k-correction is 0 (Section~\ref{sec:intro}).

\bibliographystyle{apj}
\begin{footnotesize}
\bibliography{ms.bib}

\begin{thebibliography}{39}
\expandafter\ifx\csname natexlab\endcsname\relax\def\natexlab#1{#1}\fi

\bibitem[{{Abazajian} {et~al.}(2009){Abazajian}, {Adelman-McCarthy},
  {Ag{\"u}eros}, {Allam}, {Allende Prieto}, {An}, {Anderson}, {Anderson},
  {Annis}, {Bahcall}, {Bailer-Jones}, {Barentine}, {Bassett}, {Becker}, \&
  {et~al. (SDSS collaboration)}}]{SDSS7}
{Abazajian}, K.~N., {Adelman-McCarthy}, J.~K., {Ag{\"u}eros}, M.~A., {Allam},
  S.~S., {Allende Prieto}, C., {An}, D., {Anderson}, K.~S.~J., {Anderson},
  S.~F., {Annis}, J., {Bahcall}, N.~A., {Bailer-Jones}, C.~A.~L., {Barentine},
  J.~C., {Bassett}, B.~A., {Becker}, A.~C., \& {et~al. (SDSS collaboration)}.
  2009, \apjs, 182, 543

\bibitem[{Anderson {et~al.}(1999)Anderson, Bai, Bischof, Blackford, Demmel,
  Dongarra, Du~Croz, Greenbaum, Hammarling, McKenney, \& Sorensen}]{lapack}
Anderson, E., Bai, Z., Bischof, C., Blackford, S., Demmel, J., Dongarra, J.,
  Du~Croz, J., Greenbaum, A., Hammarling, S., McKenney, A., \& Sorensen, D.
  1999, {LAPACK} Users' Guide, 3rd edn. (Philadelphia, PA: Society for
  Industrial and Applied Mathematics)

\bibitem[{{Balogh} {et~al.}(1999){Balogh}, {Morris}, {Yee}, {Carlberg}, \&
  {Ellingson}}]{Balogh99}
{Balogh}, M.~L., {Morris}, S.~L., {Yee}, H.~K.~C., {Carlberg}, R.~G., \&
  {Ellingson}, E. 1999, \apj, 527, 54

\bibitem[{{Bessell}(1990)}]{Bessell90}
{Bessell}, M.~S. 1990, \pasp, 102, 1181

\bibitem[{{Blanton} {et~al.}(2001){Blanton}, {Dalcanton}, {Eisenstein},
  {Loveday}, {Strauss}, {SubbaRao}, {Weinberg}, {Anderson}, {Annis}, {Bahcall},
  {Bernardi}, {Brinkmann}, {Brunner}, {Burles}, {Carey}, {Castander},
  {Connolly}, {Csabai}, {Doi}, {Finkbeiner}, {Friedman}, {Frieman}, {Fukugita},
  {Gunn}, {Hennessy}, {Hindsley}, {Hogg}, {Ichikawa}, {Ivezi{\'c}}, {Kent},
  {Knapp}, {Lamb}, {Leger}, {Long}, {Lupton}, {McKay}, {Meiksin}, {Merelli},
  {Munn}, {Narayanan}, {Newcomb}, {Nichol}, {Okamura}, {Owen}, {Pier}, {Pope},
  {Postman}, {Quinn}, {Rockosi}, {Schlegel}, {Schneider}, {Shimasaku},
  {Siegmund}, {Smee}, {Snir}, {Stoughton}, {Stubbs}, {Szalay}, {Szokoly},
  {Thakar}, {Tremonti}, {Tucker}, {Uomoto}, {Vanden Berk}, {Vogeley},
  {Waddell}, {Yanny}, {Yasuda}, \& {York}}]{Blanton01}
{Blanton}, M.~R., {Dalcanton}, J., {Eisenstein}, D., {Loveday}, J., {Strauss},
  M.~A., {SubbaRao}, M., {Weinberg}, D.~H., {Anderson}, Jr., J.~E., {Annis},
  J., {Bahcall}, N.~A., {Bernardi}, M., {Brinkmann}, J., {Brunner}, R.~J.,
  {Burles}, S., {Carey}, L., {Castander}, F.~J., {Connolly}, A.~J., {Csabai},
  I., {Doi}, M., {Finkbeiner}, D., {Friedman}, S., {Frieman}, J.~A.,
  {Fukugita}, M., {Gunn}, J.~E., {Hennessy}, G.~S., {Hindsley}, R.~B., {Hogg},
  D.~W., {Ichikawa}, T., {Ivezi{\'c}}, {\v Z}., {Kent}, S., {Knapp}, G.~R.,
  {Lamb}, D.~Q., {Leger}, R.~F., {Long}, D.~C., {Lupton}, R.~H., {McKay},
  T.~A., {Meiksin}, A., {Merelli}, A., {Munn}, J.~A., {Narayanan}, V.,
  {Newcomb}, M., {Nichol}, R.~C., {Okamura}, S., {Owen}, R., {Pier}, J.~R.,
  {Pope}, A., {Postman}, M., {Quinn}, T., {Rockosi}, C.~M., {Schlegel}, D.~J.,
  {Schneider}, D.~P., {Shimasaku}, K., {Siegmund}, W.~A., {Smee}, S., {Snir},
  Y., {Stoughton}, C., {Stubbs}, C., {Szalay}, A.~S., {Szokoly}, G.~P.,
  {Thakar}, A.~R., {Tremonti}, C., {Tucker}, D.~L., {Uomoto}, A., {Vanden
  Berk}, D., {Vogeley}, M.~S., {Waddell}, P., {Yanny}, B., {Yasuda}, N., \&
  {York}, D.~G. 2001, \aj, 121, 2358

\bibitem[{{Blanton} \& {Roweis}(2007)}]{Blanton07}
{Blanton}, M.~R. \& {Roweis}, S. 2007, \aj, 133, 734

\bibitem[{{Brown} {et~al.}(2008){Brown}, {Kewley}, \& {Geller}}]{Brown08}
{Brown}, W.~R., {Kewley}, L.~J., \& {Geller}, M.~J. 2008, \aj, 135, 92

\bibitem[{Bruzual(2006)}]{Bruzual07}
Bruzual, A.~G. 2006, Proceedings of the International Astronomical Union, 2,
  125

\bibitem[{{Bruzual}(1983)}]{Bruzual83}
{Bruzual}, G. 1983, \apj, 273, 105

\bibitem[{{Bruzual} \& {Charlot}(2003)}]{Bruzual03}
{Bruzual}, G. \& {Charlot}, S. 2003, \mnras, 344, 1000

\bibitem[{{Calzetti} {et~al.}(2000){Calzetti}, {Armus}, {Bohlin}, {Kinney},
  {Koornneef}, \& {Storchi-Bergmann}}]{Calzetti00}
{Calzetti}, D., {Armus}, L., {Bohlin}, R.~C., {Kinney}, A.~L., {Koornneef}, J.,
  \& {Storchi-Bergmann}, T. 2000, \apj, 533, 682

\bibitem[{{Charlot} \& {Fall}(2000)}]{Charlot00}
{Charlot}, S. \& {Fall}, S.~M. 2000, \apj, 539, 718

\bibitem[{{Chilingarian} {et~al.}(2010){Chilingarian}, {Melchior}, \&
  {Zolotukhin}}]{Chilingarian10}
{Chilingarian}, I., {Melchior}, A., \& {Zolotukhin}, I. 2010, ArXiv e-prints

\bibitem[{{Fabricant} {et~al.}(2005){Fabricant}, {Fata}, {Roll}, {Hertz},
  {Caldwell}, {Gauron}, {Geary}, {McLeod}, {Szentgyorgyi}, {Zajac}, {Kurtz},
  {Barberis}, {Bergner}, {Brown}, {Conroy}, {Eng}, {Geller}, {Goddard},
  {Honsa}, {Mueller}, {Mink}, {Ordway}, {Tokarz}, {Woods}, {Wyatt}, {Epps}, \&
  {Dell'Antonio}}]{Fabricant05}
{Fabricant}, D., {Fata}, R., {Roll}, J., {Hertz}, E., {Caldwell}, N., {Gauron},
  T., {Geary}, J., {McLeod}, B., {Szentgyorgyi}, A., {Zajac}, J., {Kurtz}, M.,
  {Barberis}, J., {Bergner}, H., {Brown}, W., {Conroy}, M., {Eng}, R.,
  {Geller}, M., {Goddard}, R., {Honsa}, M., {Mueller}, M., {Mink}, D.,
  {Ordway}, M., {Tokarz}, S., {Woods}, D., {Wyatt}, W., {Epps}, H., \&
  {Dell'Antonio}, I. 2005, \pasp, 117, 1411

\bibitem[{{Fabricant} {et~al.}(1998){Fabricant}, {Hertz}, {Szentgyorgyi},
  {Fata}, {Roll}, \& {Zajac}}]{Fabricant98}
{Fabricant}, D.~G., {Hertz}, E.~N., {Szentgyorgyi}, A.~H., {Fata}, R.~G.,
  {Roll}, J.~B., \& {Zajac}, J.~M. 1998, in Presented at the Society of
  Photo-Optical Instrumentation Engineers (SPIE) Conference, Vol. 3355, Society
  of Photo-Optical Instrumentation Engineers (SPIE) Conference Series, ed.
  S.~{D'Odorico}, 285--296

\bibitem[{{Fabricant} {et~al.}(2008){Fabricant}, {Kurtz}, {Geller}, {Caldwell},
  {Woods}, \& {Dell'Antonio}}]{Fabricant08}
{Fabricant}, D.~G., {Kurtz}, M.~J., {Geller}, M.~J., {Caldwell}, N., {Woods},
  D., \& {Dell'Antonio}, I. 2008, \pasp, 120, 1222

\bibitem[{{Fioc} \& {Rocca-Volmerange}(1997)}]{Fioc97}
{Fioc}, M. \& {Rocca-Volmerange}, B. 1997, \aap, 326, 950

\bibitem[{{Geller} {et~al.}(2005){Geller}, {Dell'Antonio}, {Kurtz}, {Ramella},
  {Fabricant}, {Caldwell}, {Tyson}, \& {Wittman}}]{Geller05}
{Geller}, M.~J., {Dell'Antonio}, I.~P., {Kurtz}, M.~J., {Ramella}, M.,
  {Fabricant}, D.~G., {Caldwell}, N., {Tyson}, J.~A., \& {Wittman}, D. 2005,
  \apjl, 635, L125

\bibitem[{{Geller} {et~al.}(2010){Geller}, {Kurtz}, {Dell'Antonio}, {Ramella},
  \& {Fabricant}}]{Geller10}
{Geller}, M.~J., {Kurtz}, M.~J., {Dell'Antonio}, I.~P., {Ramella}, M., \&
  {Fabricant}, D.~G. 2010, \apj, 709, 832

\bibitem[{{Hogg} {et~al.}(2002){Hogg}, {Baldry}, {Blanton}, \&
  {Eisenstein}}]{Hogg02}
{Hogg}, D.~W., {Baldry}, I.~K., {Blanton}, M.~R., \& {Eisenstein}, D.~J. 2002,
  ArXiv Astrophysics e-prints

\bibitem[{{Hubble}(1936)}]{Hubble36}
{Hubble}, E. 1936, \apj, 84, 517

\bibitem[{{Kauffmann} {et~al.}(2003){Kauffmann}, {Heckman}, {White}, {Charlot},
  {Tremonti}, {Brinchmann}, {Bruzual}, {Peng}, {Seibert}, {Bernardi},
  {Blanton}, {Brinkmann}, {Castander}, {Cs{\'a}bai}, {Fukugita}, {Ivezic},
  {Munn}, {Nichol}, {Padmanabhan}, {Thakar}, {Weinberg}, \&
  {York}}]{Kauffmann03}
{Kauffmann}, G., {Heckman}, T.~M., {White}, S.~D.~M., {Charlot}, S.,
  {Tremonti}, C., {Brinchmann}, J., {Bruzual}, G., {Peng}, E.~W., {Seibert},
  M., {Bernardi}, M., {Blanton}, M., {Brinkmann}, J., {Castander}, F.,
  {Cs{\'a}bai}, I., {Fukugita}, M., {Ivezic}, Z., {Munn}, J.~A., {Nichol},
  R.~C., {Padmanabhan}, N., {Thakar}, A.~R., {Weinberg}, D.~H., \& {York}, D.
  2003, \mnras, 341, 33

\bibitem[{{Kewley} {et~al.}(2005){Kewley}, {Jansen}, \& {Geller}}]{Kewley05}
{Kewley}, L.~J., {Jansen}, R.~A., \& {Geller}, M.~J. 2005, \pasp, 117, 227

\bibitem[{{Kurtz} {et~al.}(2007){Kurtz}, {Geller}, {Fabricant}, {Wyatt}, \&
  {Dell'Antonio}}]{Kurtz07}
{Kurtz}, M.~J., {Geller}, M.~J., {Fabricant}, D.~G., {Wyatt}, W.~F., \&
  {Dell'Antonio}, I.~P. 2007, \aj, 134, 1360

\bibitem[{{Kurtz} \& {Mink}(1998)}]{Kurtz98}
{Kurtz}, M.~J. \& {Mink}, D.~J. 1998, \pasp, 110, 934

\bibitem[{{Lawrence} {et~al.}(2007){Lawrence}, {Warren}, {Almaini}, {Edge},
  {Hambly}, {Jameson}, {Lucas}, {Casali}, {Adamson}, {Dye}, {Emerson},
  {Foucaud}, {Hewett}, {Hirst}, {Hodgkin}, {Irwin}, {Lodieu}, {McMahon},
  {Simpson}, {Smail}, {Mortlock}, \& {Folger}}]{UKIDSS5}
{Lawrence}, A., {Warren}, S.~J., {Almaini}, O., {Edge}, A.~C., {Hambly}, N.~C.,
  {Jameson}, R.~F., {Lucas}, P., {Casali}, M., {Adamson}, A., {Dye}, S.,
  {Emerson}, J.~P., {Foucaud}, S., {Hewett}, P., {Hirst}, P., {Hodgkin}, S.~T.,
  {Irwin}, M.~J., {Lodieu}, N., {McMahon}, R.~G., {Simpson}, C., {Smail}, I.,
  {Mortlock}, D., \& {Folger}, M. 2007, \mnras, 379, 1599

\bibitem[{{Mink} {et~al.}(2007){Mink}, {Wyatt}, {Caldwell}, {Conroy}, {Furesz},
  \& {Tokarz}}]{Mink07}
{Mink}, D.~J., {Wyatt}, W.~F., {Caldwell}, N., {Conroy}, M.~A., {Furesz}, G.,
  \& {Tokarz}, S.~P. 2007, in Astronomical Society of the Pacific Conference
  Series, Vol. 376, Astronomical Data Analysis Software and Systems XVI, ed.
  R.~A. {Shaw}, F.~{Hill}, \& D.~J. {Bell}, 249--+

\bibitem[{{Muller} {et~al.}(1998){Muller}, {Reed}, {Armandroff}, {Boroson}, \&
  {Jacoby}}]{Muller98}
{Muller}, G.~P., {Reed}, R., {Armandroff}, T., {Boroson}, T.~A., \& {Jacoby},
  G.~H. 1998, in Presented at the Society of Photo-Optical Instrumentation
  Engineers (SPIE) Conference, Vol. 3355, Society of Photo-Optical
  Instrumentation Engineers (SPIE) Conference Series, ed. S.~{D'Odorico},
  577--585

\bibitem[{{Oke} \& {Gunn}(1983)}]{Oke83}
{Oke}, J.~B. \& {Gunn}, J.~E. 1983, \apj, 266, 713

\bibitem[{{Pascual} {et~al.}(2007){Pascual}, {Gallego}, \&
  {Zamorano}}]{Pascual07}
{Pascual}, S., {Gallego}, J., \& {Zamorano}, J. 2007, \pasp, 119, 30

\bibitem[{{Roll} {et~al.}(1998){Roll}, {Fabricant}, \& {McLeod}}]{Roll98}
{Roll}, J.~B., {Fabricant}, D.~G., \& {McLeod}, B.~A. 1998, in Society of
  Photo-Optical Instrumentation Engineers (SPIE) Conference Series, Vol. 3355,
  Society of Photo-Optical Instrumentation Engineers (SPIE) Conference Series,
  ed. S.~{D'Odorico}, 324--332

\bibitem[{{Steidel} {et~al.}(1995){Steidel}, {Pettini}, \&
  {Hamilton}}]{Steidel95}
{Steidel}, C.~C., {Pettini}, M., \& {Hamilton}, D. 1995, \aj, 110, 2519

\bibitem[{{Tremonti} {et~al.}(2004){Tremonti}, {Heckman}, {Kauffmann},
  {Brinchmann}, {Charlot}, {White}, {Seibert}, {Peng}, {Schlegel}, {Uomoto},
  {Fukugita}, \& {Brinkmann}}]{Tremonti04}
{Tremonti}, C.~A., {Heckman}, T.~M., {Kauffmann}, G., {Brinchmann}, J.,
  {Charlot}, S., {White}, S.~D.~M., {Seibert}, M., {Peng}, E.~W., {Schlegel},
  D.~J., {Uomoto}, A., {Fukugita}, M., \& {Brinkmann}, J. 2004, \apj, 613, 898

\bibitem[{{Waller}(1990)}]{Waller90}
{Waller}, W.~H. 1990, \pasp, 102, 1217

\bibitem[{{Westra} {et~al.}(2010){Westra}, {Geller}, {Kurtz}, {Fabricant}, \&
  {Dell'Antonio}}]{Westra10}
{Westra}, E., {Geller}, M.~J., {Kurtz}, M.~J., {Fabricant}, D.~G., \&
  {Dell'Antonio}, I. 2010, \apj, 708, 534

\bibitem[{{Wittman} {et~al.}(2006){Wittman}, {Dell'Antonio}, {Hughes},
  {Margoniner}, {Tyson}, {Cohen}, \& {Norman}}]{Wittman06}
{Wittman}, D., {Dell'Antonio}, I.~P., {Hughes}, J.~P., {Margoniner}, V.~E.,
  {Tyson}, J.~A., {Cohen}, J.~G., \& {Norman}, D. 2006, \apj, 643, 128

\bibitem[{{Wittman} {et~al.}(2002){Wittman}, {Tyson}, {Dell'Antonio}, {Becker},
  {Margoniner}, {Cohen}, {Norman}, {Loomba}, {Squires}, {Wilson}, {Stubbs},
  {Hennawi}, {Spergel}, {Boeshaar}, {Clocchiatti}, {Hamuy}, {Bernstein},
  {Gonzalez}, {Guhathakurta}, {Hu}, {Seljak}, \& {Zaritsky}}]{Wittman02}
{Wittman}, D.~M., {Tyson}, J.~A., {Dell'Antonio}, I.~P., {Becker}, A.,
  {Margoniner}, V., {Cohen}, J.~G., {Norman}, D., {Loomba}, D., {Squires}, G.,
  {Wilson}, G., {Stubbs}, C.~W., {Hennawi}, J., {Spergel}, D.~N., {Boeshaar},
  P., {Clocchiatti}, A., {Hamuy}, M., {Bernstein}, G., {Gonzalez}, A.,
  {Guhathakurta}, P., {Hu}, W., {Seljak}, U., \& {Zaritsky}, D. 2002, in
  Society of Photo-Optical Instrumentation Engineers (SPIE) Conference Series,
  Vol. 4836, Society of Photo-Optical Instrumentation Engineers (SPIE)
  Conference Series, ed. J.~A. {Tyson} \& S.~{Wolff}, 73--82

\bibitem[{{Woods} {et~al.}(2010){Woods}, {Geller}, {Kurtz}, {Westra},
  {Fabricant}, \& {Dell'Antonio}}]{Woods10}
{Woods}, D.~F., {Geller}, M.~J., {Kurtz}, M.~J., {Westra}, E., {Fabricant},
  D.~G., \& {Dell'Antonio}, I. 2010, \aj, 139, 1857

\bibitem[{{Yasuda} {et~al.}(2001){Yasuda}, {Fukugita}, {Narayanan}, {Lupton},
  {Strateva}, {Strauss}, {Ivezi{\'c}}, {Kim}, {Hogg}, {Weinberg}, {Shimasaku},
  {Loveday}, {Annis}, {Bahcall}, {Blanton}, {Brinkmann}, {Brunner}, {Connolly},
  {Csabai}, {Doi}, {Hamabe}, {Ichikawa}, {Ichikawa}, {Johnston}, {Knapp},
  {Kunszt}, {Lamb}, {McKay}, {Munn}, {Nichol}, {Okamura}, {Schneider},
  {Szokoly}, {Vogeley}, {Watanabe}, \& {York}}]{Yasuda01}
{Yasuda}, N., {Fukugita}, M., {Narayanan}, V.~K., {Lupton}, R.~H., {Strateva},
  I., {Strauss}, M.~A., {Ivezi{\'c}}, {\v Z}., {Kim}, R.~S.~J., {Hogg}, D.~W.,
  {Weinberg}, D.~H., {Shimasaku}, K., {Loveday}, J., {Annis}, J., {Bahcall},
  N.~A., {Blanton}, M., {Brinkmann}, J., {Brunner}, R.~J., {Connolly}, A.~J.,
  {Csabai}, I., {Doi}, M., {Hamabe}, M., {Ichikawa}, S., {Ichikawa}, T.,
  {Johnston}, D.~E., {Knapp}, G.~R., {Kunszt}, P.~Z., {Lamb}, D.~Q., {McKay},
  T.~A., {Munn}, J.~A., {Nichol}, R.~C., {Okamura}, S., {Schneider}, D.~P.,
  {Szokoly}, G.~P., {Vogeley}, M.~S., {Watanabe}, M., \& {York}, D.~G. 2001,
  \aj, 122, 1104

\end{thebibliography}
\end{footnotesize}

{\it Facilities:} \facility{MMT (Hectospec)}

\begin{deluxetable}{rcc}
  \tablewidth{0pt}

  \tablecaption{\dn{} determined from the models.\label{tab:d4000models}}

  \tablehead{Age & $A_V$ & \dn}
  \startdata
     5\,\Myr & 0.0 & 0.94 \\
    25\,\Myr & 0.0 & 0.97 \\
   100\,\Myr & 0.0 & 1.05 \\
   250\,\Myr & 0.0 & 1.13 \\
   500\,\Myr & 0.0 & 1.26 \\
   1.0\,\Gyr & 0.0 & 1.41 \\
   1.4\,\Gyr & 0.0 & 1.45 \\
   2.5\,\Gyr & 0.0 & 1.58 \\
   5.0\,\Gyr & 0.0 & 1.77 \\
  10.0\,\Gyr & 0.0 & 1.96 \\
     5\,\Myr & 1.0 & 0.98 \\
  10.0\,\Gyr & 0.5 & 2.00 \\
  \enddata

  \tablecomments{These models are indicated in
  Fig.~\ref{fig:modeltracks}.}
\end{deluxetable}

\begin{deluxetable}{ccccccc}
  \tablewidth{0pt}

  \tablecaption{Statistics for the analytic approximations for $g$ and
  $r$.\label{tab:stats}}

  \tablehead{& \multicolumn{3}{c}{\dn} & \multicolumn{3}{c}{\grCol}\\\multicolumn{1}{c}{filter} & $\sigma$ & rms & 68.3\,\% & $\sigma$ & rms & 68.3\,\%}
  \startdata
  \multicolumn{7}{c}{Spectra}\\
  $g$ & 0.08 & 0.17 & 0.11 & 0.12 & 0.25 & 0.17\\
  $r$ & 0.08 & 0.12 & 0.09 & 0.07 & 0.11 & 0.08\\
  \multicolumn{7}{c}{Models}\\
  $g$ & 0.09 & 0.14 & 0.11 & 0.12 & 0.21 & 0.16\\
  $r$ & 0.07 & 0.10 & 0.08 & 0.06 & 0.09 & 0.07\\
  \enddata

  \tablecomments{$\sigma$ is the $\sigma$ of the Gaussian fit to the
   residuals for the analytic approximations to the k-corrections
   derived from the spectra and the models for $g$ and $r$ with $z \le
   0.68$ and 0.33, respectively, as a function of \dn{} and
   \grCol{}; rms is the rms of the residuals; 68.3\,\% is
   the range around the mean of the residuals.
   Figures~\ref{fig:residuals_d4000_g} and \ref{fig:residuals_gr_g}
   show the residuals and the Gaussian fits to the residuals.}
\end{deluxetable}

\begin{deluxetable}{cccccccccccccccccc}
  \rotate
  \tabletypesize{\footnotesize}
  \tablewidth{0pt}

  \tablecaption{Example uncertainties for typical values and
  uncertainties in \dn{} and corresponding
  \grCol{}.\label{tab:exampleunc}}

  \tablehead{filter & $z$ & \dn{} & $\sigma_\mathrm{petro}$ & $\sigma_\mathrm{fib}$ & $k_\dn$ & $\sigma_\dn$ & $\sigma_k$ & $\sigma_\dn$ & $\sigma_k$ & $\sigma_\dn$ & $\sigma_k$ & \grCol{} & $k_{(g-r)}$ & $\sigma_{(g-r),\mathrm{petro}}$ & $\sigma_k$ & $\sigma_{(g-r),\mathrm{fib}}$ & $\sigma_k$}
  \startdata
  $g$ & 0.30 & 1.25 & 0.15 & 0.09 & 0.61 & 0.21 & $^{+0.33} _{-0.61}$ & 0.11 & $^{+0.20} _{-0.27}$ & 0.07 & $^{+0.13} _{-0.16}$ & 0.97 & 0.61 & 0.17 & $^{+0.16} _{-0.16}$ & 0.10 & $^{+0.10} _{-0.09}$\\
  $g$ & 0.30 & 1.65 & 0.18 & 0.11 & 1.15 & 0.21 & $^{+0.20} _{-0.18}$ & 0.11 & $^{+0.09} _{-0.09}$ & 0.07 & $^{+0.06} _{-0.06}$ & 1.48 & 1.12 & 0.20 & $^{+0.10} _{-0.15}$ & 0.11 & $^{+0.06} _{-0.08}$\\
	     		    
  $r$ & 0.30 & 1.25 & 0.09 & 0.06 & 0.21 & 0.21 & $^{+0.14} _{-0.33}$ & 0.11 & $^{+0.09} _{-0.14}$ & 0.07 & $^{+0.06} _{-0.08}$ & 0.97 & 0.21 & 0.17 & $^{+0.08} _{-0.10}$ & 0.10 & $^{+0.05} _{-0.06}$\\
  $r$ & 0.30 & 1.65 & 0.08 & 0.04 & 0.41 & 0.21 & $^{+0.03} _{-0.04}$ & 0.11 & $^{+0.01} _{-0.01}$ & 0.07 & $^{+0.01} _{-0.01}$ & 1.48 & 0.42 & 0.20 & $^{+0.02} _{-0.05}$ & 0.11 & $^{+0.02} _{-0.02}$\\

  $g$ & 0.55 & 1.25 & 0.29 & 0.15 & 1.08 & 0.21 & $^{+0.65} _{-0.99}$ & 0.11 & $^{+0.37} _{-0.47}$ & 0.07 & $^{+0.25} _{-0.28}$ & 1.04 & 1.29 & 0.32 & $^{+0.45} _{-0.51}$ & 0.17 & $^{+0.25} _{-0.27}$\\
  $g$ & 0.55 & 1.65 & 0.47 & 0.24 & 2.24 & 0.21 & $^{+0.41} _{-0.44}$ & 0.11 & $^{+0.21} _{-0.22}$ & 0.07 & $^{+0.13} _{-0.14}$ & 1.65 & 2.06 & 0.49 & $^{+0.24} _{-0.59}$ & 0.25 & $^{+0.18} _{-0.27}$\\

  \enddata

  \tablecomments{$\sigma_\mathrm{petro}$ and $\sigma_\mathrm{fib}$ are
  the typical for the SDSS Petrosian and fiber magnitude
  of a galaxy with chosen \dn{} and redshift $z$. The uncertainties
  $\sigma_\dn$ are those typical for very low, moderate and high S/N
  spectra. $\sigma_k$ is the range in the calculated k-correction
  corresponding to the uncertainty in \dn{} (or \grCol{}). \grCol{} is
  the typical galaxy color for the chosen \dn{} and
  redshift, and $\sigma_{(g-r),\mathrm{petro}}$ and
  $\sigma_{(g-r),\mathrm{fib}}$ are the typical uncertainties in the
  SDSS Petrosian and fiber color, respectively, for such a galaxy.}

\end{deluxetable}

\begin{deluxetable}{cccccccc}
  \tablewidth{0pt}

  \tablecaption{Statistics for the analytic approximations for $UBVRI$
  and $ugriz$.\label{tab:sigmas}}

  \tablehead{& \multicolumn{3}{c}{\dn} & & \multicolumn{3}{c}{Color}\\\multicolumn{1}{c}{filter} & $\sigma$ & rms & 68.3\,\% & color & $\sigma$ & rms & 68.3\,\%}
  \startdata
  $u$ & 0.16 & 0.30 & 0.22 & $(u-r)$ & 0.30 & 0.48 & 0.55 \\
  $g$ & 0.09 & 0.11 & 0.14 & $(g-r)$ & 0.12 & 0.22 & 0.16 \\
  $r$ & 0.10 & 0.15 & 0.12 & $(g-r)$ & 0.09 & 0.17 & 0.12 \\
  $i$ & 0.10 & 0.15 & 0.12 & $(g-i)$ & 0.09 & 0.14 & 0.11 \\
  $z$ & 0.09 & 0.14 & 0.11 & $(r-z)$ & 0.09 & 0.14 & 0.11 \\
  $U$ & 0.16 & 0.30 & 0.22 & \ldots  & \ldots & \ldots & \ldots\\
  $B$ & 0.10 & 0.17 & 0.13 & \ldots  & \ldots & \ldots & \ldots\\
  $V$ & 0.10 & 0.14 & 0.12 & \ldots  & \ldots & \ldots & \ldots\\
  $R$ & 0.10 & 0.15 & 0.12 & \ldots  & \ldots & \ldots & \ldots\\
  $I$ & 0.09 & 0.15 & 0.12 & \ldots  & \ldots & \ldots & \ldots\\
  \enddata

  \tablecomments{The table headings are the same as
  Table~\ref{tab:stats}. The figures at
  \url{http://tdc-www.cfa.harvard.edu/instruments/hectospec/progs/EOK/figures.shtml}
  show the residuals and the Gaussian fits to the
  residuals. Figure~\ref{fig:sdss_u} shows the figures for $u$ and
  $V$.}
\end{deluxetable}

\clearpage

\setcounter{table}{0}
\renewcommand*\thetable{\Alph{section}.\arabic{table}}

\begin{kxtable}{Coefficients for the SDSS $g$ band as a function of redshift and \dn{} with k-corrections derived from the spectra.\label{app:coefffirst_d}}{(\dn)}
$z^1$ & 29.9754 & -78.4292 & 64.0436 & -15.8884 \\
$z^2$ & -368.572 & 818.057 & -581.89 & 134.427 \\
$z^3$ & 490.834 & -1084.35 & 768.318 & -176.591 \\
\end{kxtable}

\begin{kxtable}{Coefficients for the SDSS $g$ band as a function of redshift and \dn{} with k-corrections derived from the models.}{(\dn)}
$z^1$ & 6.6812 & -27.6858 & 27.7299 & -7.319 \\
$z^2$ & -231.313 & 517.183 & -364.586 & 82.7252 \\
$z^3$ & 330.803 & -735.254 & 517.408 & -117.366 \\
\end{kxtable}

\begin{kxtable}{Coefficients for the SDSS $r$ band as a function of redshift and \dn{} with k-corrections derived from the spectra.}{(\dn)}
$z^1$ & 25.4725 & -52.2513 & 35.6903 & -7.85585 \\
$z^2$ & -177.822 & 336.276 & -203.045 & 39.6419 \\
$z^3$ & 66.6927 & -82.7486 & 7.78744 & 9.65736 \\
\end{kxtable}

\begin{kxtable}{Coefficients for the SDSS $r$ band as a function of redshift and \dn{} with k-corrections derived from the models.\label{app:coefflast_d}}{(\dn)}
$z^1$ & -2.4549 & 3.08855 & -0.85258 & 0.169083 \\
$z^2$ & 31.7806 & -96.6607 & 92.3012 & -26.9024 \\
$z^3$ & -334.912 & 776.204 & -593.901 & 148.071 \\
\end{kxtable}

\begin{kxtable}{Coefficients for the SDSS $g$ band as a function of redshift and \grCol{} with k-corrections derived from the spectra.\label{app:coefffirst_c}}{\grCol}
$z^1$ & 0.126452 & 0.0619892 & 6.90796 & -2.98031 \\
$z^2$ & 3.51286 & -9.40049 & -9.17117 & 6.52303 \\
$z^3$ & -8.56012 & 24.3121 & -2.88499 & -3.57257 \\
\end{kxtable}

\begin{kxtable}{Coefficients for the SDSS $g$ band as a function of redshift and \grCol{} with k-corrections derived from the model fits.}{\grCol}
$z^1$ & -0.160448 & 1.12881 & 5.93095 & -2.68308 \\
$z^2$ & 4.04345 & -11.8962 & -6.0987 & 5.37315 \\
$z^3$ & -9.01521 & 26.6088 & -6.50937 & -2.17746 \\
\end{kxtable}

\begin{kxtable}{Coefficients for the SDSS $r$ band as a function of redshift and \grCol{} with k-corrections derived from the spectra.}{\grCol}
$z^1$ & 2.30349 & -7.69531 & 11.1784 & -4.33398 \\
$z^2$ & -20.1966 & 77.9084 & -88.5243 & 31.1532 \\
$z^3$ & 21.5438 & -130.494 & 156.555 & -54.8953 \\
\end{kxtable}

\begin{kxtable}{Coefficients for the SDSS $r$ band as a function of redshift and \grCol{} with k-corrections derived from the model fits.\label{app:coefflast_c}}{\grCol}
$z^1$ & -0.823562 & 1.8179 & 1.8466 & -1.21972 \\
$z^2$ & 6.54473 & -13.4719 & 3.54192 & 1.11608 \\
$z^3$ & -28.7761 & 59.0299 & -39.6835 & 8.96288 \\
\end{kxtable}

\clearpage
\begin{kxtable}{Coefficients for the SDSS $u$ band as a function of redshift ($\le 0.7$) and \dn{} with k-corrections derived from the models.\label{tab:mod_kx_u}}{(\dn)}
$z^1$ & 2.43154 & -16.3322 & 20.0889 & -6.24137 \\
$z^2$ & -52.0421 & 120.631 & -101.738 & 31.5506 \\
$z^3$ & 147.367 & -326.482 & 247.473 & -66.0534 \\
\end{kxtable}

\begin{kxtable}{Coefficients for the SDSS $g$ band as a function of redshift ($\le 0.7$) and \dn{} with k-corrections derived from the models.\label{tab:mod_kx_g}}{(\dn)}
$z^1$ & 7.12187 & -28.8064 & 28.6961 & -7.57514 \\
$z^2$ & -234.056 & 523.817 & -370.133 & 84.1717 \\
$z^3$ & 333.917 & -742.667 & 523.643 & -118.999 \\
\end{kxtable}

\begin{kxtable}{Coefficients for the SDSS $r$ band as a function of redshift ($\le 0.7$) and \dn{} with k-corrections derived from the models.\label{tab:mod_kx_r}}{(\dn)}
$z^1$ & 3.90822 & -11.2489 & 10.1242 & -2.60069 \\
$z^2$ & -134.442 & 274.564 & -182.224 & 39.2976 \\
$z^3$ & 165.269 & -339.815 & 227.173 & -48.4853 \\
\end{kxtable}

\begin{kxtable}{Coefficients for the SDSS $i$ band as a function of redshift ($\le 0.7$) and \dn{} with k-corrections derived from the models.\label{tab:mod_kx_i}}{(\dn)}
$z^1$ & 24.114 & -53.0651 & 38.4687 & -8.97268 \\
$z^2$ & -231.287 & 488.825 & -334.771 & 74.5591 \\
$z^3$ & 292.203 & -618.56 & 423.008 & -93.5608 \\
\end{kxtable}

\begin{kxtable}{Coefficients for the SDSS $z$ band as a function of redshift ($\le 0.7$) and \dn{} with k-corrections derived from the models.\label{tab:mod_kx_z}}{(\dn)}
$z^1$ & 14.5525 & -32.5735 & 24.1581 & -5.70902 \\
$z^2$ & -203.691 & 433.748 & -298.226 & 66.4327 \\
$z^3$ & 272.826 & -580.065 & 397.471 & -87.9759 \\
\end{kxtable}

\begin{kxtable}{Coefficients for the Johnson-Cousins $U$ band as a function of redshift ($\le 0.7$) and \dn{} with k-corrections derived from the models.\label{tab:mod_kx_U}}{(\dn)}
$z^1$ & -9.40727 & 10.8489 & 0.644962 & -1.85733 \\
$z^2$ & -16.6686 & 36.2742 & -38.3145 & 16.2542 \\
$z^3$ & 113.958 & -246.016 & 185.69 & -50.5656 \\
\end{kxtable}

\begin{kxtable}{Coefficients for the Johnson-Cousins $B$ band as a function of redshift ($\le 0.7$) and \dn{} with k-corrections derived from the models.\label{tab:mod_kx_B}}{(\dn)}
$z^1$ & -30.9335 & 56.0725 & -30.7097 & 5.99606 \\
$z^2$ & -61.7054 & 140.634 & -100.957 & 22.3406 \\
$z^3$ & 165.231 & -369.545 & 261.328 & -58.2109 \\
\end{kxtable}

\begin{kxtable}{Coefficients for the Johnson-Cousins $V$ band as a function of redshift ($\le 0.7$) and \dn{} with k-corrections derived from the models.\label{tab:mod_kx_V}}{(\dn)}
$z^1$ & 12.8942 & -33.7534 & 26.7974 & -6.63462 \\
$z^2$ & -190.187 & 396.174 & -262.704 & 57.944 \\
$z^3$ & 217.218 & -449.526 & 298.181 & -65.3664 \\
\end{kxtable}

\begin{kxtable}{Coefficients for the Johnson-Cousins $R$ band as a function of redshift ($\le 0.7$) and \dn{} with k-corrections derived from the models.\label{tab:mod_kx_R}}{(\dn)}
$z^1$ & 12.4246 & -29.1592 & 22.4394 & -5.40824 \\
$z^2$ & -175.969 & 366.836 & -248.343 & 54.7488 \\
$z^3$ & 218.881 & -457.84 & 310.701 & -68.0011 \\
\end{kxtable}

\begin{kxtable}{Coefficients for the Johnson-Cousins $I$ band as a function of redshift ($\le 0.7$) and \dn{} with k-corrections derived from the models.\label{tab:mod_kx_I}}{(\dn)}
$z^1$ & 15.6639 & -34.3478 & 25.0727 & -5.82947 \\
$z^2$ & -196.21 & 412.661 & -280.861 & 61.9077 \\
$z^3$ & 256.842 & -540.797 & 367.306 & -80.3907 \\
\end{kxtable}

\begin{kxtable}{Coefficients for the SDSS $u$ band as a function of redshift ($\le 0.7$) and $(u-r)$ with k-corrections derived from the models.\label{tab:mod_kx_u_col}}{(u-r)}
$z^1$ & 2.80197 & -5.1719 & 3.01032 & -0.447618 \\
$z^2$ & -16.0821 & 25.7573 & -10.8789 & 1.60111 \\
$z^3$ & 17.9556 & -26.6165 & 11.3526 & -1.79009 \\
\end{kxtable}

\begin{kxtable}{Coefficients for the SDSS $g$ band as a function of redshift ($\le 0.7$) and $(g-r)$ with k-corrections derived from the models.\label{tab:mod_kx_g_col}}{(g-r)}
$z^1$ & 0.184129 & 1.00938 & 5.40158 & -2.41043 \\
$z^2$ & 0.112299 & -6.17196 & -7.21778 & 4.89703 \\
$z^3$ & -2.37339 & 15.3331 & -1.95179 & -2.36049 \\
\end{kxtable}

\begin{kxtable}{Coefficients for the SDSS $r$ band as a function of redshift ($\le 0.7$) and $(g-r)$ with k-corrections derived from the models.\label{tab:mod_kx_r_col}}{(g-r)}
$z^1$ & -0.196478 & 3.11481 & 0.486552 & -1.12639 \\
$z^2$ & -5.4512 & -6.547 & 1.09495 & 2.90943 \\
$z^3$ & 7.81183 & 8.41136 & -3.05571 & -2.2226 \\
\end{kxtable}

\begin{kxtable}{Coefficients for the SDSS $i$ band as a function of redshift ($\le 0.7$) and $(g-i)$ with k-corrections derived from the models.\label{tab:mod_kx_i_col}}{(g-i)}
$z^1$ & 0.0678646 & 2.034 & -1.09155 & 0.110666 \\
$z^2$ & -7.44388 & 4.05582 & 0.521655 & -0.136389 \\
$z^3$ & 10.1914 & -9.5645 & 2.07254 & -0.160024 \\
\end{kxtable}

\begin{kxtable}{Coefficients for the SDSS $z$ band as a function of redshift ($\le 0.7$) and $(r-z)$ with k-corrections derived from the models.\label{tab:mod_kx_z_col}}{(r-z)}
$z^1$ & 0.269465 & -0.496672 & 1.43921 & -0.220175 \\
$z^2$ & 0.935489 & 3.91534 & -5.03895 & 0.274978 \\
$z^3$ & -2.15467 & -3.71225 & 5.19059 & -0.292333 \\
\end{kxtable}

\clearpage

\clearpage

\end{document}